\documentclass[pdflatex,sn-mathphys-num]{sn-jnl}

\usepackage{graphicx}%
\usepackage{multirow}%
\usepackage{amsmath,amssymb,amsfonts}%
\usepackage{mathrsfs}%
\usepackage[title]{appendix}%
\usepackage{xcolor}%
\usepackage{textcomp}%
\usepackage{manyfoot}%
\usepackage{booktabs}%
\usepackage{algorithmicx}%
\usepackage{algpseudocode}%
\usepackage{listings}%
\usepackage{hyperref}
\usepackage{float}
\usepackage{verbatim} 
\usepackage{multirow}
\usepackage{booktabs}
\restylefloat{figure}
\restylefloat{table}

\usepackage{fancyhdr}
\usepackage{graphicx}
\usepackage{afterpage}
\usepackage{booktabs}
\usepackage{longtable}
\usepackage{adjustbox}
\usepackage{multicol}
\usepackage{amsmath}
\usepackage{ntheorem}
\theoremseparator{:}
\usepackage{url}
\usepackage{colortbl,longtable}
\usepackage[stable]{footmisc}
\usepackage{graphicx}
\usepackage{subcaption}
\usepackage{multirow}
\usepackage{algorithm2e}

\usepackage{amssymb}
\usepackage{xcolor}

\usepackage{listings}
\usepackage[utf8]{inputenc}
\usepackage{titlesec}
\usepackage{graphicx}
\usepackage{lipsum}
\usepackage{subcaption}
\usepackage{caption}
\usepackage{lipsum}
\usepackage{hyperref}
\usepackage{float}
\usepackage{booktabs}
\usepackage{longtable}
\usepackage{algorithm2e}
\usepackage{xcolor}
\usepackage{svg}

\theoremstyle{thmstyleone}%
\theoremstyle{thmstyletwo}%

\theoremstyle{thmstylethree}%

\begin{document}

\title[Article Title]{Exploring social bots: A feature-based approach to improve bot detection in social networks}


\author*[1,2]{\fnm{Salvador} \sur{Lopez-Joya}}\email{slopezjoya@ugr.es}

\author[1,2]{\fnm{Jose A.} \sur{Diaz-Garcia}}\email{jagarcia@decsai.ugr.es}
\author[1,2]{\fnm{M. } \sur{Dolores Ruiz}}\email{mdruiz@decsai.ugr.es}

\author[1,2]{\fnm{Maria J.} \sur{Martin-Bautista}}\email{mbautis@decsai.ugr.es}

\affil*[1]{\orgdiv{Department of Computer Science and A.I,}, \orgname{University of Granada,}, \orgaddress{\street{C. Periodista Daniel Saucedo Aranda}, \city{Granada}, \postcode{18014}, \country{Spain}}}

\affil[2]{\orgname{Research Centre for Information and Communications Technologies}, \orgaddress{\street{C. Periodista Rafael Gómez Montero}, \city{Granada}, \postcode{18014}, \country{Spain}}}


\abstract{The importance of social media in our daily lives has unfortunately led to an increase in the spread of misinformation, political messages and malicious links. One of the most popular ways of carrying out those activities is using automated accounts, also known as bots, which makes the detection of such accounts a necessity. This paper addresses that problem by investigating features based on the user account profile and its content, aiming to understand the relevance of each feature as a basis for improving future bot detectors. Through an exhaustive process of research, inference and feature selection, we are able to surpass the state of the art on several metrics using classical machine learning algorithms and identify the types of features that are most important in detecting automated accounts. }




\maketitle

\section{Introduction} 
\label{Introduction}
The proliferation of social media has undeniably transformed our daily lives, becoming an integral part of our communication with family and friends, a source of information on various topics \cite{almars2022users}, a platform for work, and a means of entertainment. However, this remarkable success has also given rise to malicious activities, such as the deliberate dissemination of misinformation. Many nations have raised concerns about foreign interference in their electoral processes and social movements, often orchestrated by other countries or organisations \cite{linvill2020troll, nisbet2021presumed, kennedy2022repeat, freelon2022black}. A significant portion of this disinformation is propagated by social bots, automated accounts that mimic human behaviour on social networks, creating and sharing content while interacting with unsuspecting users who are typically unaware that they are engaging with artificial entities. Detecting and stopping the activities of these bots is critical to maintaining the integrity of online information and preserving the authenticity of public discourse \cite{shao2017spread}. The presence of bots on social media can also harm online ecosystems by engaging in malicious activities such as spamming, phishing, and cyber attacks \cite{deseriis2017hacktivism, hammi2019empirical}.

\renewcommand{\footnotesize}{\tiny} 
Effective bot detection plays a crucial role in safeguarding online platforms, creating a secure and reliable environment for users. However, the constant news reports \footnote{\url{https://www.theguardian.com/us-news/2024/feb/26/ai-deepfakes-disinformation-election} \\ \url{https://abcnews.go.com/Politics/pro-trump-bots-sowing-division-republican-party-report/story?id=97997613} \\ \url{https://www.bloomberg.com/opinion/articles/2023-11-03/2024-campaign-don-t-let-chinese-bots-influence-the-next-us-election?embedded-checkout=true}} about the presence of bots in various aspects of people's lives suggest that there is still work to be done in the area of bot detection. Artificial Intelligence has emerged as one of the most promising avenues to address this challenge \cite{hajli2022social, miller2016role, nguyen2023supervised}.

Exploring the literature we can see two primary avenues of research in the realm of bot detection based on AI systems: one rooted in graph theory and network metrics, and the other centred on account-based and content-based metrics. Numerous authors address the issue of bots in social media across diverse domains such as public health, politics, and stock markets \cite{gorwa2020unpacking, fan2020social, weng2022public}. These authors propose novel approaches, predominantly defined or guided by characteristics related to account behaviour or content. Motivated by the premise that bots can be defined by their characteristics, this paper focuses on leveraging both account and content-based features. We understand the combination of these two kinds of features encompassed the term \emph{user-profile} features. This study introduces a thorough feature engineering process to combat bots by leveraging user-profile measures. Our research aims to answer the following questions:

\begin{quote}\emph{\textbf{RQ. 1:} What features define a social bot?}\end{quote}

\begin{quote}\emph{\textbf{RQ. 2:} Which source of features holds greater importance in social bot detection, account-based or content-based features?}\end{quote}

\begin{quote}\emph{\textbf{RQ. 3:} Can a social bot be identified based on user-profile features? Are they enough?}\end{quote}

To answer the research questions, we have designed an experimental framework over three different datasets. In summary, our paper contributes significantly to the current state-of-the-art in several ways:

\begin{itemize}
  \item We conduct a comprehensive review, bringing together the features proposed in the literature addressing social bot detection. As far as we know, this paper provides the most extensive analysis, using more features suggested in the literature and testing them on a wider range of datasets. We also consider and compare the most diverse set of models to date.
  \item We provide a detailed analysis, identifying the features that have the most impact on social bot classification. To the best of our knowledge, this is the most thorough and complete analysis outlining the features that affect the categorisation and classification of social bots.
  \item We introduce a set of new features that, in addition to those collected from the literature, have served to surpass the state of the art in social bot detection using classical machine learning algorithms. This has been achieved through a feature selection process, comparing the results with other methods in the literature using different metrics.
\end{itemize}

These contributions are intended to provide insights into bot detection in order to improve the accuracy and efficiency of automated detection systems. This study focuses on $\mathbb{X}$ (formerly Twitter) with a specific emphasis on three widely recognised datasets commonly employed for benchmarking social bot detection. 

The structure of the paper is organised as follows: Section~\ref{sec:related} provides an in-depth exploration of related works in the field. Our proposed framework for enhancing bot detection through feature engineering is described in Section~\ref{sec:proposal}. The experimental process is comprehensively detailed in Section~\ref{sec:experiments}. The subsequent section, Section~\ref{sec:discussion}, evaluates and interprets the results. Final remarks and potential extensions are considered in Section~\ref{sec:Conclusions}.

\section{Related works}
\label{sec:related}

This section aims to provide context and some of the related work in the literature. It starts with an introduction to the concept of a bot in social media, followed by a categorisation of feature-based bot detection methods, and finally an in-depth look at feature engineering and selection for bot detection.

\subsection{Bots in social media}
Although the authors generally agree that a bot in social networks is an account with a certain degree of automation, there is no extended definition that covers all the details related to these accounts. This is due to the speed at which technology advances and the doubts that exist when attributing certain characteristics to a bot. One of these characteristics is the level of automation required for an account to transition from a human-managed account to a bot; there are accounts that are partially automated, and establishing a threshold to differentiate between accounts that are not bots and accounts that are bots is complex. Examples of this can be seen in the work conducted by Pastor et al. \cite{pastor2022profiling}, where they carried out a thorough experimentation focused on profiling bots according to their level of automation and behaviour across social media platforms, utilising social network analysis and graph theory.

Another of these features is the similarity to human behaviour. Some authors pay particular attention to this aspect by defining a bot in a social network context as an account that attempts to mimic human behaviour to a greater or lesser extent \cite{abokhodair2015dissecting}. The last point to consider is the different fields of study from which these bots are studied; computer scientists tend to give more importance to more technical characteristics, while social scientists focus more on the social implications \cite{cresci2020decade}.

Depending on how we value these characteristics we can give a more lax or restrictive definition. For example, as mentioned above, \cite{abokhodair2015dissecting} considers a social media bot any program that acts in the same way as a person in a social space, \cite{morstatter2016new} considers a bot as any account controlled by software within the social media, others like \cite{yang2020scalable, assenmacher2020demystifying} emphasize that this account can be only partially automated. Among the most restrictive definitions we can find the one given by \cite{ferrara2016rise} that considers a social media bot as a program that interacts with humans in a social environment and produces content automatically, adding that their intention is to mimic and perhaps alter human behaviour.

The definition followed in this study, as stated in \cite{lopez2023bot}, is as follows: \textit{``a Social Media Bot is an account that is automated enough to produce content and/or interact with other accounts within a social media context.''}
\subsection{Feature-based bot detection}

There are three types of approaches for bot detection in the literature: feature-based, graph-based and crowdsourcing techniques. The most popular bot detection approaches are feature-based methods. Feature-based methods attempt to leverage the data contained in both the account metadata and the user-written text itself. Most methods based on deep learning and machine learning techniques fall into this category. These methods are divided into three categories: account-based, content-based and hybrid \cite{lopez2023bot}.

\begin{itemize}
    \item \textbf{Account-based.} Account-based bot detection techniques use the information from the user's account as features or to infer new ones, e.g., account age, username length, number of retweets, number of followers, or follower growth rate. An example of such a method is given in \cite{hayawi2022deeprobot}. In this study, the authors use only account-based features, making use of feature engineering and feature selection techniques. They provide a hybrid deep learning architecture divided into two layers, one for the most relevant numerical variables and another for the description of the profile using embeddings. This study achieves a good generalisation and competitive results compared to other baselines.
    \item \textbf{Content-based.} 
    Content-based bot detection techniques use information from the content of tweets as features, e.g., the number of URLs, the number of hashtags, the sentiment or the length of the tweet. An interesting example that falls into this category is presented in \cite{mazza2019rtbust}. 
    Their approach relies on analysing the temporal retweet activity among $\mathbb{X}$ accounts. They employ an LSTM   
    \footnote{A Long Short-Term Memory (LSTM) is a neural network belonging to the family of recurrent neural networks (RNNs). Unlike conventional RNNs, LSTM networks can learn short-term dependencies in sequential data while also possessing a long-term memory useful for learning broader dependencies.} 
    variational autoencoder, a specialised neural network combining LSTM's sequential data modelling with variational autoencoders' probabilistic distributions. This fusion allows the extraction of latent features from the retweet time series of individual accounts.
    Finally, they use a clustering algorithm for bot detection. Notably, this study stands out not only for its competitive results but also for its use of graphical representations. These graphs facilitate the visual exploration of the temporal patterns within each account, enhancing the interpretability of the results compared to other studies.
    \item \textbf{Hybrid.} Hybrid bot detection techniques use a combination of features from the user's account and its content. An example of a hybrid approach is highlighted in \cite{heidari2020deep}. They used word embeddings from tweets, employing GloVe (Global Vectors)\cite{pennington2014glove} and ELMo (Embeddings from Language Models)\cite{sarzynska2021detecting} for a contextualised semantic representation of the text. Following this, they trained eight neural networks based on user profiling techniques, using characteristics such as gender, age, personality and education. By segmenting the dataset based on similar profiles, they enhanced classification accuracy. In the last step, the authors implemented a final model that has as input the values resulting from all previous models. Finally, to optimise the results, they explored different architectures for this final model, ultimately identifying the Feedforward Neural Network (FNN) as the most effective.
\end{itemize}

\subsection{Feature engineering and feature selection for bot detection}

The process of creating, selecting, or transforming attributes or features that machine learning models use for making predictions is known as \textbf{feature engineering}. It involves extracting relevant information from the raw data and creating informative features that can improve the performance of the model \cite{yin2024person, wang2024feature}.

In the context of bot detection in social media, feature engineering entails crafting features that capture the distinctive behaviour and characteristics of bots. These features may include:

\renewcommand{\footnotesize}{\small} 
\begin{itemize}
    \item \textbf{Temporal patterns:} Metrics related to the frequency and timing of the user activity.
    \item \textbf{Social interactions:} Measures of the user's interactions, such as the number of followers, friends, and mentions.
    \item \textbf{Platform attributes:} Platform-dependent features, such as the presence of a profile picture, profile background colour or source of the user activity. 
    \item \textbf{Language and content:} Features that describe the content and language used in tweets or posts, including sentiment analysis, stylometry, and linguistic complexity.
    \item \textbf{Network features:} Attributes related to the user's connections and network structure, such as centrality measures.
\end{itemize}

Effective feature engineering aids in identifying the underlying patterns that separate real users from automated programs. These features serve as a critical input to machine learning models and contribute to the ability of the model to accurately classify and detect bots.

In the same way, the process of choosing a subset of the most relevant features from the available feature set is known as \textbf{feature selection}. It is aimed at reducing dimensionality, improving model interpretability, and potentially enhancing model performance. Feature selection techniques are particularly valuable when working with high-dimensional datasets or when dealing with noisy and irrelevant features. Types of feature selection methods include:

\begin{itemize}
    \item \textbf{Filter methods:} These methods evaluate the relevance of features independently of the machine learning model. Filter methods are efficient, but don't consider the relationship between features. Examples include Chi-Square \cite{peters1940chi}, Mutual Information \cite{shannon1948mathematical}, and Fisher's Score \cite{fisher1936use}.
    \item \textbf{Wrapper methods:} These methods use a machine learning model to evaluate different subsets of features. They include techniques such as Recursive Feature Elimination \cite{guyon2002gene} and Forward/Backward Selection \cite{ferri1994comparative}.
    \item \textbf{Embedded methods:} Feature selection is integrated into the model training process. These methods are able to capture feature dependencies. L1 regularisation \cite{tibshirani1996regression} in linear models and Random Forest importance \cite{breiman2001random} are examples of embedded methods.
\end{itemize}

In the realm of bot detection leveraging feature engineering and selection, we found recent studies that overlap with our research \cite{mbona2022feature, ilias2021detecting, cardaioli2021sa, wu2021novel}. 

In \cite{mbona2022feature}, Mbona and Eloff proposed leveraging Benford's Law to identify the most accurate features for bot categorisation and the development of a bot detection system. They laid the foundation for further exploration in feature selection methodologies for bot categorisation, emphasising that a well-conceived approach for selection can yield superior results. However, the paper is predominantly focused on feature selection and does not provide a comparative analysis of their results in terms of bot identification against the established baselines in the literature. Moreover, it lacks utilisation of benchmark datasets commonly employed in bot categorisation research.

Cardaioli et al. introduced the utilisation of writing style crafted features to improve bot categorisation and detection in their work \cite{cardaioli2021sa}. Although the proposed features are interesting, their contribution is somewhat limited in terms of exploring interaction results among features and their study focused solely on the Cresci-17 dataset, warranting further investigation for broader applicability and robustness. In contrast, the work presented in \cite{ilias2021detecting} introduces a deep-learning system that is fed with a comprehensive set of 66 newly crafted features encompassing both account-based and content-based aspects. The authors further proposed a meticulous feature engineering process, conducting a comparative set of experiments across two benchmark datasets, Cresci-17 and the Social Honeypot Dataset \cite{lee2011seven}, achieving state-of-the-art results in both datasets. 

Our research builds upon this feature engineering methodology, merging features from various papers and introducing a total of 19 features. Through an exhaustive feature selection process and leveraging classical classification models (in particular Random Forest), we demonstrate across three diverse benchmark datasets that our set of features surpasses existing literature in terms of precision, recall, F1 score, and accuracy.

\section{Enhancing feature engineering and feature selection for bot detection}
\label{sec:proposal}

\begin{figure}[H]
  \centering
  \includegraphics[width=0.82\textwidth]{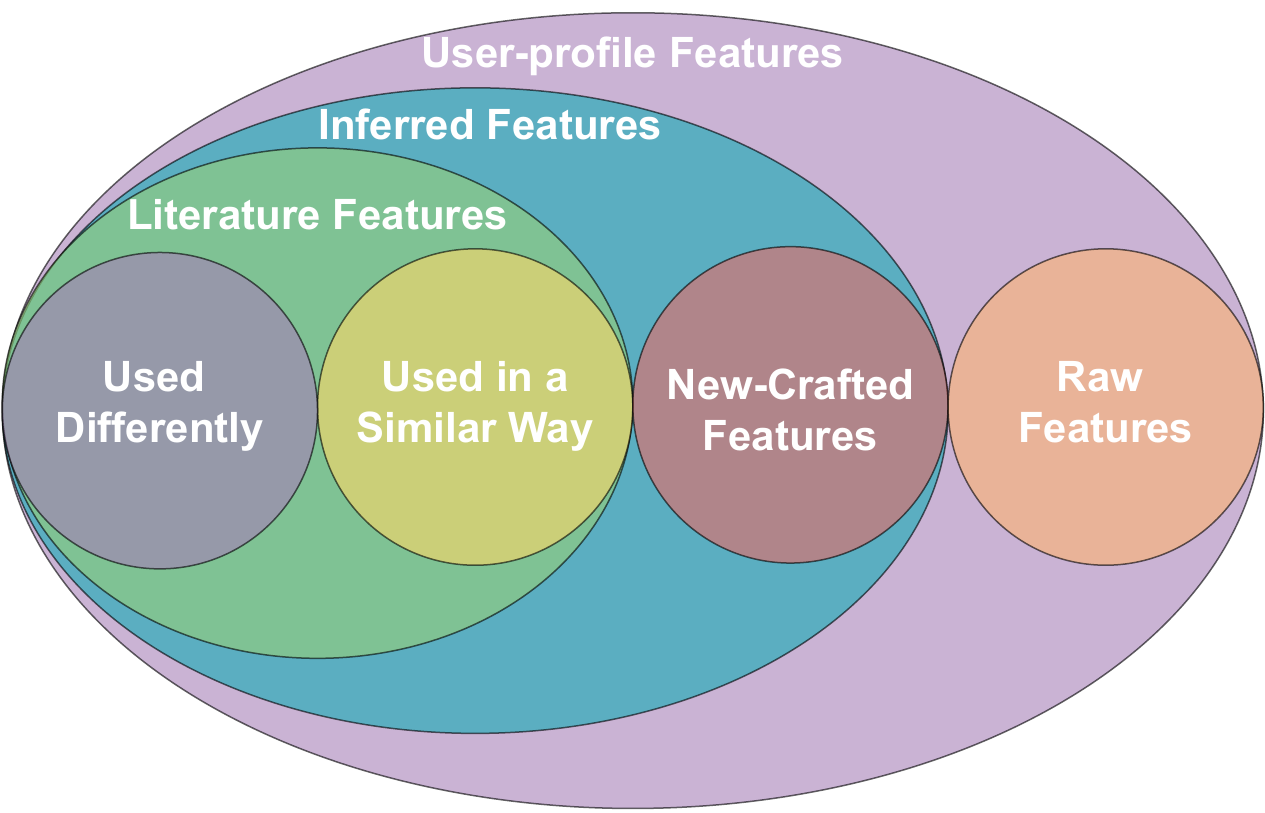}
  \caption{Feature engineering diagram}
  \label{fig:diagrama_features_g}
\end{figure}

A feature engineering approach has been followed to identify the features that are useful for differentiating genuine accounts from automated accounts. 
Three of the most widely spread datasets on bot detection have been chosen: Cresci-15 \cite{cresci2015fame}, Cresci-17 \cite{cresci2017paradigm} and TwiBot-20 \cite{feng2021twibot}. These datasets have been analysed by selecting useful features from them to serve as a basis for this process. The datasets do not have exactly the same features so not all inferred features can be obtained in all datasets. A table with all the raw and inferred features, in which dataset they can be calculated and whether they have appeared in the literature can be found in Appendix \ref{final-user-based-features-from-each-dataset} and \ref{final-content-based-features-from-each-dataset}.
Three types of features of different nature have been obtained: raw features, literature features and new-crafted features. The hierarchical diagram of these features can be seen in Figure \ref{fig:diagrama_features_g}. 

Among the characteristics of the elements in each dataset, features that are inherently useful without requiring any inference process can be extracted. We have called these features \textit{raw features} and, among them, we can find the number of followers, the number of favourites, the account verification or the number of lists in which the user appears. These characteristics will be the basis by which we can infer new ones.

From the \textit{raw features} new features have been inferred that could be of value for bot detection. These features are divided into features from the literature that have been explored by other authors and new features proposed in this study. We have calculated some of the classic meta-features of feature engineering such as mean, ratios, minimum, maximum, among others, including them in the features coming from the literature if other authors have used them for the task of bot detection.
An exhaustive study of the literature was carried out with the aim of finding calculable characteristics that other authors have proposed. Two different databases, Web of Science and Google Scholar, were used in this process. With $\mathbb{X}$ as the focus, different terms and queries were used to perform the search, including: \textit{``bot detection in Twitter"}, \textit{``bot detection feature engineering"}, \textit{``identifying bots in Twitter"}, \textit{``automated account detection in social media"}, \textit{``bot characteristics"}, etc. The results have been sorted by relevance and selecting the top 20 of each search if any. From this collection of articles, we proceeded to read each one and discarded those that did not propose new features for our problem. To finish this process, we have selected from the remaining articles those that propose original and possible features to be calculated in our dataset. Semantic features have not been included due to the desire to maintain explainability and many of these features are computed using deep learning models, nor have graph-based features been included as many of the available datasets do not have the network architecture information and the computational time is high if the network is large enough. 

\begin{table}[H]

\begin{center}

\resizebox{\textwidth}{!}{\begin{tabular}{llllll}

\hline
Feature Name                      & Ref.                                        & Description              &  Use     & Type \\ \hline
Followers\_growth\_rate          & \cite{hayawi2022deeprobot}    &  $n\_followers / user\_age$              &   S           & \multirow{8}{*}{Social Based}\\
Friends\_growth\_rate            & \cite{hayawi2022deeprobot}    &  $n\_friends / user\_age$              &   S           & \\
Favourites\_growth\_rate         & \cite{hayawi2022deeprobot}    &  $n\_favourites / user\_age$              &   S           & \\
Listed\_growth\_rate             & \cite{hayawi2022deeprobot}    &  $n\_listed / user\_age$              &   S           & \\
Followers\_friends\_ratio        & \cite{hayawi2022deeprobot}    &  $n\_followers / n\_friends$              &   S           & \\
Average\_favorites               & \cite{daouadi2019bot} & $n\_favorites / n\_followers$ & S & \\ 
Average\_retweets                & \cite{daouadi2019bot} & $n\_retweets / n\_followers$ & S & \\ 
Reputation        & \cite{ilias2021detecting} & Reputation of the user & S & \\ \hline

User\_age                        & \cite{hayawi2022deeprobot} & The age of the account in days & S           & \multirow{2}{*}{Temporal}\\ 
Tweet\_freq                      & \cite{hayawi2022deeprobot}    &  $n\_tweets / user\_age$    &   S
& \\ 
\hline
Description\_flesch\_reading\_ease & \cite{cardaioli2021sa} & Flesch Reading Ease Score of description& D &\multirow{9}{*}{Readability}\\ 
Description\_flesch\_kincaid\_grade & \cite{cardaioli2021sa} & Flesch-Kincaid Grade of description&D &\\
Description\_smog\_index & \cite{cardaioli2021sa} & SMOG index of description&D &\\
Description\_coleman\_liau\_index & \cite{cardaioli2021sa} & Coleman–Liau index of description&D &\\
Description\_automated\_readability\_index & \cite{cardaioli2021sa} & Automated Readability Index of description&D &\\
Description\_dale\_chall\_readability\_score & \cite{cardaioli2021sa} & Grade level using the New Dale-Chall Formula in description&D &\\
Description\_difficult\_words & \cite{cardaioli2021sa} & Number of difficult words in description&D &\\
Description\_linsear\_write\_formula &  \cite{cardaioli2021sa} & Grade level using Linsear Write Formula of description&D &\\
Description\_gunning\_fog & \cite{cardaioli2021sa} & Gunning fog index of description&D &\\
\hline 
Screen\_name\_length             & \cite{hayawi2022deeprobot}    &  Length of screen name    &   S           & \multirow{28}{*}{Stylometry}\\
Name\_length                     & \cite{hayawi2022deeprobot}    &  Length of name               &  S            & \\
Description\_length              & \cite{hayawi2022deeprobot}    &  Length of description              &   S           & \\
Description\_digits\_count       & \cite{hayawi2022deeprobot}    &  Count of digits in description              &   D           & \\
Description\_mean\_bigram\_freq  & \cite{hayawi2022deeprobot}    &  Mean bigram freq. in description             &   D           & \\
Screen\_name\_digits\_count      & \cite{hayawi2022deeprobot}    &  Count of digits in screen name              &   S           & \\
Name\_digits\_count              & \cite{hayawi2022deeprobot}    &  Count of digits in name              &   S           & \\
Screen\_name\_mean\_bigram\_freq & \cite{hayawi2022deeprobot}    &  Mean bigram freq. in screen name             &   S           & \\
Screen\_name\_entropy            & \cite{hayawi2022deeprobot}    &  Entropy of screen name              &   S           & \\
Name\_mean\_bigram\_freq         & \cite{hayawi2022deeprobot}    &  Mean bigram freq. in name            &   D           & \\
Name\_entropy                    & \cite{hayawi2022deeprobot}    &  Entropy of name              &   S           & \\
Description\_entropy             & \cite{hayawi2022deeprobot}    &  Entropy of description              &   S           & \\
Name\_sim                        & \cite{hayawi2022deeprobot}    &  Name and screen name similarity              &   S           & \\
Name\_ratio                      & \cite{hayawi2022deeprobot}    &  Name and screen name length ratio              &   S           & \\ 
Name\_contains\_bot              & \cite{beskow2018bot} & If name contains ``bot'' & D & \\   
Screen\_name\_contains\_bot      & \cite{beskow2018bot} & If screen name contains ``bot'' & D & \\ 
Description\_contains\_bot       & \cite{beskow2018bot} & If description contains ``bot''& D & \\ 
Description\_hashtag\_count      & \cite{ilias2021detecting} & Hashtags count in description & S & \\ 
Description\_url\_count          & \cite{ilias2021detecting} & URLs count in description & S & \\ 
Description\_unique\_url\_count  & \cite{ilias2021detecting} & Unique URLs count in description & D & \\ 
Description\_unique\_mention\_count & \cite{ilias2021detecting} & Unique mentions count in description & D & \\ 
Description\_fraction\_of\_words\_lowercase   & \cite{przybyla2021classification} & Fraction of lowercase words in description & D & \\
Description\_fraction\_of\_words\_uppercase   & \cite{przybyla2021classification} & Fraction of uppercase words in description & D & \\
Description\_fraction\_of\_words\_tilecase   & \cite{przybyla2021classification} & Fraction of tilecase words in description & D & \\
Description\_word\_count   & \cite{przybyla2021classification} & Number of words in description & D & \\
Description\_sentence\_count   & \cite{przybyla2021classification} & Number of sentences in description & D & \\
Description\_average\_word\_length   & \cite{przybyla2021classification} & Average length of words in description & D & \\
Description\_average\_words\_per\_sentence   & \cite{przybyla2021classification} & Description avg. words per sent.& D &\\ \hline

\end{tabular}}
\caption{\label{profile-literature} Inferred literature features from account. }
\end{center}
\end{table}

Tables \ref{profile-literature} and \ref{content-literature} show the features derived from both the account and the content, grouped by type. In addition, they have been marked in the \textit{Use} column with an \textit{S} if they were used in the same fields as in the original study, or with a \textit{D} if they were used in different fields.

\begin{table}[H]

\begin{center}

\resizebox{\textwidth}{!}{\begin{tabular}{llllll}

\hline
Feature Name                     & Ref.                                        & Description              &  Use   & Type  \\ \hline
Ratio\_retweet                   & \cite{ilias2021detecting} & $n\_retweets / n\_tweets$  & S &\multirow{1}{*}{Social Based}\\ 

\hline 

Average\_time\_between\_tweets   & \cite{ilias2021detecting} & Average time between tweets& S &\multirow{8}{*}{Temporal}\\
Idle\_hours                      & \cite{ilias2021detecting} & Max time without activity& S &\\
Size\_DNA\_type                  & \cite{cresci2017social} & Size of DNA type before compression & S &\\
Compress\_size\_DNA\_type        & \cite{cresci2017social} & Size of DNA type after compression & S &\\
Compression\_ratio\_type         & \cite{cresci2017social} & Ratio of DNA type sizes & S &\\
Size\_DNA\_content               & \cite{cresci2017social} & Size of DNA content before compression & S &\\
Compress\_size\_DNA\_content     & \cite{cresci2017social} & Size of DNA content after compression & S &\\
Compression\_ratio\_content      & \cite{cresci2017social} & Ratio of DNA content sizes & S &\\
\hline 

Flesch\_reading\_ease & \cite{cardaioli2021sa} & Average Flesch Reading Ease Score in tweets & S & \multirow{9}{*}{Readability}\\
Flesch\_kincaid\_grade & \cite{cardaioli2021sa} & Average Flesch-Kincaid Grade in tweets & S &\\
Smog\_index & \cite{cardaioli2021sa} & Average SMOG index in tweets & S &\\
Coleman\_liau\_index & \cite{cardaioli2021sa} & Average Coleman–Liau index in tweets & S &\\
Automated\_readability\_index & \cite{cardaioli2021sa} & Average Automated Readability Index in tweets & S &\\
Dale\_chall\_readability\_score & \cite{cardaioli2021sa} & Average grade level using the New Dale-Chall Formula in tweets & S &\\
Difficult\_words & \cite{cardaioli2021sa} & Average number of difficult words in tweets & S &\\
Linsear\_write\_formula & \cite{cardaioli2021sa} & Average grade level using Linsear Write Formula in tweets & S &\\
Gunning\_fog & \cite{cardaioli2021sa} & Average Gunning fog index in tweets & S &\\
\hline
Different\_sources               & \cite{ilias2021detecting} & $n\_sources\_used / n\_total\_sources$  & S & \multirow{14}{*}{Platform Based}\\ 
Source\_tweetadder\_percentage & \cite{wu2020using} & Percentage of tweets from tweetadder & S &\\ 
Source\_iphone\_percentage & \cite{wu2020using} & Percentage of tweets from iphone & S &\\ 
Source\_android\_percentage & \cite{wu2020using} & Percentage of tweets from android & S &\\ 
Source\_twitter\_percentage & \cite{wu2020using} & Percentage of tweets from twitter & S &\\ 
Source\_tweetdeck\_percentage & \cite{wu2020using} & Percentage of tweets from tweetdeck & S &\\ 
Source\_ipad\_percentage & \cite{wu2020using} & Percentage of tweets from ipad & S &\\ 
Source\_web\_percentage & \cite{wu2020using} & Percentage of tweets from web & S &\\ 
Source\_facebook\_percentage & \cite{wu2020using} & Percentage of tweets from facebook & S &\\ 
Source\_instagram\_percentage & \cite{wu2020using} & Percentage of tweets from instagram & S &\\ 
Source\_api\_percentage & \cite{wu2020using} & Percentage of tweets from API & S &\\ 
Source\_web\_api\_percentage & \cite{wu2020using} & Percentage of tweets from web API & S &\\ 
Source\_mobile\_percentage & \cite{wu2020using} & Percentage of tweets from mobile & S &\\ 
Source\_other\_percentage & \cite{wu2020using} & Percentage of tweets from other & S &\\ 
\hline 
Bot\_reference\_mean             & \cite{beskow2018bot} & References mean to ``bot'' in tweets & S & \multirow{18}{*}{Stylometry}\\ 
Average\_tweet\_length           & \cite{ilias2021detecting} & Average length of tweets  & D &\\ 
Num\_unique\_urls\_mean          & \cite{ilias2021detecting} & Unique URLs count in tweets  & D &\\ 
Num\_unique\_mentions\_mean      & \cite{ilias2021detecting} & Unique mentions count in tweets  & D &\\

Max\_urls\_in\_a\_tweet          & \cite{ilias2021detecting} & Max number of URLs in a tweet& S &\\
Max\_hashtags\_in\_a\_tweet      & \cite{ilias2021detecting} & Max number of hashtags in a tweet& S &\\
Max\_mentions\_in\_a\_tweet      & \cite{ilias2021detecting} & Max number of mentions in a tweet& S &\\

Average\_tweets\_only\_url       & \cite{ilias2021detecting} & Average tweets with one URL& S &\\
Average\_elongated\_words        & \cite{ilias2021detecting} & Average elongated words in tweets& S &\\
Num\_unique\_langs               & \cite{ilias2021detecting} & Number of unique langs in tweets& S &\\ 

Word\_count\_mean   & \cite{przybyla2021classification} & Mean of word count in tweets & S &\\
Sentence\_count\_mean   & \cite{przybyla2021classification} & Mean of sentence count in tweets & S &\\
Average\_word\_length   & \cite{przybyla2021classification} & Average length of words in tweets & S &\\
Average\_words\_lowercase   & \cite{przybyla2021classification} & Average number of lowercase words in tweets & S &\\
Average\_words\_uppercase   & \cite{przybyla2021classification} & Average number of uppercase words in tweets & S &\\
Average\_words\_titlecase   & \cite{przybyla2021classification} & Average number of tilecase words in tweets & S &\\
Tweets\_sim\_length   & \cite{wu2020using} & Similarity of tweet lengths & S &\\
Tweets\_sim\_punctuation   & \cite{wu2020using} & Similarity of tweet punctuation & S &\\ 
\hline

\end{tabular}}
\caption{\label{content-literature} Inferred literature features from content. }

\end{center}
\end{table}

\subsection{New-Crafted features}

Two new sets of features have been proposed. The first one is account-based and aims to take advantage of the information that can give us the level of personalisation of the user's account when detecting bots, in the second one we recover the measures exposed in \cite{diaz2022noface} that allow modelling the credibility and engagement of a user based on the metadata of the tweets.

\begin{table}[H]

\begin{center}

\resizebox{\textwidth}{!}{\begin{tabular}{llll}
\hline

Feature Name  & Description    & Type           \\ \hline

Profile\_background\_color\_is\_default  & If profile background colour is default&\multirow{17}{*}{Platform Based}\\
Profile\_background\_color\_is\_uncommon  & If profile background colour is uncommon&\\
Profile\_background\_color\_is\_common  & If profile background colour is common&\\
Profile\_background\_image\_url\_default\_other\_none & If profile background image exists, it is the default or other &\\
Has\_profile\_background\_tile &  If has profile background tile&\\
Profile\_link\_color\_default & If profile link colour is default&\\
Profile\_link\_color\_common & If profile link colour is common&\\
Profile\_link\_color\_uncommon & If profile link colour is uncommon&\\
Profile\_sidebar\_border\_color\_default & If profile sidebar border colour is default&\\
Profile\_sidebar\_border\_color\_common & If profile sidebar border colour is common&\\
Profile\_sidebar\_border\_color\_uncommon & If profile sidebar border colour is uncommon&\\
Profile\_sidebar\_fill\_color\_default & If profile sidebar fill colour is default&\\
Profile\_sidebar\_fill\_color\_common & If profile sidebar fill colour is common&\\
Profile\_sidebar\_fill\_color\_uncommon & If profile sidebar fill colour is uncommon&\\
Profile\_text\_color\_default & If profile text colour is default&\\
Profile\_text\_color\_common & If profile text colour is common&\\
Profile\_text\_color\_uncommon & If profile text colour is uncommon&\\ \hline

\end{tabular}}
\caption{\label{profile-new-crafted} Inferred new features based on user account. }
\end{center}
\end{table}

\begin{table}[H]

\begin{center}

\resizebox{171.9pt}{!}{\begin{tabular}{llll}
\hline
Feature Name        & Description  &     Type        \\ \hline
Credibility & Credibility of the user & \multirow{2}{*}{Social Based}\\ 
Engagement & Engagement of the user\\

\hline 
\end{tabular}}
\caption{\label{content-new-crafted} Inferred new features based on user content. }
\end{center}
\end{table}

\subsubsection{Colour binning}
Among the raw features that can be extracted from the $\mathbb{X}$ account are those that represent how the user has customised the profile using colours in hexadecimal base. To our knowledge, these features have not been exploited in the literature. Intuition tells us that these features may be relevant in identifying bots. It is expected that a high degree of personalisation of an account will tend to be more like a real user.

\begin{figure}[h]
  \centering
  \includegraphics[width=0.6\textwidth]{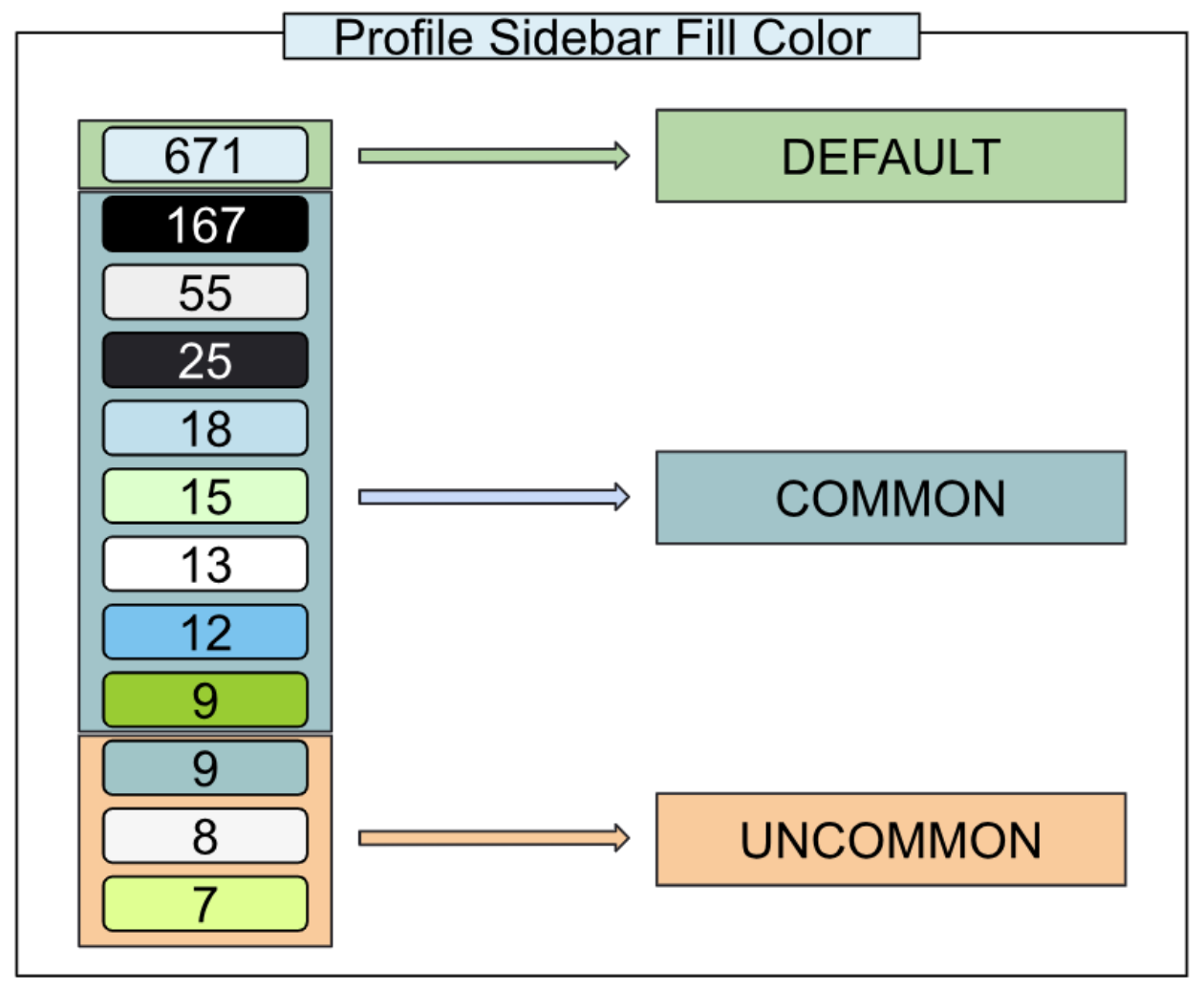}
  \caption{Profile sidebar fill colour ranking}
  \label{fig:profile_sidebar_fill_color}
\end{figure}

Three binary value categories have been created from the colours in each dataset including default (if not modified), common (among the top eight colours used), and uncommon (if not in any previous category). The process is illustrated in Figure \ref{fig:profile_sidebar_fill_color} where the colours from the profile sidebar are categorised.

\subsubsection{Credibility and engagement}
In the study conducted by the authors in \cite{diaz2022noface, diaz2021comparative}, a filter based on the assignment of credibility and knowledge of users on a specific topic is proposed. The main objective of this study is to reduce irrelevant content coming from social networks, thus allowing to filter and highlight useful and credible content.  

In this filter, engagement is mathematically modelled by relating the number of favourites and the number of retweets on a topic to the number of followers, to then establish a cut-off threshold. In the same way they model credibility relating the number of followers, the number of listings, the number of retweets and the number of favourites to establish another cut-off threshold. 

 In this work they do not use general engagement but engagement on a certain topic. For the authors, the fact that a person generates quality content on a specific topic does not mean that he/she has to do it for other topics. In the context of our study, we apply this formula in a general way, as our analysis does not delve into the problem at the topic level:

\begin{itemize}
\item Credibility:
\begin{equation}
\epsilon(u) = \frac{{\frac{{n\_Favorites}}{{n\_Followers}} + \frac{{n\_Retweets}}{{n\_Followers}}}}{{2}}
\end{equation}
where $\epsilon(u)$ denotes the engagement of a user $u$, $n\_Favorites$ is the sum of favourites of the user's tweets, $n\_Retweets$ is the sum of retweets of the user's tweets, and $n\_Followers$ is the number of followers of the user.

\item Engagement:
\begin{equation}
\zeta(u) = \frac{{n\_Followers + n\_Lists + n\_Retweets + n\_Favorites}}{4}
\end{equation}
where $\zeta(u)$ denotes the credibility of a user $u$, $n\_Lists$ is the number of public lists the user appears in, and $n\_Followers$, $n\_Retweets$ and $n\_Favorites$ are the same variables as in the engagement formula.

\end{itemize}

\section{Experiments}
\label{sec:experiments}

In this section the datasets will be briefly described, as well as the methodology used. Following this, the results obtained will be presented, and it will conclude with an ablation study comparing the model using both sources of features: account and content.

\subsection{Data}
\textbf{Cresci-17.} This is a dataset of user accounts and tweets obtained from $\mathbb{X}$ with the help of users of the CrowdFlower platform \cite{cresci2017paradigm}.

In Table \ref{tabla-cresci-17} it is shown what they are and how the classes are distributed in the Cresci-17 dataset. Note that the classes \textit{Traditional spambots \#2}, \textit{Traditional spambots \#3} and \textit{Traditional spambots \#4} do not contain any data about tweets, as the authors considered them irrelevant for their work. 

\begin{table}[htbp]
  \centering
  \adjustbox{max width=\textwidth}{
    \begin{tabular}{llccc}
      \toprule
      Class name & Accounts & Tweets \\
      \midrule
      Genuine accounts &  3,474 & 8,377,522 \\
      Social spambots  \#1 &  991 & 1,610,176 \\
      Social spambots \#2 &  3,457 & 428,542 \\
      Social spambots \#3 &  464 & 1,418,626 \\
      Traditional spambots \#1 & 1,000 & 145,094 \\
      Traditional spambots \#2 & 100 & 74,957 \\
      Traditional spambots \#3 & 433 & 5,794,931 \\
      Traditional spambots \#4 & 1,128 & 133,311 \\
      Fake followers & 3,351 & 196,027 \\
      \bottomrule
    \end{tabular}
  }
\captionsetup{font=small,position=bottom}
  \caption{Cresci-17 distribution \cite{cresci2017paradigm}.}
  \label{tabla-cresci-17}
\end{table}

\begin{itemize}
     
\item \textbf{Genuine accounts.} Verified accounts operated by humans.
\item \textbf{Social spambots \#1.} Retweeters of an Italian political candidate.
\item \textbf{Social spambots \#2.} Spammers of paid apps for mobile devices.
\item \textbf{Social spambots \#3.} Spammers of products for sale on Amazon.com.
\item \textbf{Traditional Spambots \#1}. Spammers training set used in \cite{yang2013empirical}.
\item \textbf{Traditional Spambots \#2.} Scam URL spammers.
\item \textbf{Traditional Spambots \#3.} Automated accounts that spam job offers.
\item \textbf{Traditional Spambots \#4.} Another set of automated accounts dedicated to disseminating job offers.
\item \textbf{Fake followers.} Simple accounts designed to artificially boost the follower count of another account.
\end{itemize}

The dataset is composed of two files, one for tweets and one for users. Attributes associated with this dataset can be found in Appendix \ref{Raw-user-based-features-from-each-dataset} and \ref{Raw-content-based-features-from-each-dataset}.

\textbf{Cresci-15.}
This dataset is part of the same project as the previous one and its target is the detection of fake followers \cite{cresci2015fame}. To obtain the data from legitimate accounts, they have relied on $\mathbb{X}$ and have followed two paths. The first was the creation of an account called @TheFakeProject with a bio that reads as follows ``Follow me only if you are NOT a fake''. The other way was through the hashtag \#elezioni2013 which collected $\mathbb{X}$ accounts that participated with this hashtag talking about Italian politics that subsequently passed a manual verification of their legitimacy.

In Table \ref{tabla-cresci-15} we can see what are and how are distributed the classes of this dataset.

\begin{table}[htbp]
  \centering
  \adjustbox{max width=\textwidth}{
\begin{tabular}{llcccc}
  \toprule
  Class name & Accounts & Tweets & Followers & Friends \\
  \midrule
  TFP (human) & 469 & 563,693 & 258,494 & 241,710  \\
  E13 (human) & 1,481 & 2,068,037 & 1,526,944 & 667,225  \\
  FSF (bot) & 1,169 & 22,910 & 11,893 & 253,026  \\
  INT (bot) & 1,337 & 58,925 & 23,173 & 517,485  \\
  TWT (bot) & 845 & 114,192 & 28,588 & 729,839  \\  
  \bottomrule
\end{tabular}
}
\captionsetup{font=small,position=bottom}
  \caption{Cresci-15 distribution \cite{cresci2015fame}.}
  \label{tabla-cresci-15}
\end{table}

\begin{itemize}
    \item \textbf{TFP.} Verified human-operated accounts collected via @TheFakeProject account.
    \item \textbf{E13.} Accounts collected via the hashtag \#elezioni2013. These accounts are verified and human-operated. 
    \item \textbf{FSF, INT and TWT.} Fake followers purchased from different websites.
\end{itemize}

As in the previous case, the dataset is composed of two files, one for tweets and one for users. We can see the attributes associated with each one in Appendix \ref{Raw-user-based-features-from-each-dataset} and \ref{Raw-content-based-features-from-each-dataset}.

\textbf{TwiBot-20.} This dataset was collected in 2020, and in \cite{feng2021twibot} the authors explain in detail the user selection process for this dataset. The authors want to maintain user diversity by using a strategy based on seed users. These seeds come from four different domains: politics, business, entertainment, and sports. To identify the bot accounts, they ran a crowdsourcing campaign, assigning five annotators to each account for more robust identification. 
In Table \ref{tabla-twibot-20} we can see what are and how are distributed the classes of this dataset.

\begin{table}[htbp]
  \centering
  \adjustbox{max width=\textwidth}{
\begin{tabular}{llc}
  \toprule
  Entity & N. samples \\
  \midrule
  Genuine Accounts & 5,237  \\
  Bot Accounts & 6,589   \\
  Tweets & 3,348,819    \\
  Edges & 3,371,617    \\
  \bottomrule
\end{tabular}
}
\captionsetup{font=small,position=bottom}
  \caption{TwiBot-20 distribution.}
  \label{tabla-twibot-20}
\end{table}

This dataset is available in JSON format with a section for the account features, a list with the raw text of up to 200 tweets and to which domain the account belongs. The features that appear in this dataset can be seen in Appendix \ref{Raw-user-based-features-from-each-dataset} and \ref{Raw-content-based-features-from-each-dataset}.

\renewcommand{\footnotesize}{\tiny}
Cresci-15 and Cresci-17 datasets can be downloaded from the official Botometer repository \footnote{\url{https://botometer.osome.iu.edu/bot-repository/}}, while the Twibot-20 dataset requires access to the authors of the corresponding paper.

\subsection{Methodology of experimentation}

\begin{figure}[h]
  \centering
  \includegraphics[width=\textwidth]{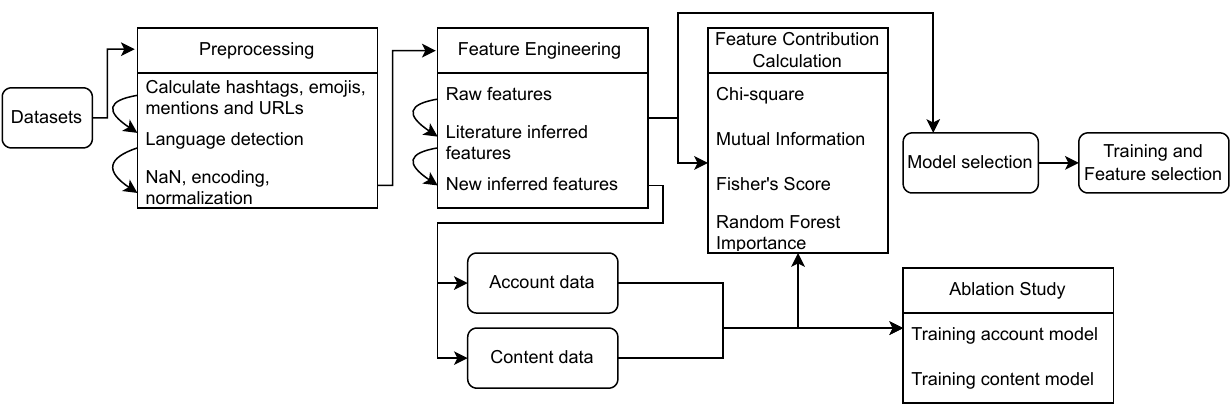}
  \caption{Experimentation flowchart}
  \label{fig:experimentation_flowchart}
\end{figure}

A rigorous methodology has been followed to provide valid and consistent results (see Figure \ref{fig:experimentation_flowchart}). As explained above, a literature review has been performed in order to find as many relevant features as possible. For each dataset we have implemented these features (see Tables \ref{profile-literature}, \ref{content-literature}) in addition to the new ones we have suggested (see Tables \ref{profile-new-crafted}, \ref{content-new-crafted}). Once the features related to hashtags, mentions, emojis and URLs were calculated, we proceeded to the elimination of these elements in the text. For language detection we used the \textit{FastText} library, specifically a language detection model trained on data from \textit{Wikipedia}, \textit{Tatoeba} and \textit{SETimes} \cite{joulin2016bag, joulin2016fasttext}. A normalisation of the data was performed prior to the experiments.

Several methods have been evaluated to perform the feature selection, specifically Chi-square, Mutual information and Fisher's Score for the filter methods and Random Forest Importance of embedded methods. It has also been represented and studied when the features come from the user's account and when from the content. 

A classification study has been carried out with the intention of comparing the accuracy obtained with each subset of features as well as the computation time required for each run. It has been chosen as a stopping criterion for the selection of features that there is no improvement in two consecutive iterations of the maximum accuracy obtained in the previous iterations.

For classification model selection, a total of 15 different baselines have been compared. Different metrics have been evaluated in each of them, as well as their training time. In each model, cross-validation has been carried out with 10 partitions and the average of the metrics obtained in each partition has been taken as the final value. 

The evaluation metrics used throughout the experiment are the usual employed in classification systems, and are defined as follows: 
\begin{equation}
Accuracy = \frac{No.\ of\ correct\ predictions}{Total\ number\ of\ predictions} = \frac{TP + TN}{TP+TN+FP+FN}
\end{equation}
\begin{equation}
Precision = \frac{TP}{TP + FP}
\end{equation}
\begin{equation}
Recall = \frac{TP}{TP + TN}
\end{equation}
\begin{equation}
F1\ Score = 2 \cdot \frac{Precision \cdot Recall}{Precision + Recall}
\end{equation}
where $TP, TN, FP$ and $FN$ represent respectively the number of true positives, true negatives, false positives and false negatives.

The experiments have been performed on a computer with the following components: AMD Ryzen 7 5800X (CPU), 32 GB DDR4 (RAM) and Samsung SSD 980 PRO 1TB M.2 (DISK). The programming language used was Python. In addition, the following libraries have been used: \textit{pycaret}, \textit{pandas}, \textit{fasttext}, \textit{emoji}, \textit{nltk} and \textit{textstats}.

\subsection{Results}

In this section, we present the outcomes and observations derived from the conducted experiments aimed at addressing the research questions outlined in the preceding sections. 

\begin{figure}[htbp]
    \centering
    \begin{subfigure}[b]{0.45\textwidth}
        \centering
        \includegraphics[width=\textwidth]{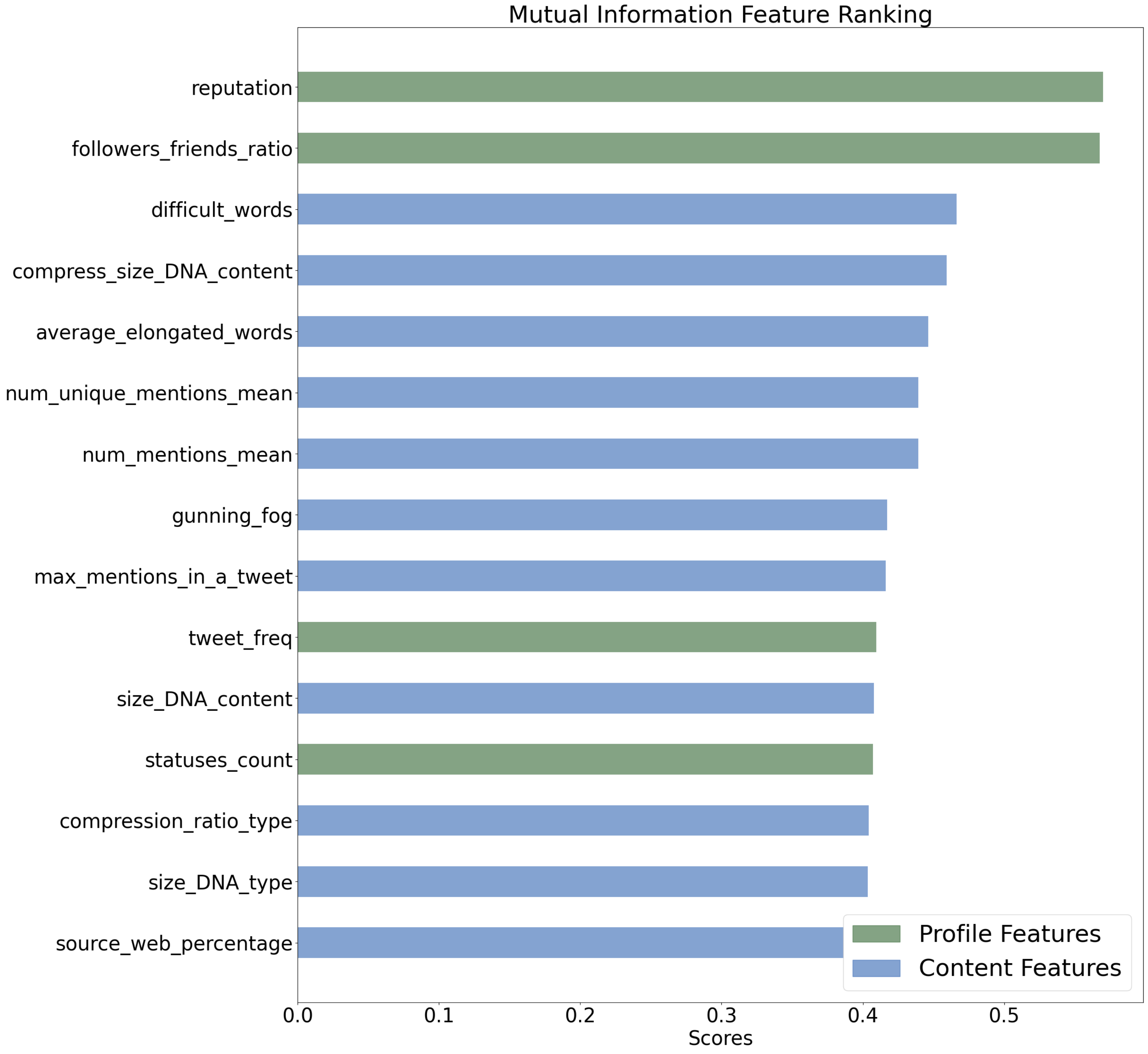}
        \caption{Mutual Information Cresci-15}
        \label{fig:img1}
    \end{subfigure}
    \hfill
    \begin{subfigure}[b]{0.45\textwidth}
        \centering
        \includegraphics[width=\textwidth]{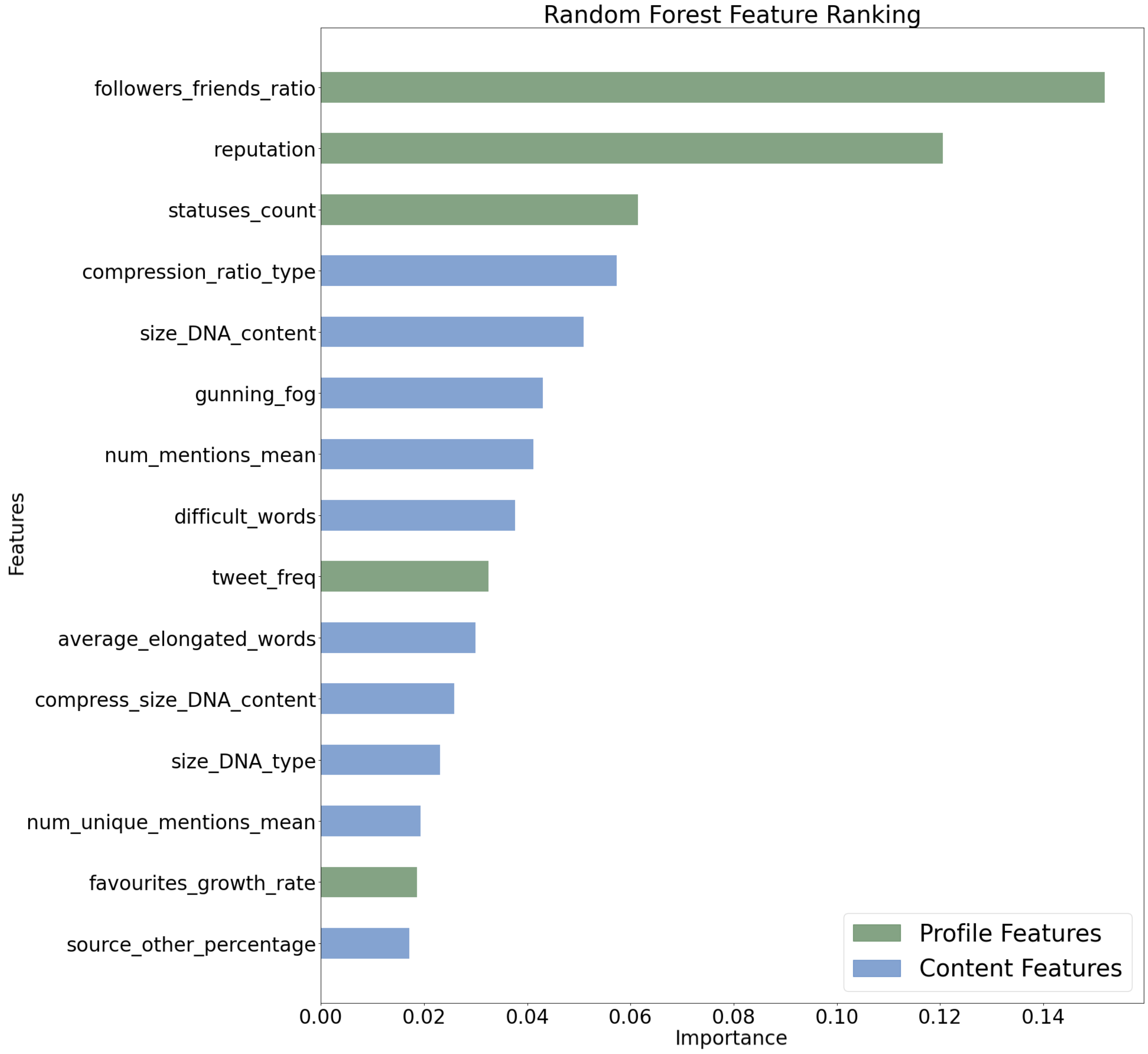}
        \caption{R.F. Importance Cresci-15}
        \label{fig:img2}
    \end{subfigure}
    \begin{subfigure}[b]{0.45\textwidth}
        \centering
        \includegraphics[width=\textwidth]{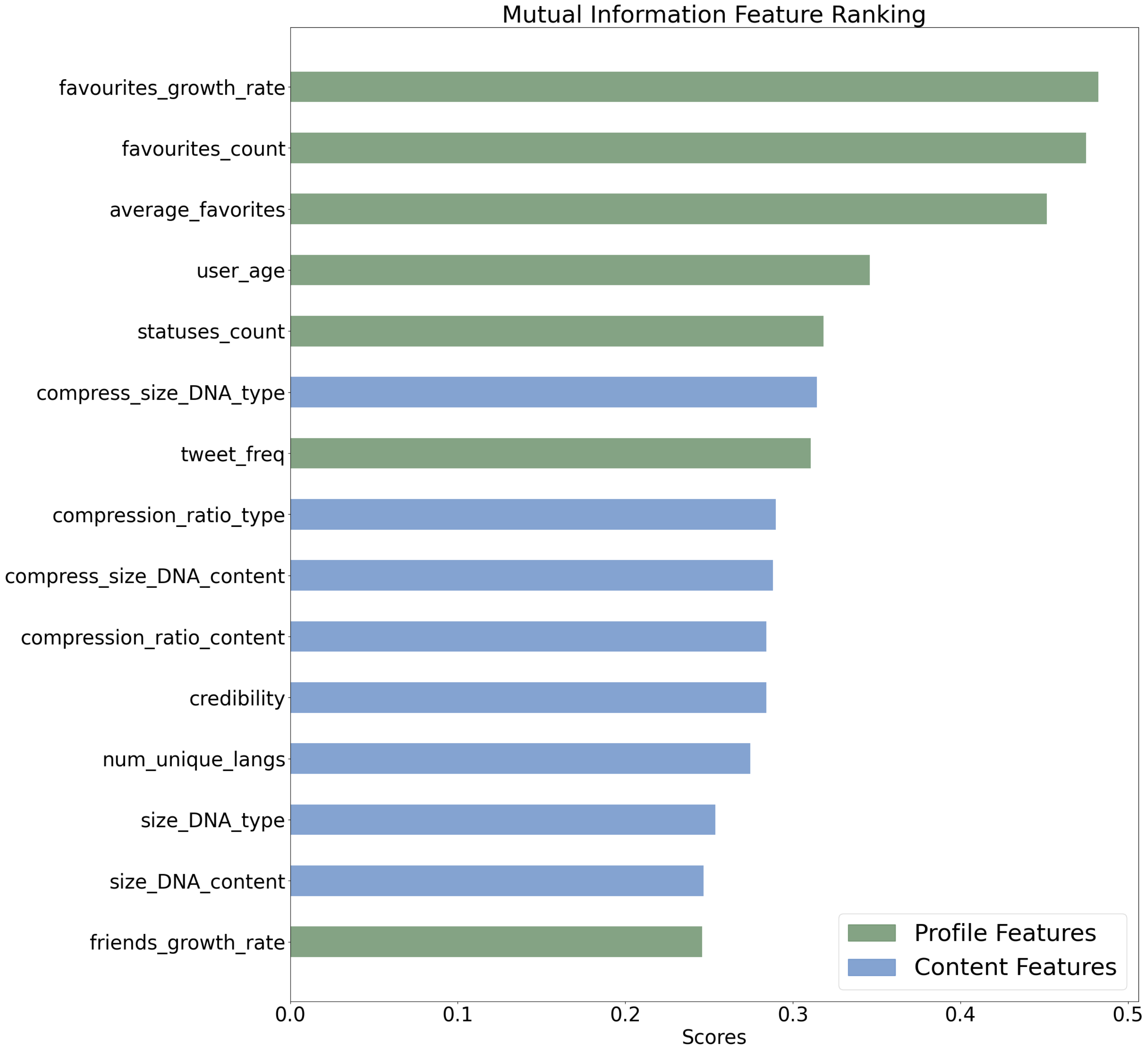}
        \caption{Mutual Information Cresci-17}
        \label{fig:img3}
    \end{subfigure}
    \hfill
    \begin{subfigure}[b]{0.45\textwidth}
        \centering
        \includegraphics[width=\textwidth]{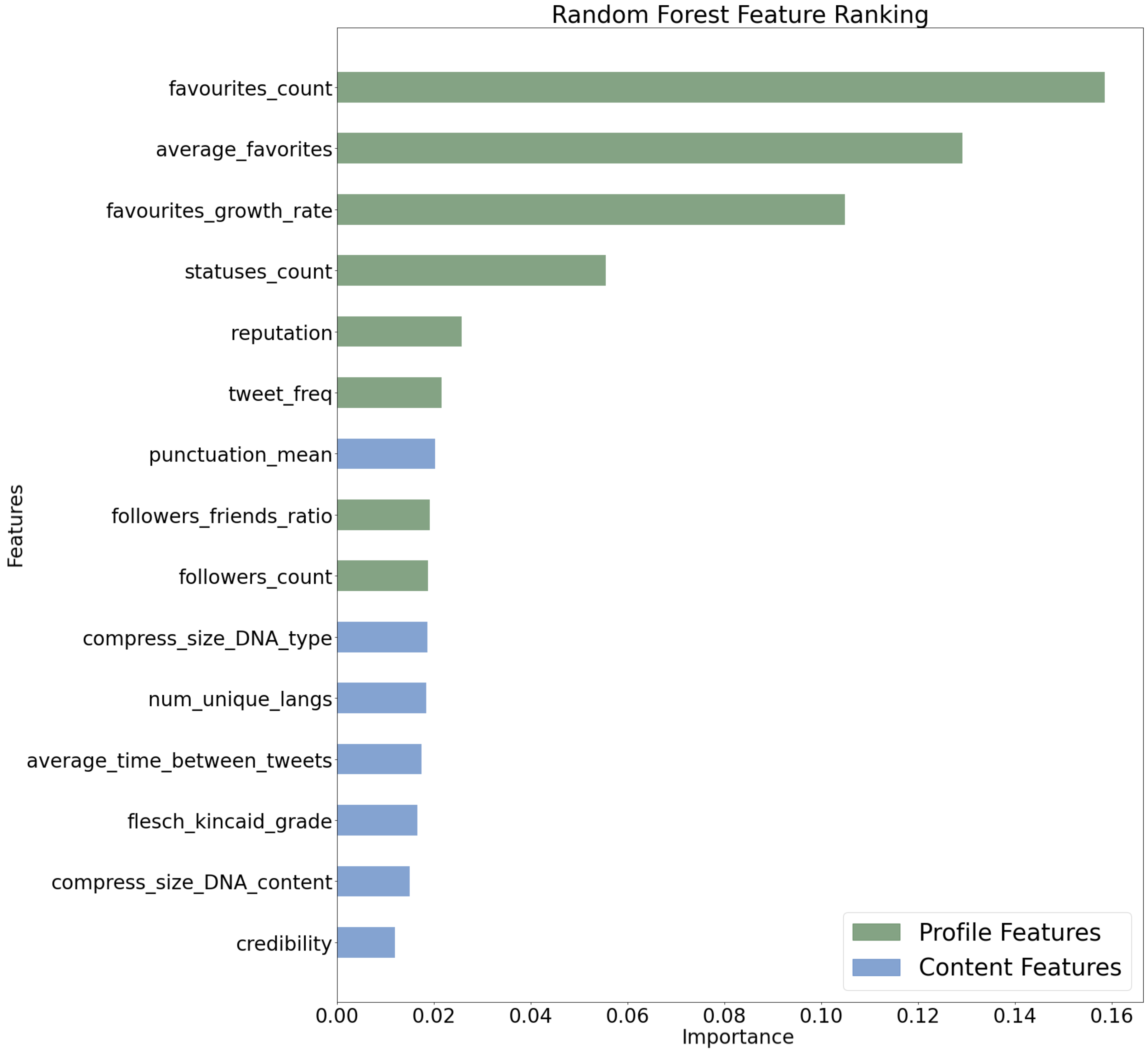}
        \caption{R.F. Importance Cresci-17}
        \label{fig:img4}
    \end{subfigure}
    \begin{subfigure}[b]{0.45\textwidth}
        \centering
        \includegraphics[width=\textwidth]{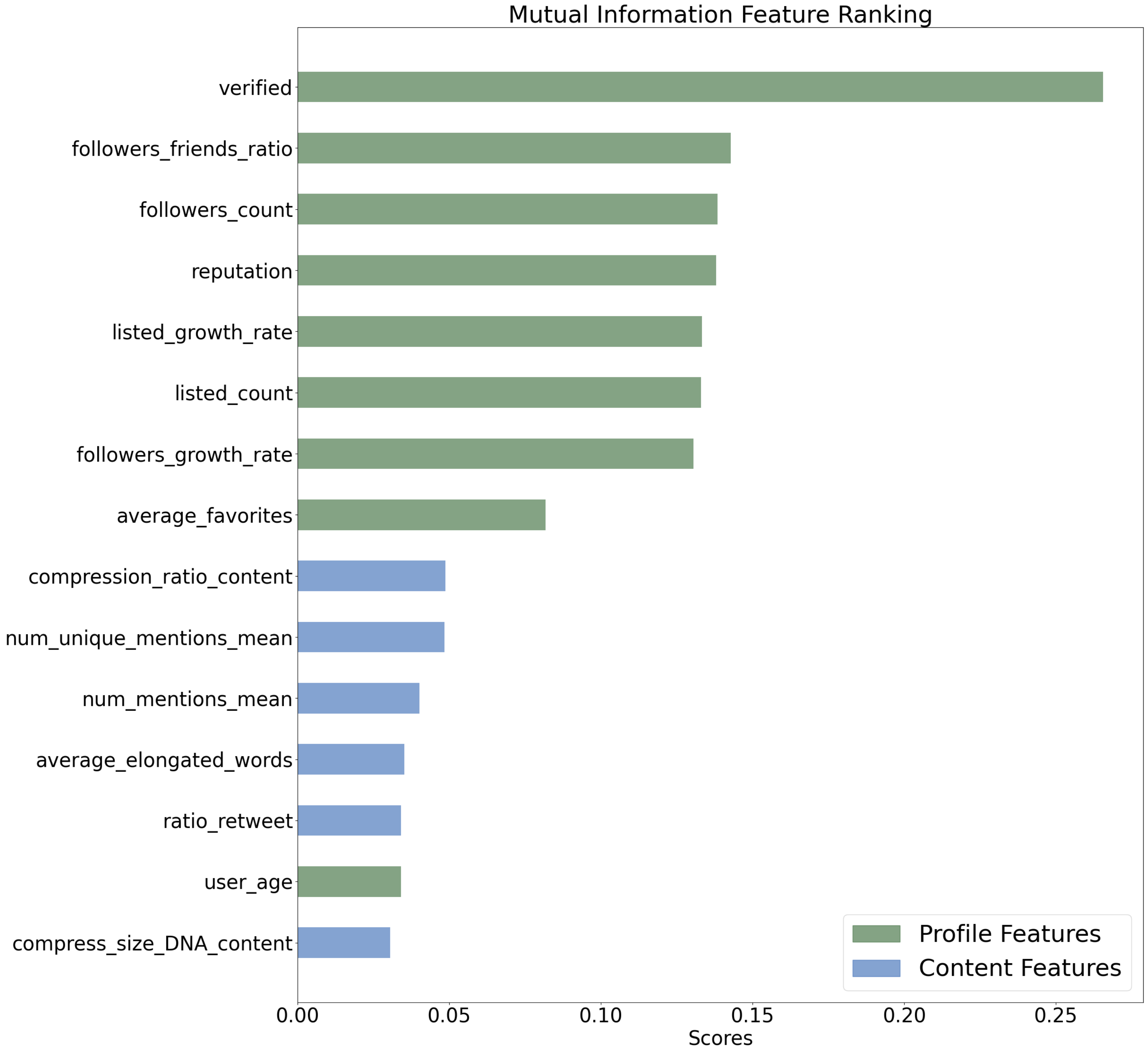}
        \caption{Mutual Information TwiBot-20}
        \label{fig:img5}
    \end{subfigure}
    \hfill
    \begin{subfigure}[b]{0.45\textwidth}
        \centering
        \includegraphics[width=\textwidth]{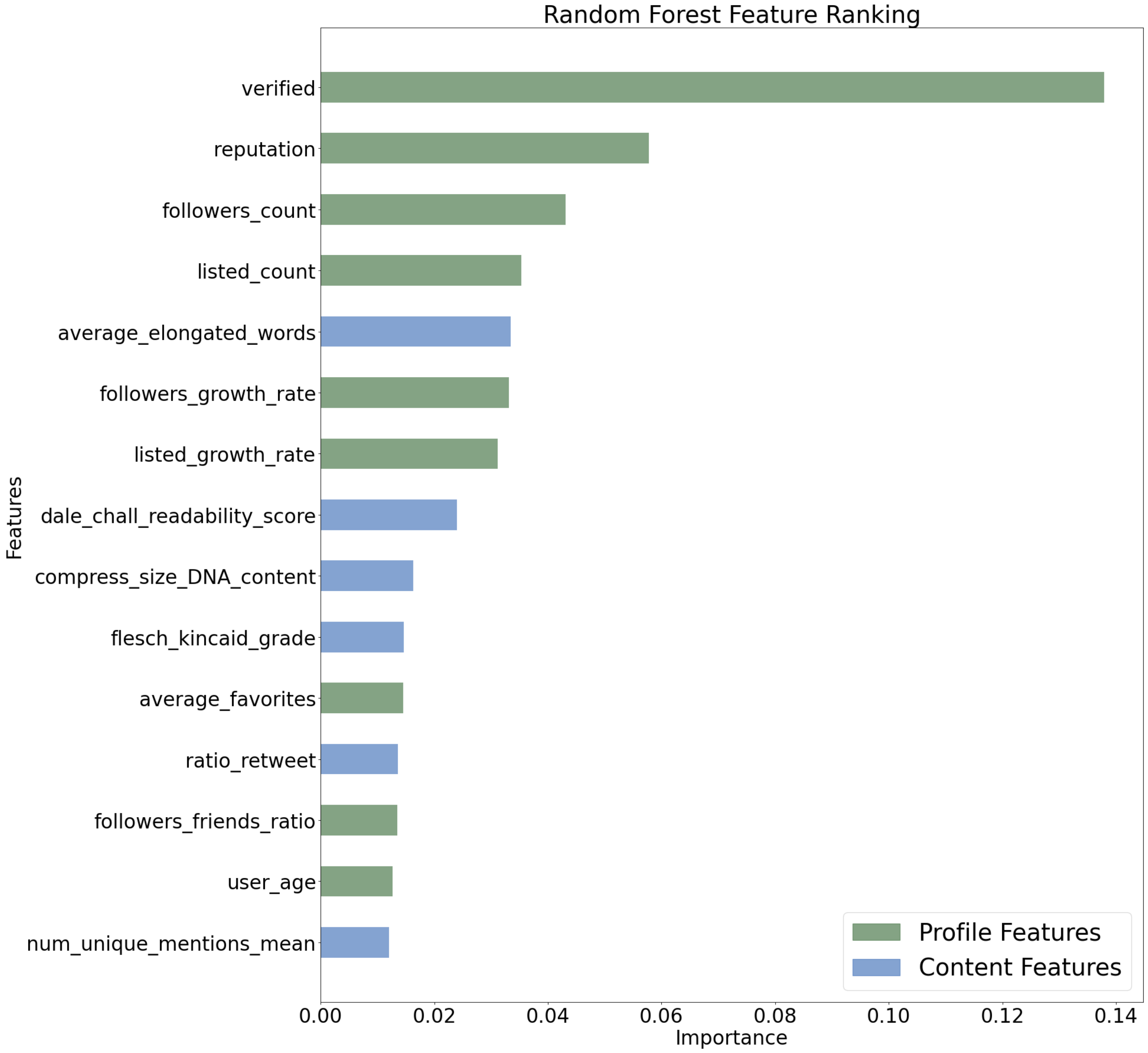}
        \caption{R.F. Importance TwiBot-20}
        \label{fig:img6}
    \end{subfigure}
    \caption{Feature ranking for all datasets}
    \label{fig:feature_importance_all_datasets}
\end{figure}

\begin{figure}[htbp]
    \centering
    \begin{subfigure}[b]{0.65\textwidth}
        \centering
        \includegraphics[width=\textwidth]{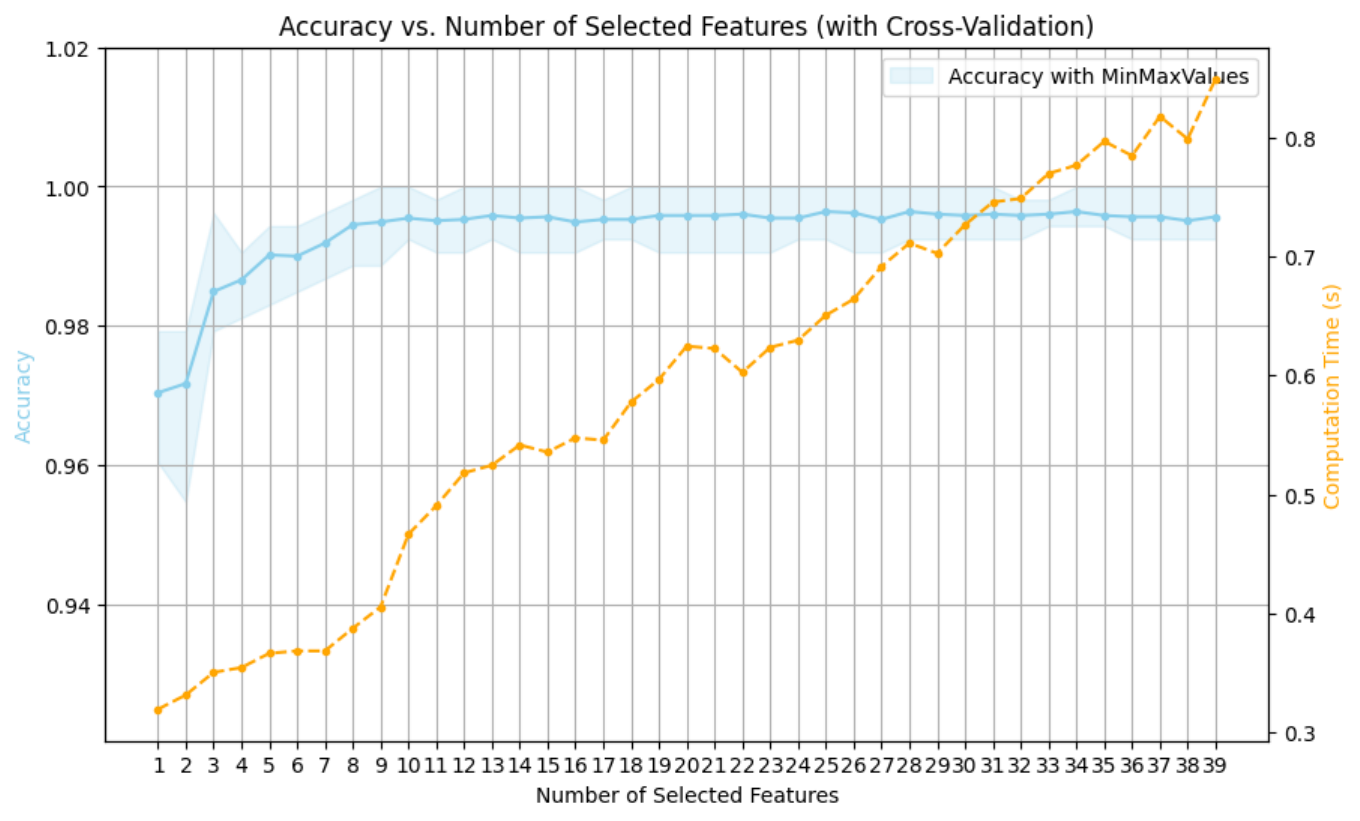}
        \caption{Cresci-15}
        \label{fig:img1}
    \end{subfigure}
    \hfill
    \begin{subfigure}[b]{0.65\textwidth}
        \centering
        \includegraphics[width=\textwidth]{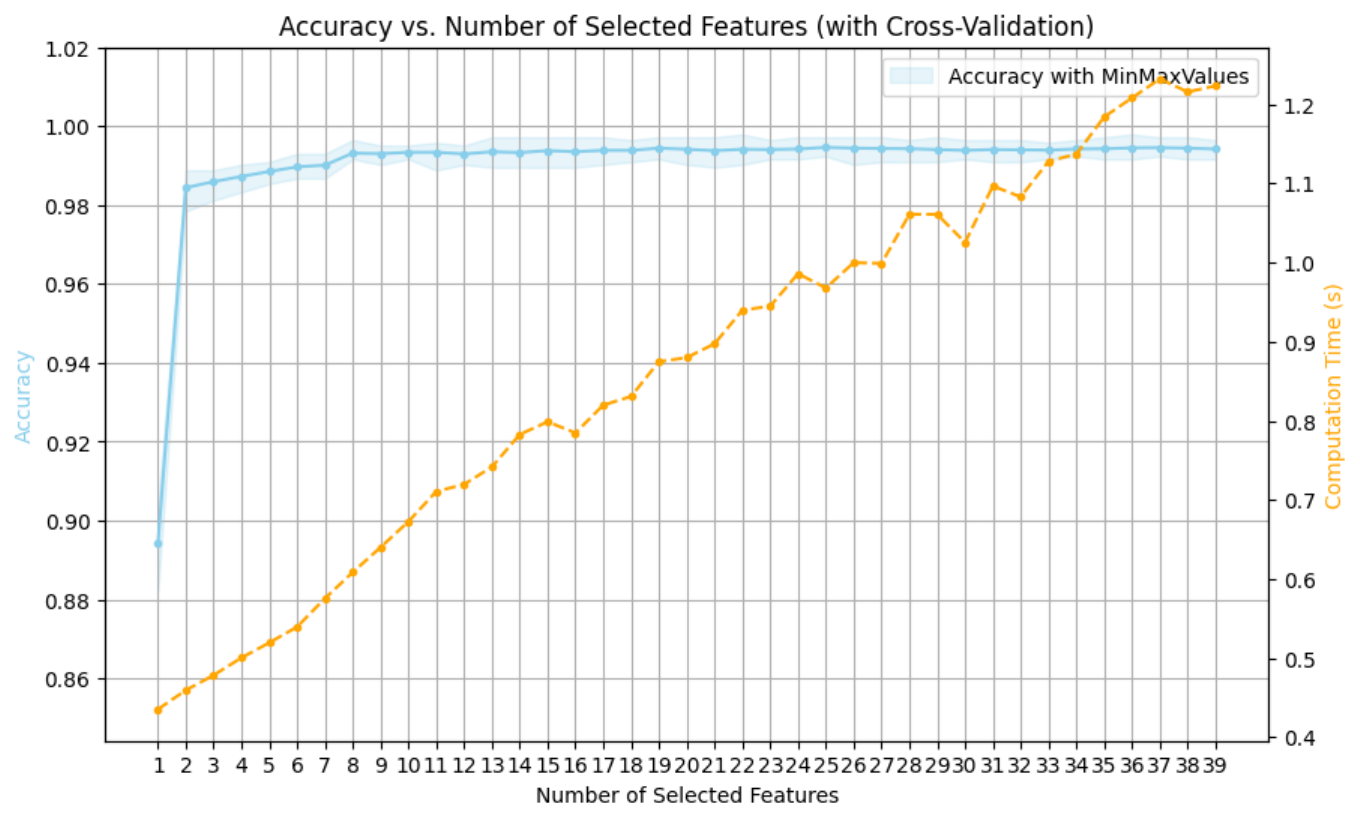}
        \caption{Cresci-17}
        \label{fig:img2}
    \end{subfigure}
    \hfill
    \begin{subfigure}[b]{0.65\textwidth}
        \centering
        \includegraphics[width=\textwidth]{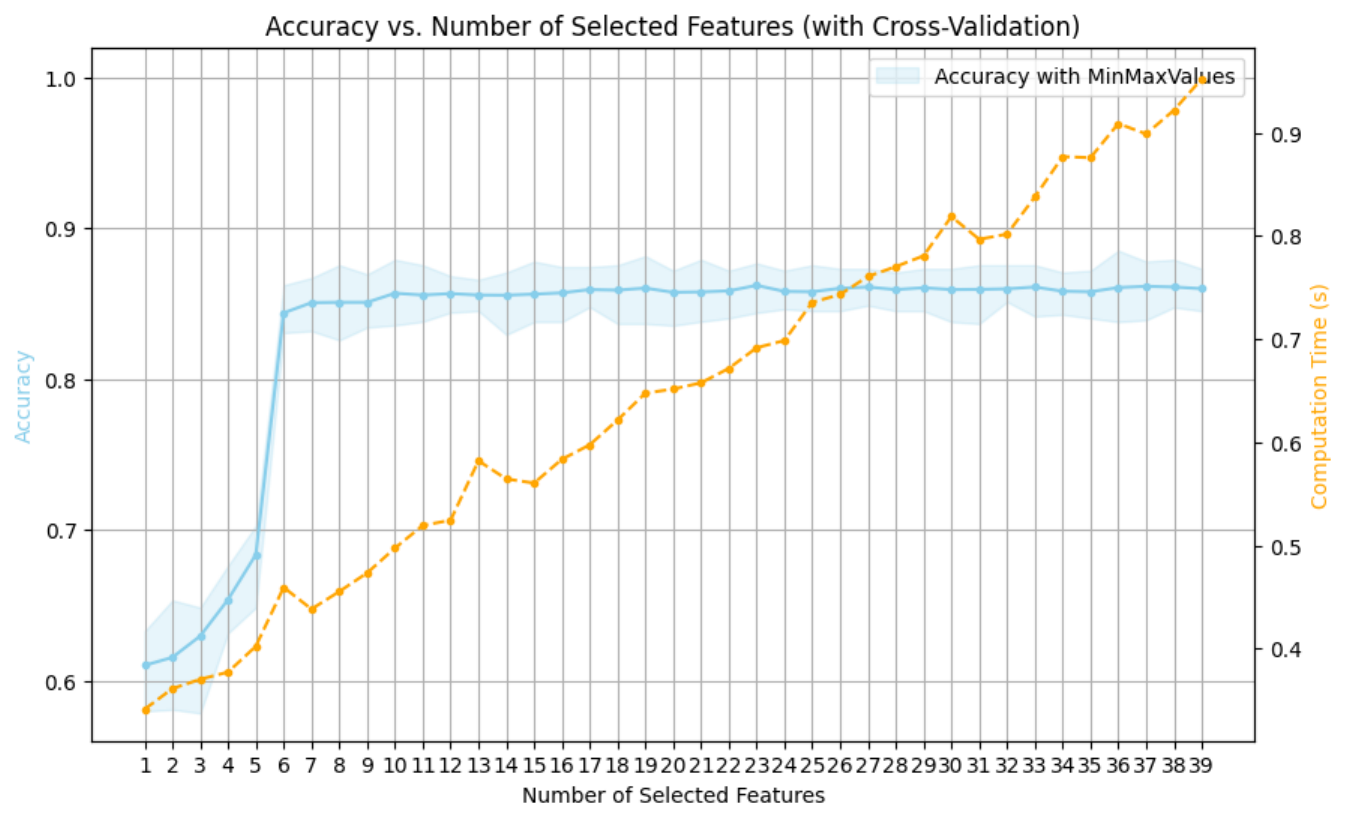}
        \caption{TwiBot-20}
        \label{fig:img3}
    \end{subfigure}
    \caption{Results of classification depending on the number of features selected and the computation time using RF.}
    \label{fig:accuracy_vs_features}
\end{figure}

In Figure \ref{fig:feature_importance_all_datasets} we can see the ranking of features in the three datasets using Mutual Information and Random Forest Importance, in addition we can see where each feature comes from, in green the features coming from the account and in blue the ones coming from the content. Only these two methods have been visualised in order to show one method that takes into account the relationships between features and one that does not. After this analysis of the features, Random Forest Importance has been chosen as the base method for feature selection. The reasons for this choice are: filtering methods do not take into account the correlation between features and, although some generalisation is lost, the embedded methods are more accurate than the methods belonging to the other two categories \cite{venkatesh2019review}. In Figure \ref{fig:accuracy_vs_features} we can see 40 runs with cross-validation of 10 of a Random Forest on each dataset comparing the accuracy obtained with each subset of features as well as the computation time required for each run. 

In Table \ref{cresci-17-model-selection} we can see the results of model selection in the Cresci-17 dataset. It should be noted that since the feature selection method is Random Forest Importance, it is expected that this same model will be among the first positions since embedded methods are dependent on the model in which they are embedded.

\definecolor{LightCyan}{rgb}{0.71, 0.94, 0.96}
\newcolumntype{x}{>{\columncolor{LightCyan}}c}
\begin{table}[!ht]
    \centering

    \resizebox{0.98\textwidth}{!}{\begin{tabular}{llllllxx}    
        \hline
        \textbf{Model} & \textbf{Accuracy} & \textbf{AUC} & \textbf{Recall} & \textbf{Prec.} & \textbf{F1} & \textbf{TT (Sec)} \\ \hline
        \textbf{Random Forest Classifier} & \textbf{0.9943} & \textbf{0.9997} & 0.9901 & \textbf{0.9865} & \textbf{0.9883} & 3.0560 \\ 
        Light Gradient Boosting Machine & 0.9939 & \textbf{0.9997} & 0.9889 & 0.9862 & 0.9875 & 2.0060 \\ 
        Extra Trees Classifier & 0.9937 & 0.9994 & \textbf{0.9951} & 0.9794 & 0.9872 & 1.8740 \\ 
        Extreme Gradient Boosting & 0.9935 & \textbf{0.9997} & 0.9893 & 0.9842 & 0.9867 & 1.5990 \\ 
        Gradient Boosting Classifier & 0.9929 & 0.9992 & 0.9856 & 0.9853 & 0.9854 & 8.7260 \\ 
        Ada Boost Classifier & 0.9916 & 0.9991 & 0.9823 & 0.9832 & 0.9827 & 2.8840 \\ 
        Decision Tree Classifier & 0.9877 & 0.9830 & 0.9741 & 0.9750 & 0.9745 & 1.5360 \\ 
        K Neighbors Classifier & 0.9824 & 0.9907 & 0.9548 & 0.9721 & 0.9633 & 1.4490 \\ 
        Linear Discriminant Analysis & 0.9691 & 0.9914 & 0.9124 & 0.9577 & 0.9342 & 1.4290 \\ 
        SVM-Linear-Kernel & 0.9665 & 0.0000 & 0.9219 & 0.9393 & 0.9298 & 1.4020 \\ 
        Ridge Classifier & 0.9627 & 0.0000 & 0.8885 & 0.9543 & 0.9198 & 1.3950 \\ 
        Logistic Regression & 0.9621 & 0.9882 & 0.8914 & 0.9488 & 0.9189 & 1.4930 \\ 
        Quadratic Discriminant Analysis & 0.8742 & 0.9748 & 0.9860 & 0.6632 & 0.7923 & 1.3580 \\ 
        Naive Bayes & 0.8623 & 0.9220 & 0.9893 & 0.6479 & 0.7809 & 1.3020 \\ 
        Dummy Classifier & 0.7582 & 0.5000 & 0.0000 & 0.0000 & 0.0000 & 1.1680 \\ \hline
    \end{tabular}}
    \caption{Cresci-17 classification model selection (top 8 features)}
    \label{cresci-17-model-selection}
\end{table}

\begin{table}[H]
    \centering
    \resizebox{1\textwidth}{!}{\begin{tabular}{l|cccc|cccc|cccc}
        \hline
        \multirow{ 2}{*}{\textbf{Method}} & \multicolumn{4}{c|}{\textbf{C-15}} & \multicolumn{4}{c|}{\textbf{C-17}} & \multicolumn{4}{c}{\textbf{T-20}} \\ 
        & \textbf{Acc.} & \textbf{Prec.} & \textbf{Rec.} & \textbf{F1} & \textbf{Acc.} & \textbf{Prec.} & \textbf{Rec.} & \textbf{F1} & \textbf{Acc.} & \textbf{Prec.} & \textbf{Rec.} & \textbf{F1} \\ \hline
        SGBot \cite{yang2020scalable} & 0.771 & \underline{0.995} & 0.637 & 0.496 & 0.921 & 0.983 & 0.909 & 0.946 & 0.816 & 0.764 & 0.949 & 0.849 \\ 
        Kudugunta et al. \cite{kudugunta2018deep} & 0.753 & \textbf{1.000} & 0.609 & 0.496 & 0.883 & 0.985 & 0.859 & 0.917 & 0.596 & \underline{0.804} & 0.335 & 0.473 \\ 
        Hayawi et al. \cite{hayawi2022deeprobot} & 0.843 & 0.930 & 0.793 & 0.205 & 0.908 & 0.955 & 0.922 & 0.938 & 0.731 & 0.716 & 0.835 & 0.771 \\ 
        BotHunter \cite{beskow2018bot} & 0.965 & 0.986 & 0.915 & 0.496 & 0.881 & 0.987 & 0.854 & 0.916 & 0.752 & 0.728 & 0.868 & 0.791 \\ 
        NameBot \cite{beskow2019its} & 0.770 & 0.768 & 0.911 & 0.385 & 0.768 & 0.804 & 0.918 & 0.857 & 0.591 & 0.587 & 0.705 & 0.651 \\ 
        Abreu et al. \cite{abreu2020twitter} & 0.757 & 0.991 & 0.621 & 0.538 & 0.927 & 0.983 & 0.920 & 0.950 & 0.734 & 0.722 & 0.828 & 0.771 \\ 
        Cresci et al. \cite{cresci2016dna} & 0.370 & 0.006 & 0.667 & 0.012 & 0.335 & 0.130 & 0.953 & 0.228 & 0.478 & 0.077 & 0.675 & 0.137 \\ 
        Wei et al. \cite{wei2019twitter} & 0.961 & 0.917 & 0.753 & 0.827 & 0.893 & 0.859 & 0.721 & 0.784 & 0.713 & 0.610 & 0.540 & 0.573 \\ 
        BGSRD \cite{guo2021social} & 0.878 & 0.865 & 0.956 & 0.130 & 0.759 & 0.759 & \textbf{1.000} & 0.863 & 0.664 & 0.676 & 0.732 & 0.701 \\ 
        RoBERTa \cite{liu2019roberta} & 0.970 & 0.976 & 0.941 & 0.959 & 0.972 & 0.924 & 0.963 & 0.943 & 0.755 & 0.739 & 0.724 & 0.731 \\ 
        T5 \cite{raffel2020exploring} & 0.923 & 0.910 & 0.877 & 0.894 & 0.964 & 0.945 & 0.902 & 0.923 & 0.735 & 0.722 & 0.691 & 0.706 \\ 
        Efthimion et al. \cite{efthimion2018supervised} & 0.925 & 0.938 & 0.944 & 0.000 & 0.880 & 0.946 & 0.892 & 0.918 & 0.628 & 0.642 & 0.706 & 0.673 \\ 
        Kantepe et al. \cite{kantepe2017preprocessing} & 0.975 & 0.813 & 0.753 & 0.782 & 0.982 & 0.830 & 0.761 & 0.794 & 0.764 & 0.634 & 0.610 & 0.622 \\ 
        Miller et al. \cite{miller2014twitter} & 0.755 & 0.721 & \textbf{1.000} & 0.838 & 0.771 & 0.772 & \underline{0.991} & 0.868 & 0.645 & 0.607 & \textbf{0.974} & 0.748 \\ 
        Varol et al. \cite{varol2017online} & 0.932 & 0.922 & 0.974 & 0.947 & - & - & - & - & 0.787 & 0.780 & 0.844 & 0.811 \\ 
        Kouvela et al. \cite{kouvela2020bot} & 0.978 & \underline{0.995} & 0.968 & 0.982 & 0.984 & 0.992 & 0.990 & \underline{0.991} & \underline{0.840} & 0.793 & \underline{0.952} & \underline{0.865} \\ 
        Santos et al. \cite{ferreira2019uncovering} & 0.708 & 0.729 & 0.858 & 0.788 & 0.738 & 0.817 & 0.844 & 0.830 & - & 0.627 & 0.581 & 0.603 \\ 
        Lee et al. \cite{lee2011seven} & 0.982 & 0.987 & 0.985 & 0.986 & 0.988 & \textbf{0.996} & \underline{0.991} & \textbf{0.993} & 0.763 & 0.766 & 0.837 & 0.800 \\ 
        LOBO \cite{echeverri2018lobo} & \underline{0.984} & 0.985 & 0.991 & \underline{0.988} & 0.966 & \underline{0.993} & 0.961 & 0.977 & 0.757 & 0.748 & 0.878 & 0.808 \\ 
        Ilias, L., \& Roussaki, I. \cite{ilias2021detecting} & - & - & - & - & 0.991 & - & - & - & - & - & - & - \\ 
        Deepsbd \cite{fazil2021deepsbd} & - & - & - & - & \underline{0.992} & - & - & - & - & - & - & - \\ 
        Bottrinet \cite{wu2023bottrinet} & - & - & - & - & 0.962 & - & - & - & - & - & - & - \\ 
        OUProfiling \cite{heidari2022online} & - & - & - & - & 0.981 & - & - & - & - & - & - & - \\ \hline
        OURS & \textbf{0.996} & 0.993 & \underline{0.995} & \textbf{0.994} & \textbf{0.994} & 0.987 & 0.990 & 0.988 & \textbf{0.854} & \textbf{0.832} & 0.936 & \textbf{0.879} \\ 
    \end{tabular}}
    \caption{Comparison between different baselines from literature \cite{feng2022twibot}}
    \label{literature-results-comparation}
\end{table}

The primary objective of this study was to make a feature engineering process to evaluate the importance of the features in bot detection and improve the performance of the existing models. In order to compare our work with the previous works located in the literature, we have selected the three more accurate approaches present in the literature for bot detection: the deep learning approach proposed by \cite{ilias2021detecting} where they use a similar approach inferring 66 features, the proposed by \cite{heidari2022online} using user profiling techniques and finally we use the results of the implementation of several baselines in the literature made in \cite{feng2022twibot} where authors provide a comparative table with several evaluation metrics. In Table \ref{literature-results-comparation} we can see all these baselines and their accuracy compared with our proposal. The following subsections detail the findings, analyses, and interpretations of the acquired results, shedding light on relevant aspects and their implications.

\subsection{Ablation study}

\begin{figure}[H]
    \centering
    \caption{Significance of 15 most crucial account features across benchmark datasets employing various selection algorithms}
    \begin{subfigure}{0.44\linewidth}
        \includegraphics[width=\linewidth]{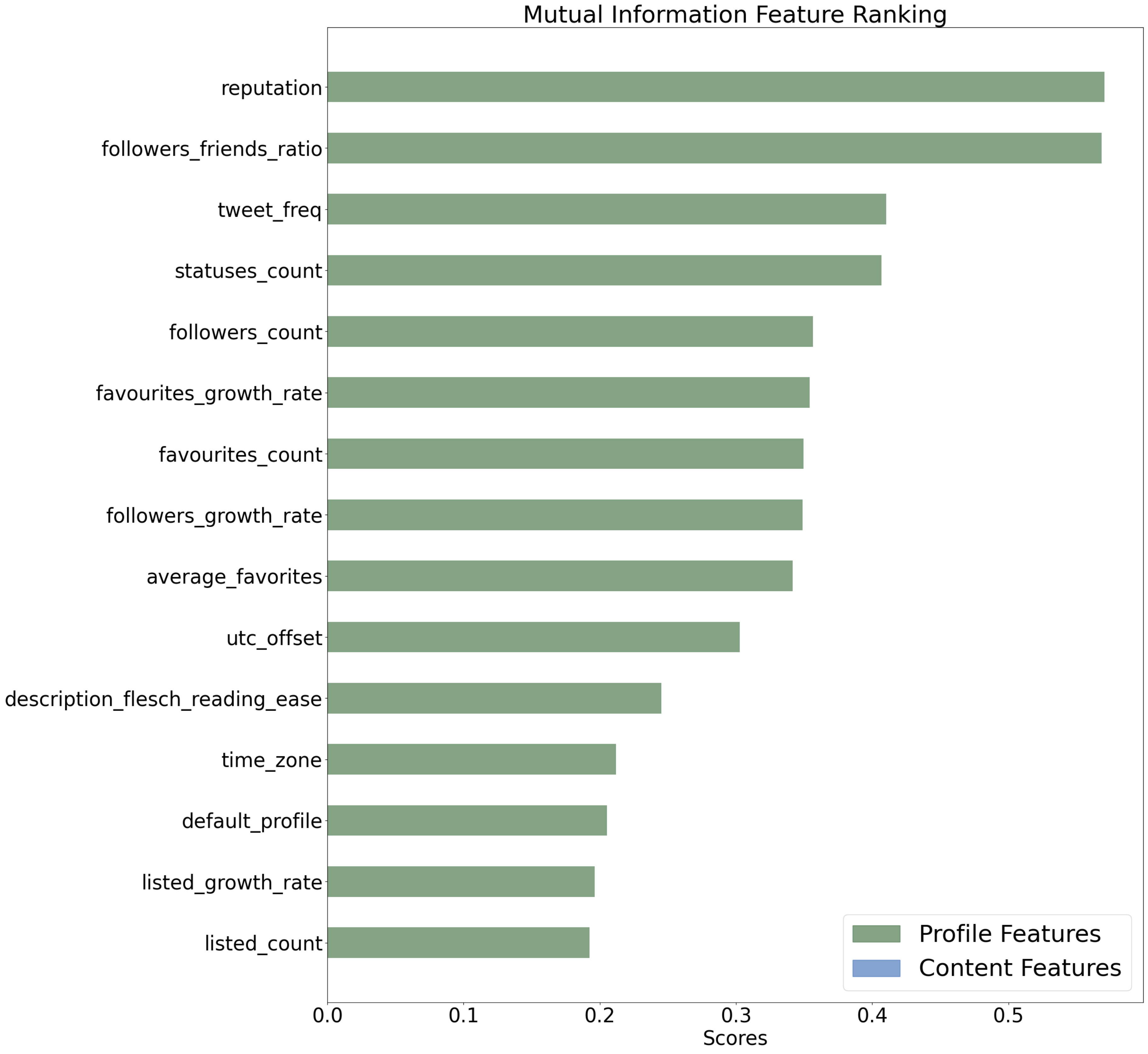}
        \caption{Mutual Information Cresci-15.}
        \label{fig:figure1}
    \end{subfigure}\hfill
    \begin{subfigure}{0.44\linewidth}
        \includegraphics[width=\linewidth]{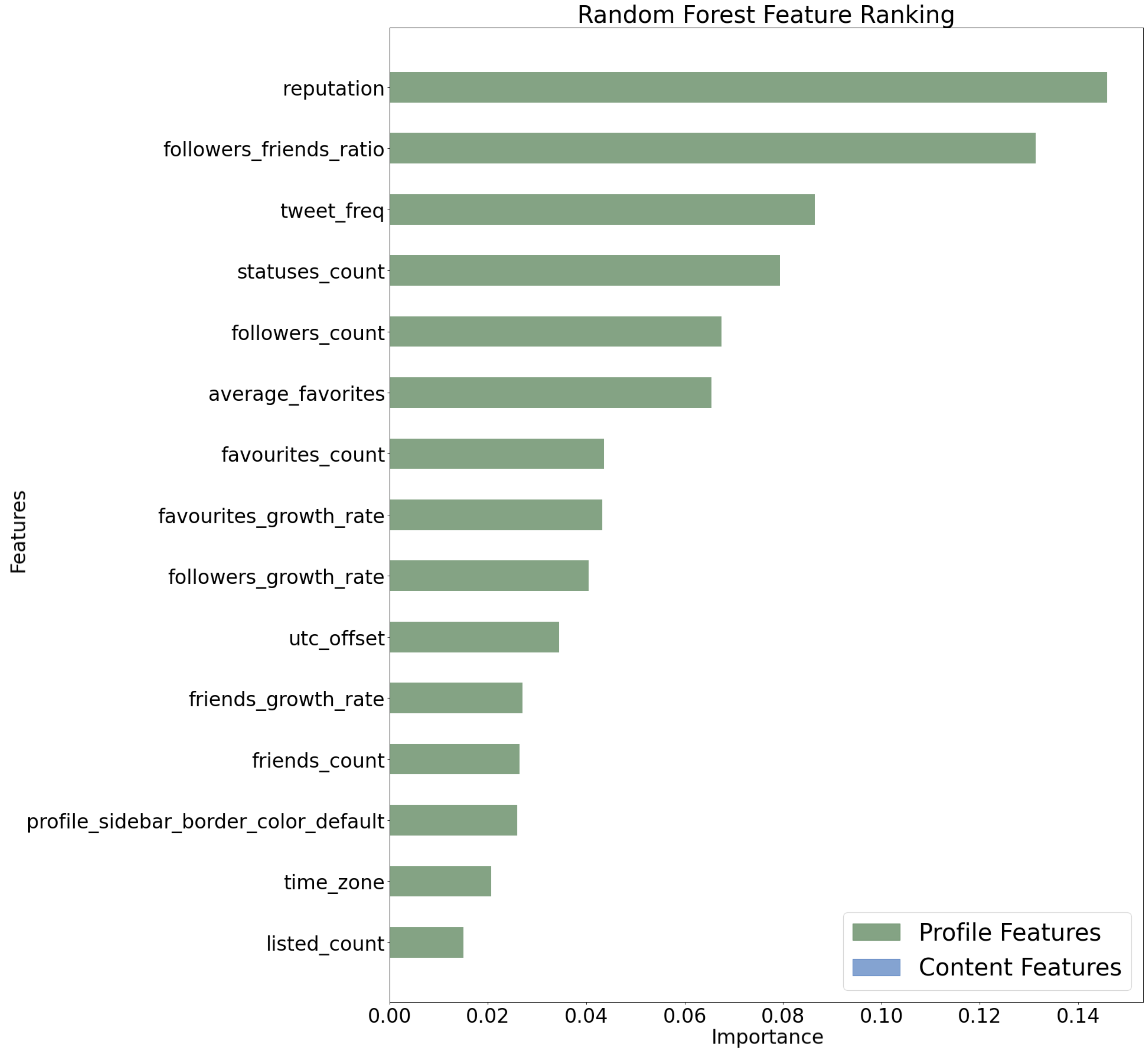}
        \caption{R.F. Importance Cresci-15.}
        \label{fig:figure2}
    \end{subfigure}\\
    \begin{subfigure}{0.44\linewidth}
        \includegraphics[width=\linewidth]{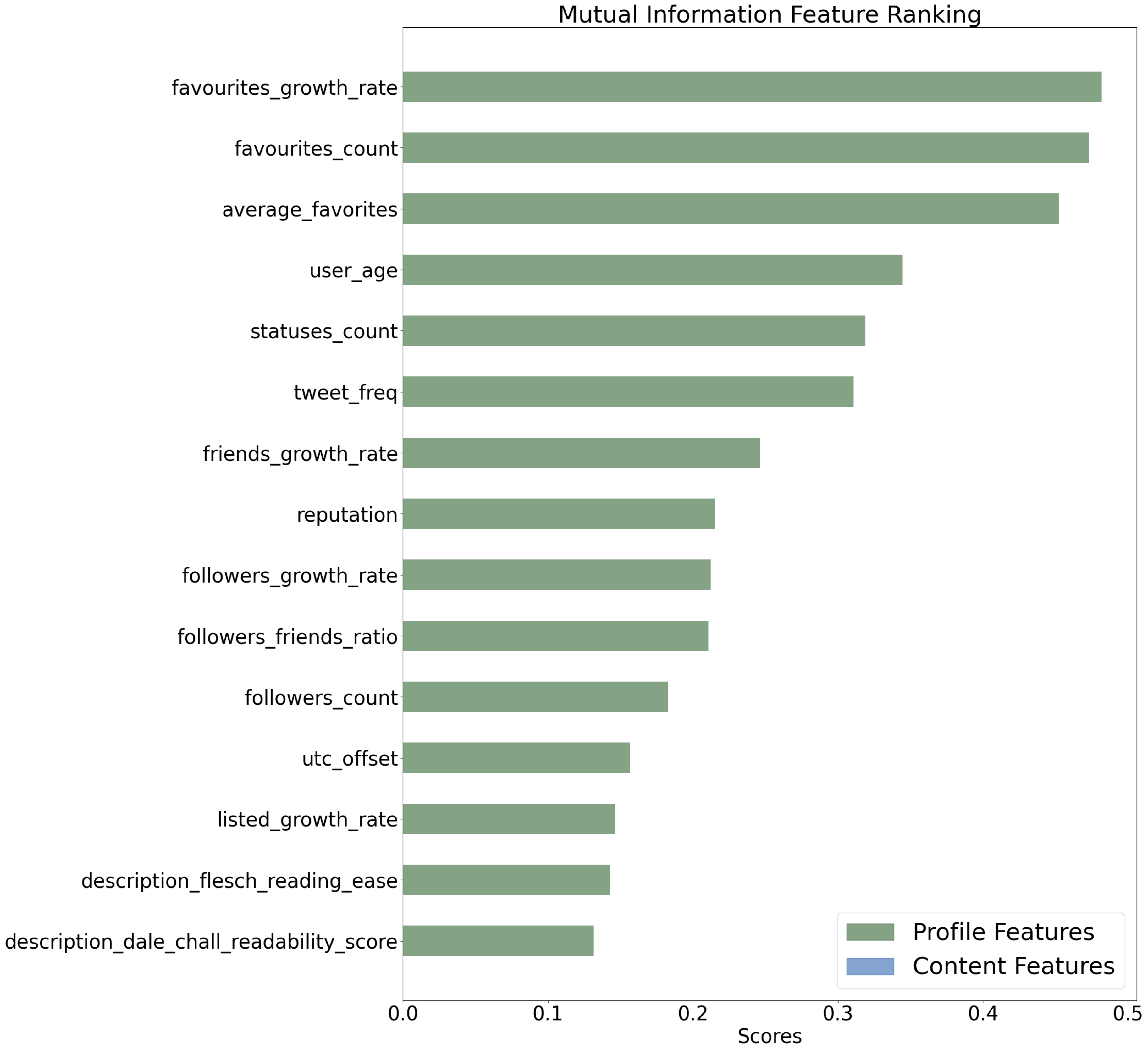}
        \caption{Mutual Information Cresci-17.}
        \label{fig:figure3}
    \end{subfigure}\hfill
    \begin{subfigure}{0.44\linewidth}
        \includegraphics[width=\linewidth]{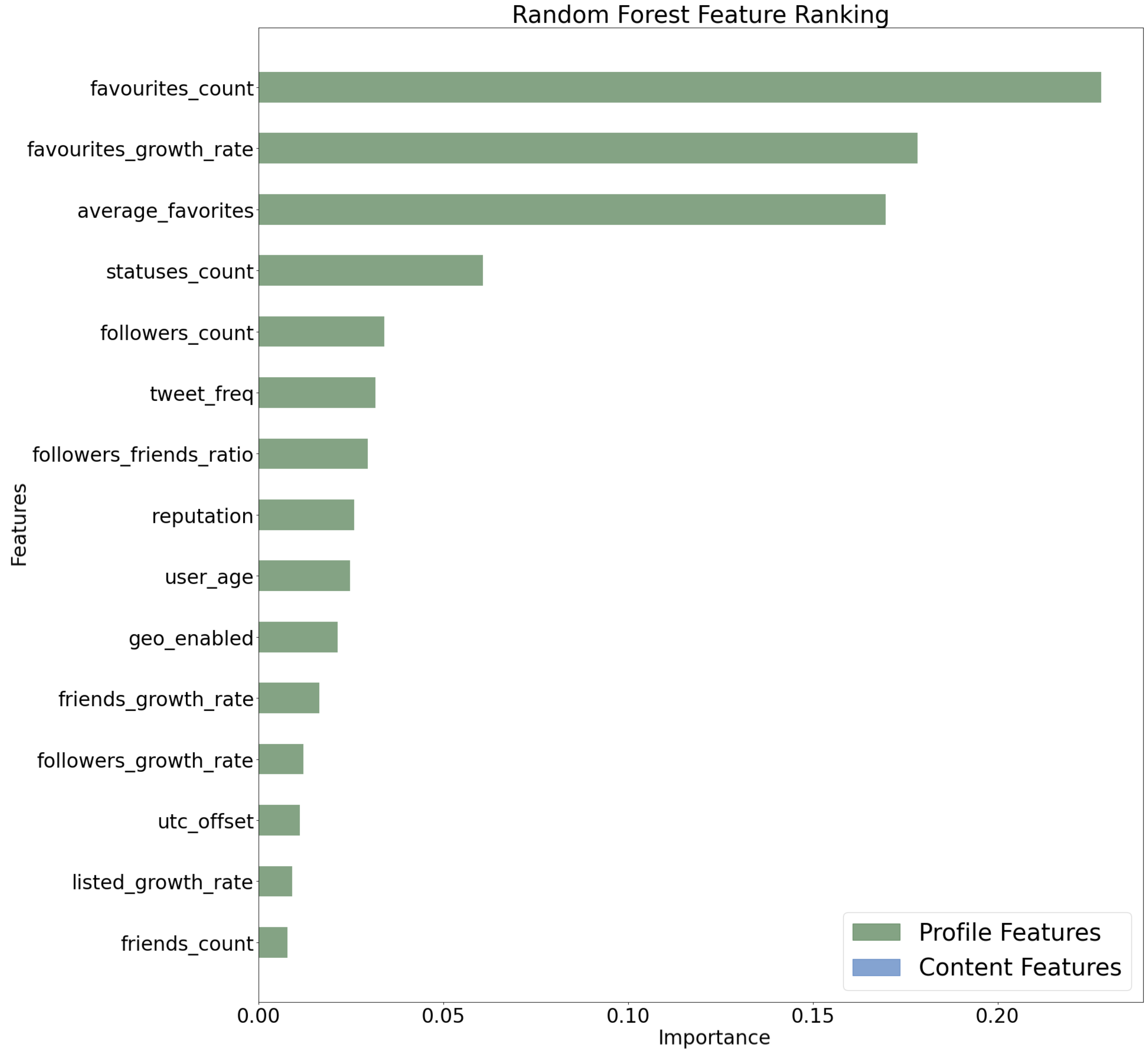}
        \caption{R.F. Importance Cresci-17.}
        \label{fig:figure4}
    \end{subfigure}
     \begin{subfigure}{0.44\linewidth}
        \includegraphics[width=\linewidth]{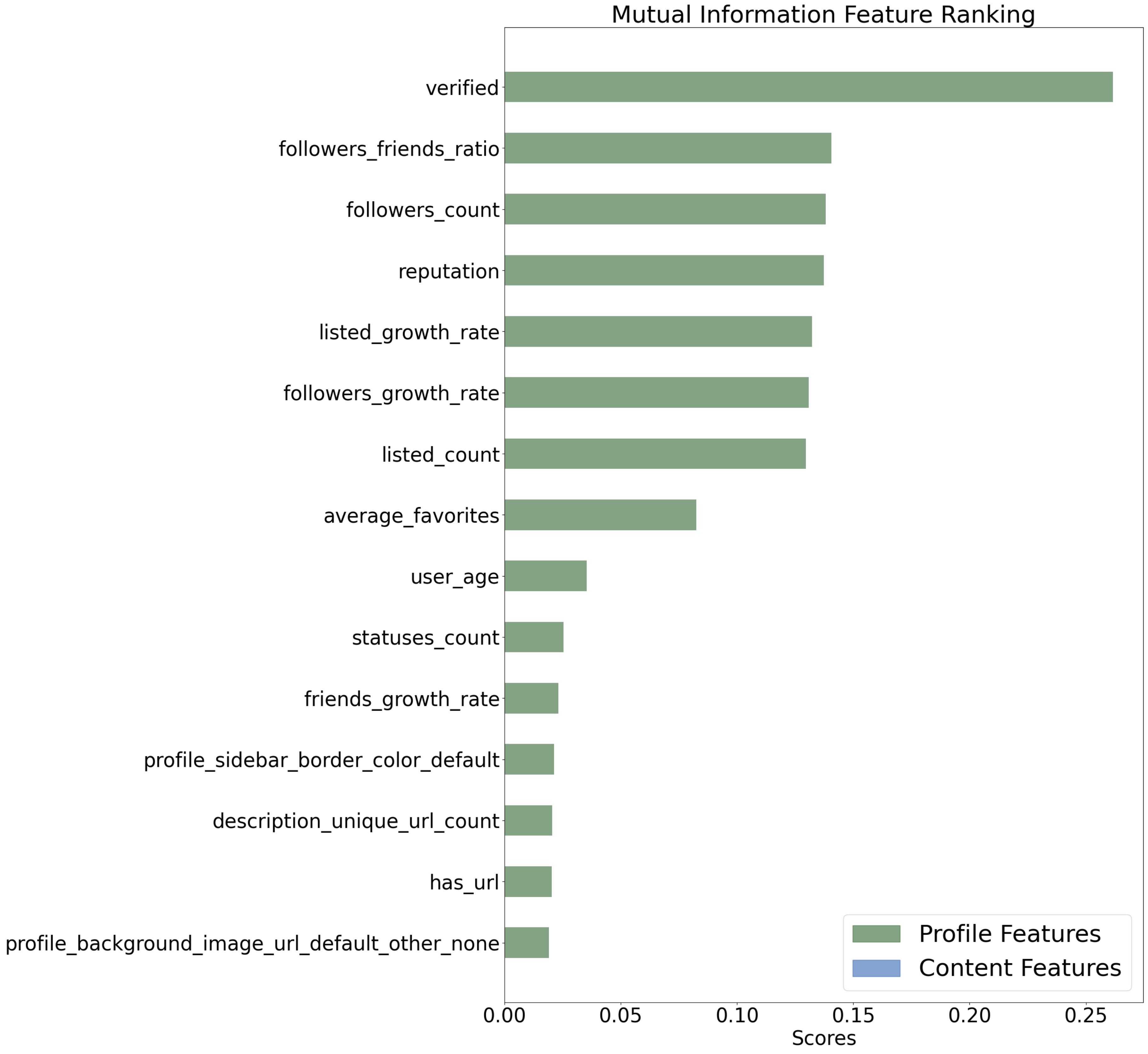}
        \caption{Mutual Information Twibot-20.}
        \label{fig:figure5}
    \end{subfigure}\hfill
    \begin{subfigure}{0.44\linewidth}
        \includegraphics[width=\linewidth]{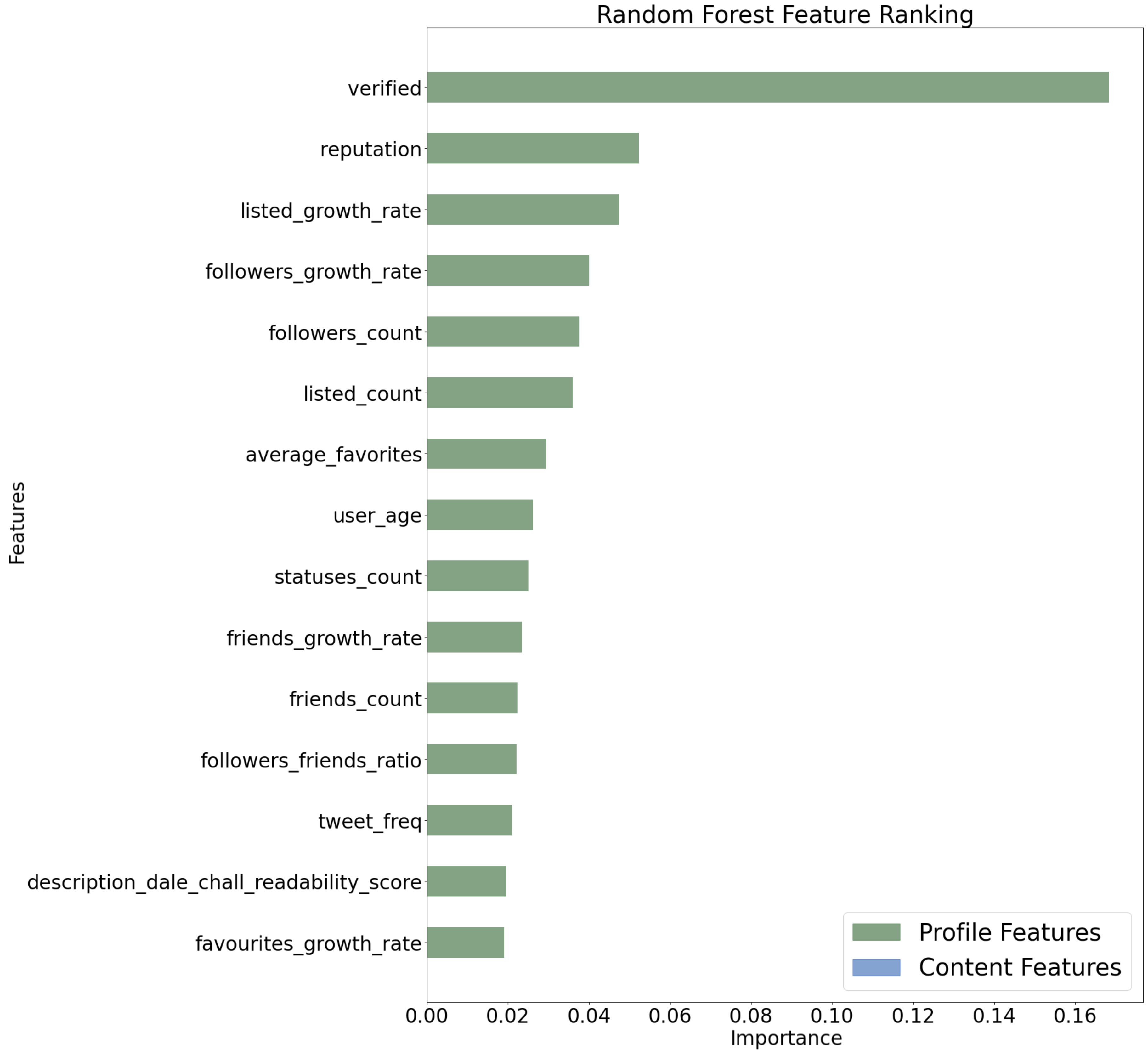}
        \caption{R.F. Importance Twibot-20.}
        \label{fig:figure6}
    \end{subfigure}
    \label{fig:ablationaccountfeatures}
\end{figure}

\begin{figure}[H]
    \centering
    \caption{Significance of 15 most crucial content features across benchmark datasets employing various selection algorithms.}
    \begin{subfigure}{0.44\linewidth}
        \includegraphics[width=\linewidth]{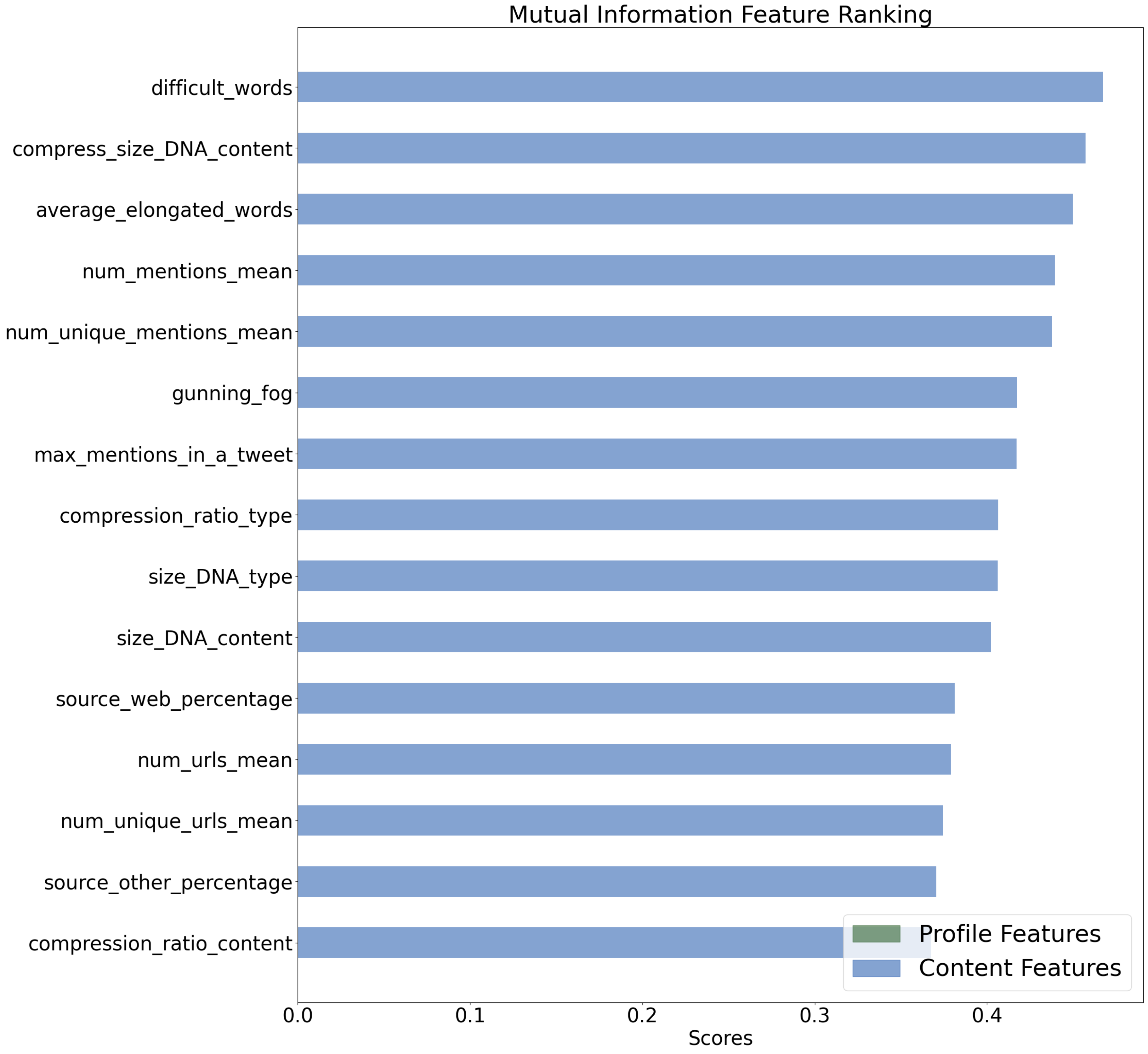}
     \caption{Mutual Information Cresci-15.}
        \label{fig:figure1}
    \end{subfigure}\hfill
    \begin{subfigure}{0.44\linewidth}
        \includegraphics[width=\linewidth]{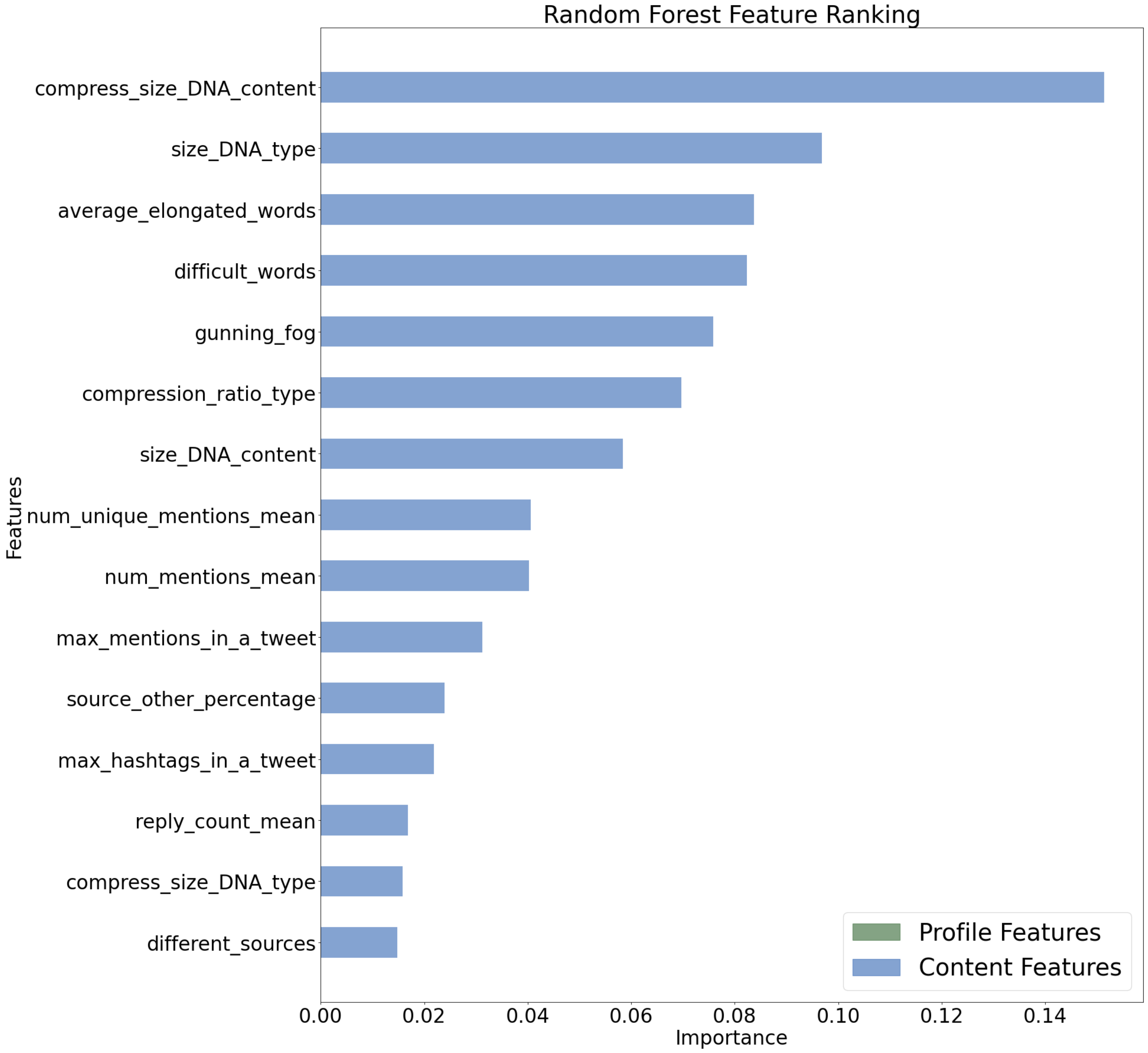}
         \caption{R.F. Importance Cresci-15.}
        \label{fig:figure2}
    \end{subfigure}\\
    \begin{subfigure}{0.44\linewidth}
        \includegraphics[width=\linewidth]{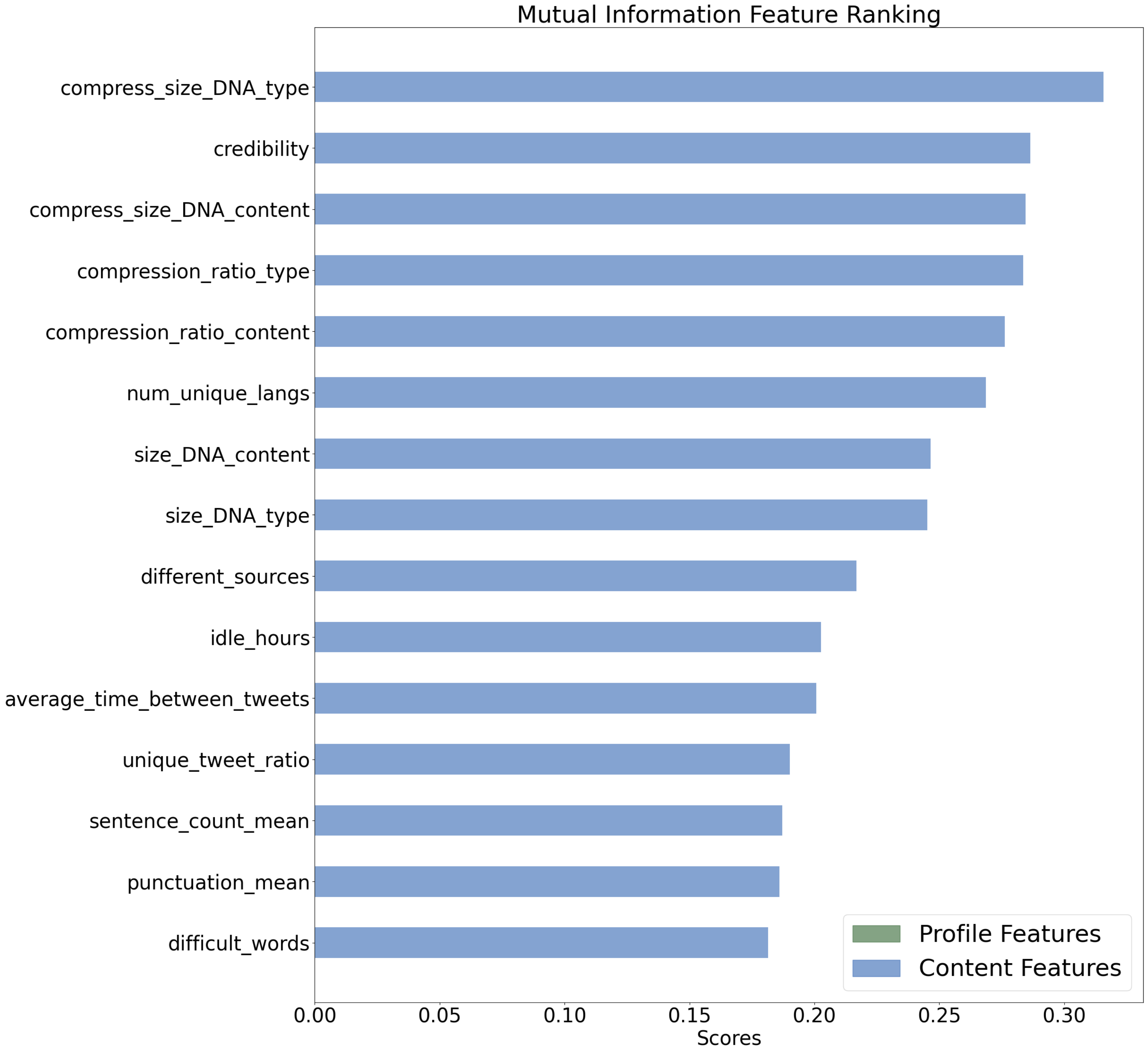}
        \caption{Mutual Information Cresci-17.}
        \label{fig:figure3}
    \end{subfigure}\hfill
    \begin{subfigure}{0.44\linewidth}
        \includegraphics[width=\linewidth]{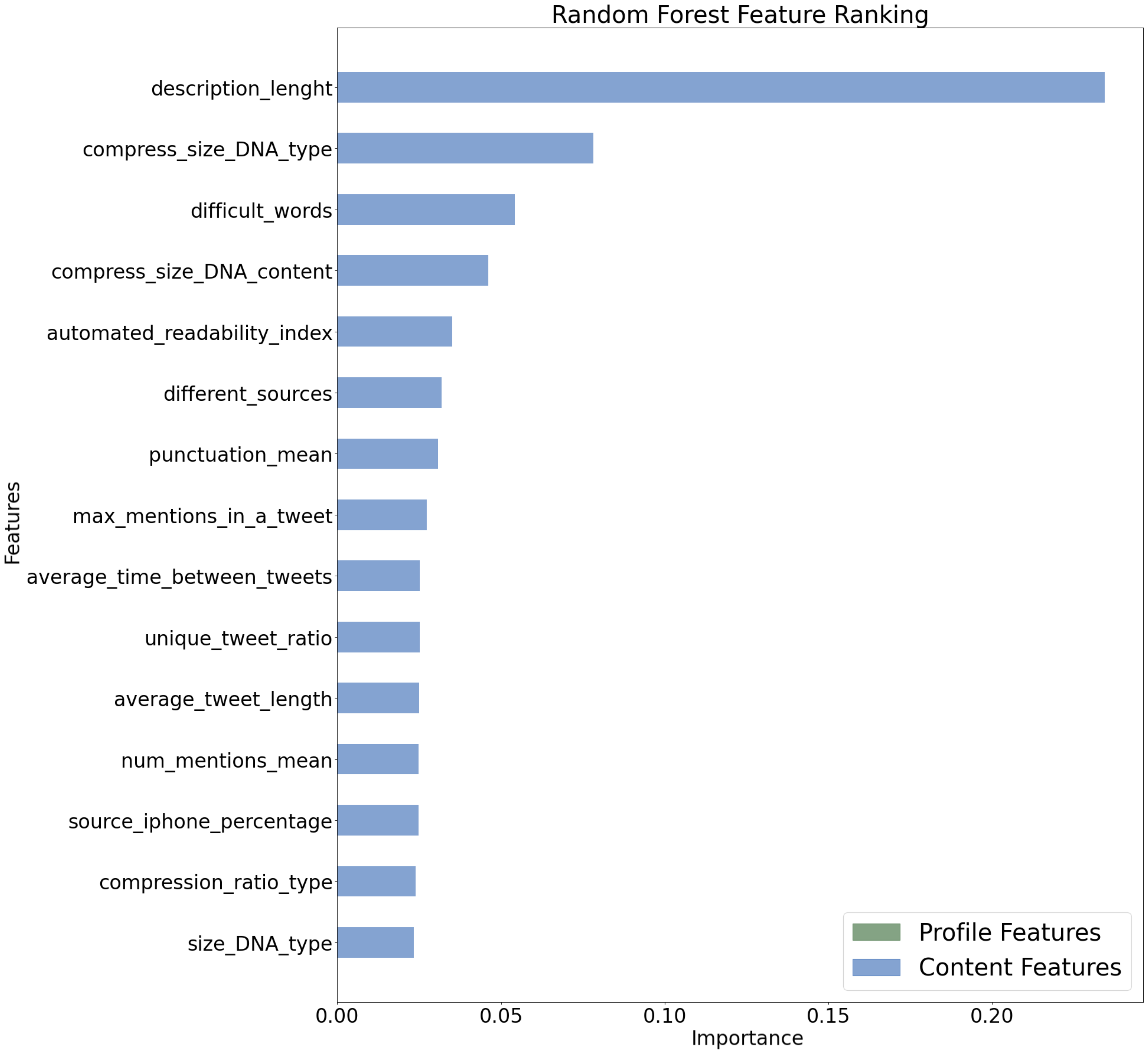}
        \caption{R.F. Importance Cresci-17.}
        \label{fig:figure4}
    \end{subfigure}
    \begin{subfigure}{0.44\linewidth}
        \includegraphics[width=\linewidth]{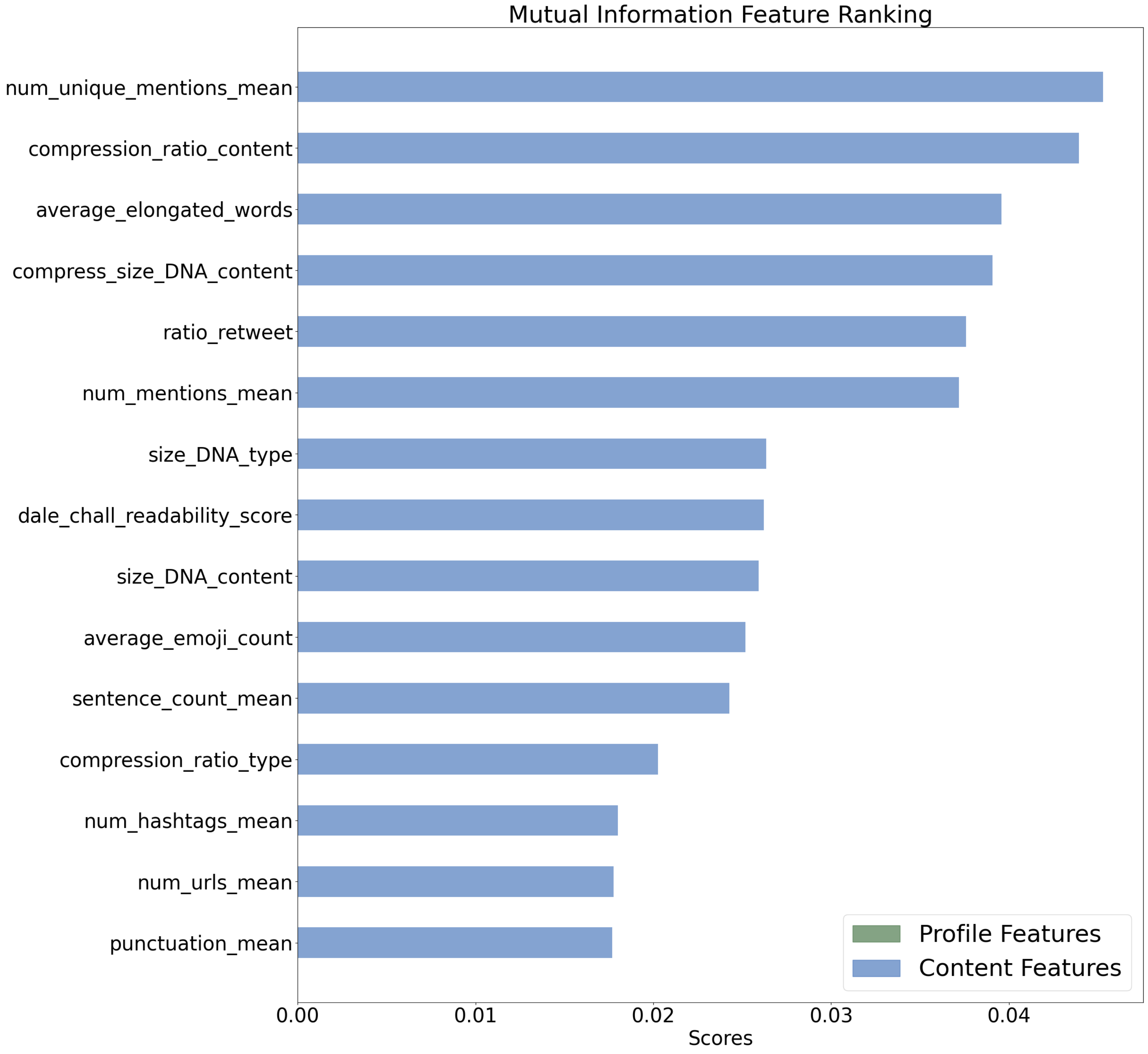}
        \caption{Mutual Information Twibot-20.}
        \label{fig:figure5}
    \end{subfigure}\hfill
    \begin{subfigure}{0.44\linewidth}
        \includegraphics[width=\linewidth]{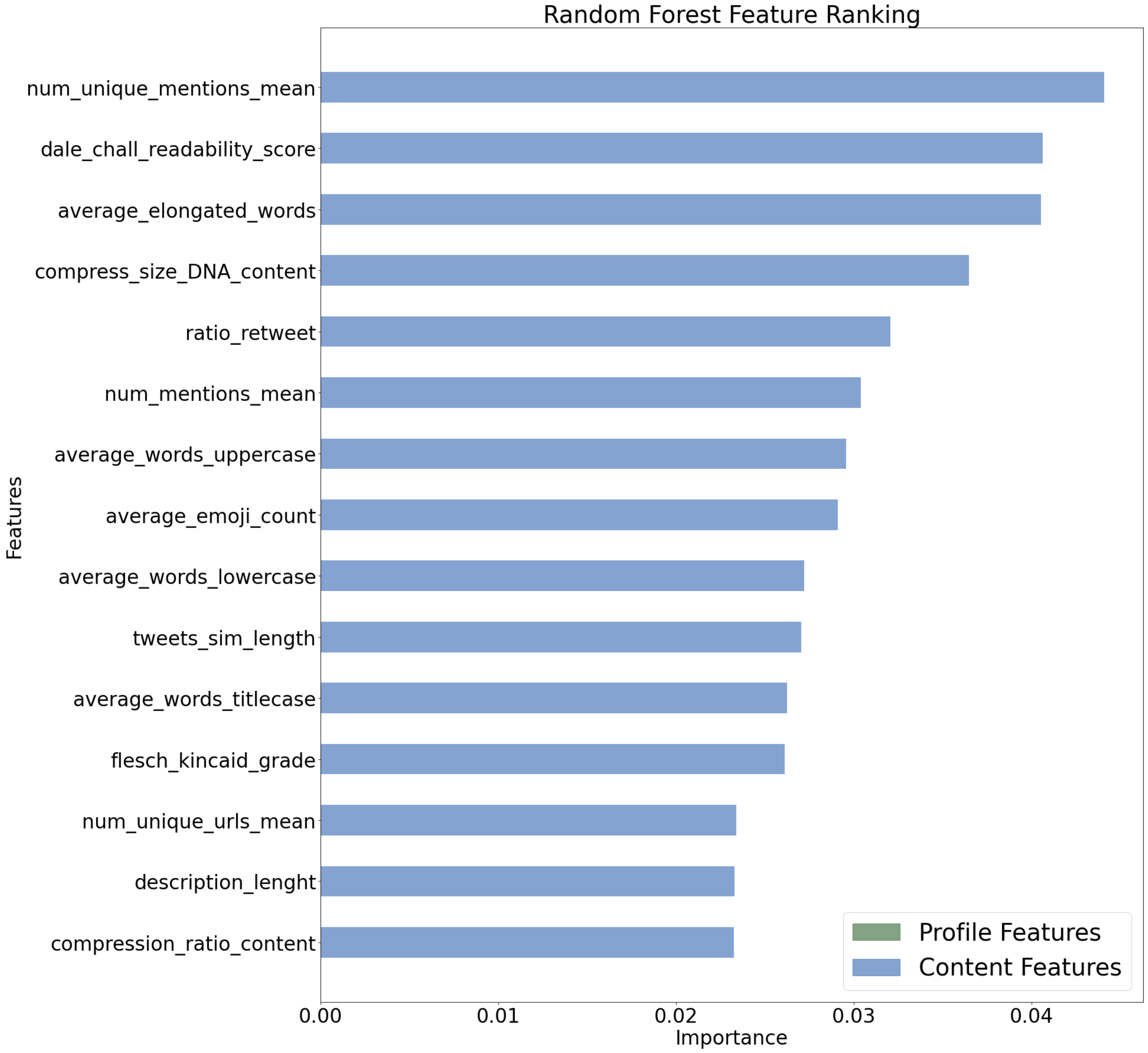}
         \caption{R.F. Importance Twibot-20.}
        \label{fig:figure6}
    \end{subfigure}

    \label{fig:ablationcontentfeatures}
\end{figure}

In this section, our objective is to determine the influence of different types of features on the classification system. To justify this, we conducted an ablation study on our sets of characteristics. This involved systematically assessing the impact first with solely content-based features, then exclusively with account-based features, and ultimately with a combined approach incorporating both types. It is important to note that, typically, ablation studies involve the addition or removal of different layers of data processing. In our case, as we delve into the distinctions among sets of features, our approach to the ablation study focuses on understanding the final behaviour of the classifier based on which features and combinations does thereof achieve optimal performance. For more robust comparison results, we performed an ablation study using two distinct feature selection methods. We selected the top-performing method among filter methods, specifically mutual information, and the most effective among embedded methods, namely random forest. In Figure \ref{fig:ablationaccountfeatures}, we present the results across different datasets and feature selection methods for the account features. In contrast, Figure \ref{fig:ablationcontentfeatures} illustrates the results focused solely on content features. 

It is worth noting that both feature selection techniques consider, albeit with some ranking variations, the same characteristics. This alignment is crucial for addressing our RQ1 on which features define a bot. Our findings suggest that concerning account-based features, those related to followers are paramount. Additionally, we observe the significance of account configuration, highlighted by the relevance of our proposed colour features.  In the realm of content features, the manner and intricacy of the writing style emerge as the most influential characteristic for defining bots. Notably, the features introduced by \cite{ilias2021detecting} prove to be particularly relevant in this context.

In determining the significance of feature types for bot categorisation on social media, Table \ref{tab:ablationcomparison} presents results for the best classification model when considering only account features, only content features, and a combination of both. The results represent the average outcomes across multiple executions. From the findings, it becomes evident that account features hold greater importance in the dataset and exhibit better capabilities for categorising bots. Notably, the most favourable result is associated with a combination of both account and content features.

From the findings, it becomes evident that account features hold greater importance in the dataset and exhibit better capabilities for categorising bots. Notably, the most favourable result is associated with a combination of both account and content features.

\begin{table}[!ht]
    \centering
    \resizebox{\textwidth}{!}{\begin{tabular}{llllll}
        \hline
        \textbf{Configuration - Dataset} & \textbf{C-15} & \textbf{C-17} & \textbf{TwiBot-20} \\ \hline
        Account Features & 0.9881 & 0.9912 &  0.7679\\
        Content Features & 0.9865 & 0.9414 & 0.6827 \\ \hline  
        Content + Account   & \textbf{0.9957} & \textbf{0.9943} & \textbf{0.8544}  \\ 
    \end{tabular}}
    \caption{Ablation study comparison in terms of accuracy}
    \label{tab:ablationcomparison}
\end{table}

\section{Discussion}
\label{sec:discussion}

The application of deep learning models has significantly advanced the state-of-the-art performance across various domains, including computer vision, natural language processing, and pattern recognition. However, despite their remarkable success, these models often suffer from a critical limitation: interpretability. In contrast, employing non-deep learning approaches or simpler machine learning models, such as decision trees, linear models, or rule-based systems, often results in more interpretable models. These traditional models operate on explicit rules or features, enabling users to comprehend how specific inputs influence the final prediction. The transparency offered by these models provides insights into the decision-making process, facilitating model debugging, error analysis, and feature importance identification.

In this study we have relied on feature engineering, feature selection and traditional machine learning techniques not only gaining in interpretability but also surpassing the state-of-the-art approaches that are based on both the user's account and its content. This is especially relevant in the practical approach because, even if you do not want to follow a detection algorithm based on machine learning, this study can serve as a basis to identify which features you should pay special attention to when facing the task of identifying bots. Having said this, we proceed to answer the research questions posed above:

\paragraph{\emph{\textbf{RQ. 1:} What features define a social bot?}}
    
\begin{figure}[ht!]
    \centering
    \begin{subfigure}{0.45\linewidth}
        \includegraphics[width=\linewidth]{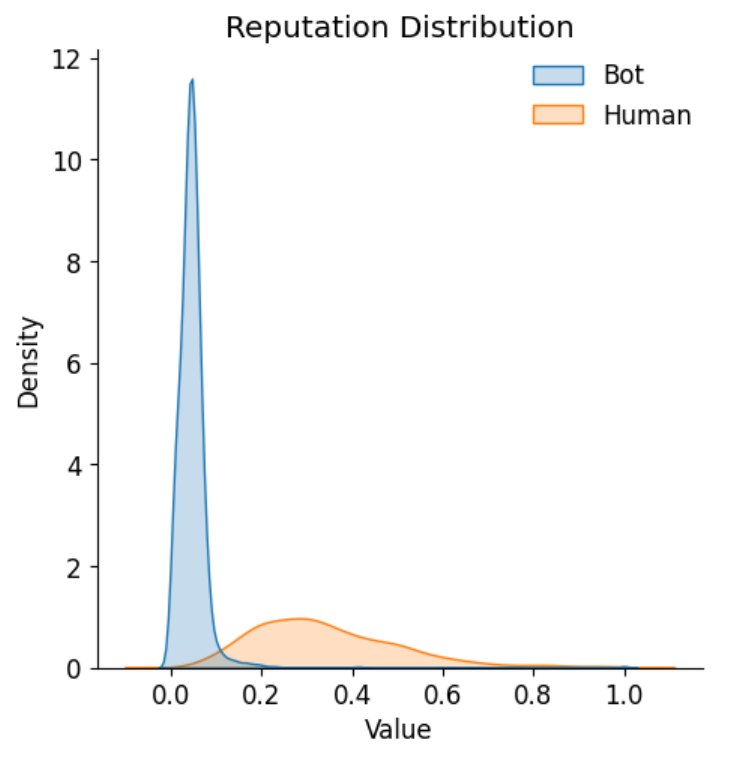}
        \caption{Reputation in Cresci-15}
        \label{fig:reputation_distribution_cresci15}
    \end{subfigure}\hfill
    \begin{subfigure}{0.45\linewidth}
        \includegraphics[width=\linewidth]{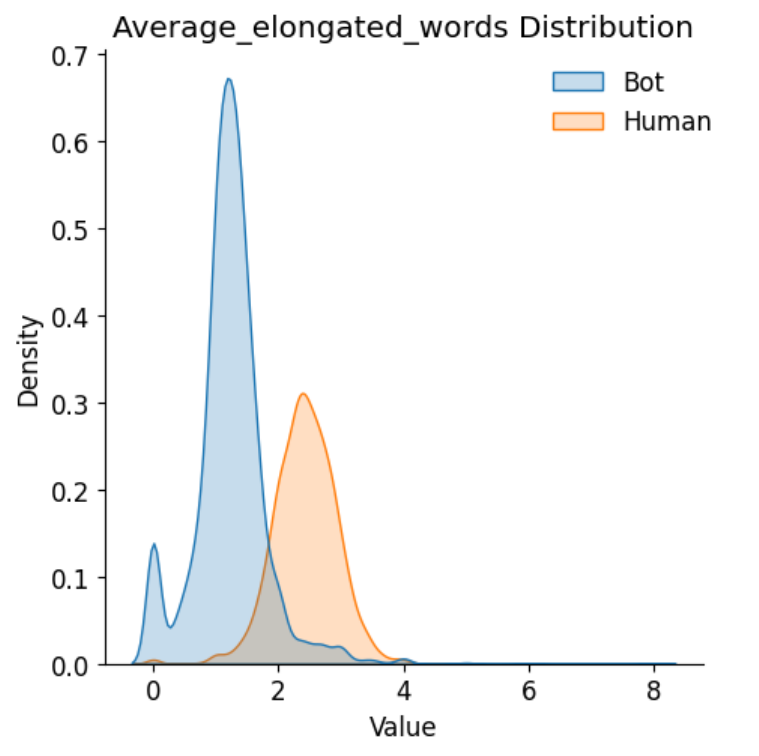}
        \caption{Average elongated words in Cresci-15}
        \label{fig:average_elongated_words_cresci15_nonorm}
    \end{subfigure}\\
    \begin{subfigure}{0.45\linewidth}
        \includegraphics[width=\linewidth]{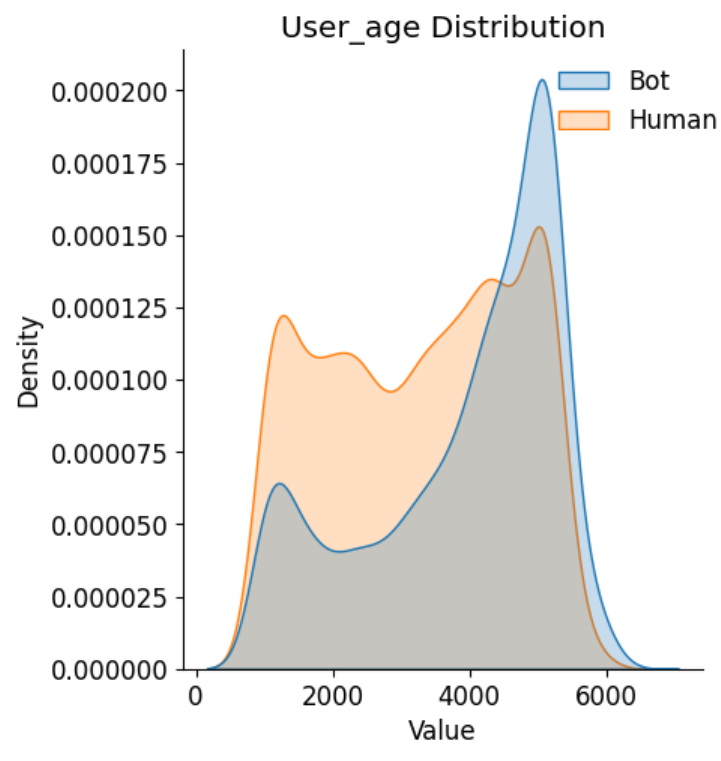}
        \caption{User age in Twibot-20}
        \label{fig:user_age_distribution_twibot20}
    \end{subfigure}\hfill
    \begin{subfigure}{0.45\linewidth}
        \includegraphics[width=\linewidth]{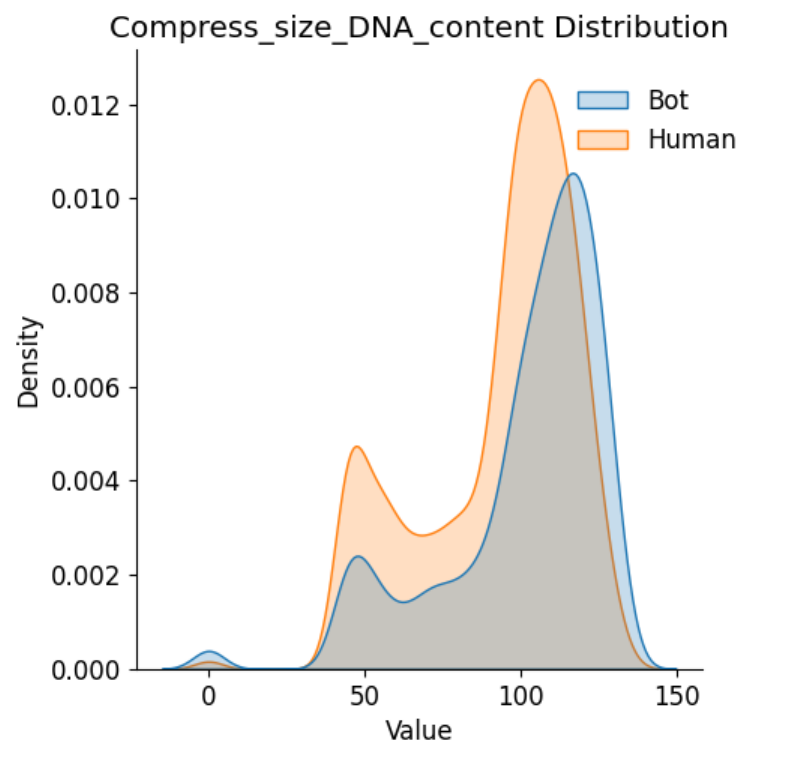}
        \caption{Compress size DNA content in Twibot-20}
        \label{fig:compress_size_DNA_content_distribution_twibot20}
    \end{subfigure}
    \caption{Feature distribution across the datasets.}
    \label{fig:feature_distribution}
\end{figure}

    To answer this question we will look at Figure \ref{fig:feature_importance_all_datasets}, as can be seen, there are several features that appear at the top of most of the graphs. First we look at the features related to social interactions with other users. We can see that these occupy the first positions in the 6 graphs, among them we can highlight the \textit{reputation} and the \textit{followers\_friends\_ratio}, these features are very similar and appear in the top in 5 of the 6 graphs. The importance of favourites, the number of followers, the number of listings, the number of mentions and the growth of these features are also striking. Furthermore, we can see that in Cresci-17 dataset the proposed measure of \textit{credibility}, also based on these interactions, is considered within the top. With these observations in mind we can intuit that it is the same people within the social network who, with their activity and interactions, give us the most important clues to differentiate an automatic account from a non-automatic one. Figure \ref{fig:reputation_distribution_cresci15} shows how automatic accounts tend to have low values of reputation, as opposed to genuine accounts that are more evenly distributed across their values.

    Another striking element is the appearance of the measurements given in \cite{cresci2017social}, referred to as DNA in Table \ref{content-literature}. At least one measure from this set appears in all the graphs, and up to 6 of them appear when applying Mutual Information in Cresci-17. These measures are related to the activity patterns of the users and can give us clues that there is some automation and sequences of activities that are repeated in certain accounts. This can be seen in Figure \ref{fig:compress_size_DNA_content_distribution_twibot20} which is especially interesting as it shows that when the number of tweets is low the difference between the size of the compressed DNA of a bot and a genuine account is very small (peaks at 50) but, as can be seen, when the number of tweets of both classes increases this difference also increases (peaks at 120-140). This tells us that this feature set is interesting to use when we are dealing with a large volume of tweets per user.

    An essential observation is the inclusion of \textit{user\_age} in the graphs (see Figure \ref{fig:img6}, \ref{fig:img5}, \ref{fig:img3}). In the realm of social networks, \textit{user\_age} signifies the duration an account has been active measured from its creation to the moment the dataset is formed. Based on that, similar user ages imply that the respective accounts were created around the same time.
    This feature, a priori not very relevant, provides us with more information than it seems. As social networks have grown, the interest in creating bots has increased. As a consequence, at the beginning there were far fewer bots than before, so it is more likely that an account with a lot of age is genuine. Looking at Figure \ref{fig:user_age_distribution_twibot20} we can see how the distributions of user age in bots and genuine accounts support our theory. 

    Finally, we can see that in most of the graphs there are usually two or more features related to stylometry and readability. In particular we can see how \textit{difficult\_words}, \textit{gunning\_fog}, \textit{flesh\_kincaid\_grade}, \textit{dale\_chall\_readability\_score} and \textit{average\_elongated\_words} appear, being this last feature common in two different datasets and appearing in the 4 graphs of these datasets. An example of this can be seen in Figure \ref{fig:average_elongated_words_cresci15_nonorm} where we can see how the bots of Cresci-15 dataset tend to use less elongated words on average than real users.

\paragraph{\emph{\textbf{RQ. 2:} Which source of features holds greater importance in social bot detection, account-based or content-based features?}}

To answer this question we must look at Figure \ref{fig:feature_importance_all_datasets} and the ablation study performed in the previous section. In Figure \ref{fig:feature_importance_all_datasets} we can see how in the three datasets the first positions of the top are occupied by the features coming from the account, this agrees well with what is observed in Table \ref{tab:ablationcomparison} where we can see that only using features coming from the user's account we reach a higher accuracy than using features coming from the account. With this we could consider the question answered if it were not for a nuance, taking a closer look at Table \ref{tab:ablationcomparison} we can see that in the Cresci-15 dataset the difference between the accuracy obtained with the two sources of features is very small, but this difference is accentuated in the other two datasets. If we remember the Cresci-15 dataset consists of only one type of bot, while the other two contain more variety. So we can say that in general the features coming from the account have more relevance when it comes to giving a classification but it is possible that the performance of the features coming from the content varies according to the type of bot we are trying to detect.

Finally to solve this question is worth mentioning that in today's era, with the presence of generative AIs, it has become effortless to generate and illicitly acquire personal information, particularly related to account features like profile images or backgrounds. This ease of access empowers bots to project greater confidence by leveraging these features. Consequently, this underscores the essential need for both content and account features to collaborate, emphasizing the importance of a united approach to address this challenge.

\paragraph{\emph{\textbf{RQ. 3:} Can a social bot be identified based on user-profile features? Are they enough?}}

To answer this question we will rely once again on the results of the baselines implemented in \cite{feng2022twibot}. In Table \ref{literature-results-comparation_graph} we have collected the baselines that use graphs as well as other techniques. As can be seen, in the cases of the Cresci-15 and Cresci-17 datasets our proposal is still superior to those presented, on the other hand, in the Twibot-20 dataset our proposal lags behind the best proposal by 1.46 points in terms of accuracy, specifically it is in fifth position compared to the other methods. This means that there is information in the network structure that cannot be captured with account and content-based methods, so if this network is available, it could be interesting to use it for detection. This does not mean that methods that do not use networks are not competitive, in fact they can achieve results close to and in some cases superior to those based on networks, but there are users who are not willing to put all their data in their account, or who use the platform to promote a personal project, or who simply have not written enough posts for many of the features obtained to be relevant. In these cases, it would be interesting to support techniques based on account features with other types of techniques.

\begin{table}[ht!]
    \centering
    \resizebox{1\textwidth}{!}{\begin{tabular}{l|cccc|cccc|cccc}
        \hline
        \multirow{2}{*}{\textbf{Method}} & \multicolumn{4}{c|}{\textbf{C-15}} & \multicolumn{4}{c|}{\textbf{C-17}} & \multicolumn{4}{c}{\textbf{T-20}} \\ 
        & \textbf{Acc.} & \textbf{Prec.} & \textbf{Rec.} & \textbf{F1} & \textbf{Acc.} & \textbf{Prec.} & \textbf{Rec.} & \textbf{F1} & \textbf{Acc.} & \textbf{Prec.} & \textbf{Rec.} & \textbf{F1} \\ \hline
        Moghaddam et al. \cite{moghaddam2022friendship} & 0.736 & \underline{0.983} & 0.592 & 0.739 & - & - & - & - & 0.740 & 0.723 & 0.844 & 0.779\\
        Alhosseini et al. \cite{ali2019detect} & 0.896 & 0.877 & 0.972 & 0.922 & - & - & - & - & 0.599 & 0.578 & 0.957 & 0.721\\
        Knauth et al. \cite{knauth2019language} & 0.859 & 0.857 & 0.974 & 0.912 & \underline{0.902} & 0.916 & 0.954 & 0.934 & 0.819 & \textbf{0.966} & 0.763 & 0.852\\
        FriendBot \cite{beskow2020you} & 0.969 & 0.953 & \textbf{1.000} & 0.976 & 0.780 & 0.776 & \textbf{1.000} & 0.874 & 0.759 & 0.726 & 0.889 & 0.799\\
        SATAR \cite{feng2021satar} & 0.934 & 0.907 & \underline{0.999} & 0.951 & - & - & - & - & 0.840 & 0.815 & 0.912 & 0.861\\
        Botometer \cite{yang2022botometer} & 0.579 & 0.505 & 0.990 & 0.669 & 0.942 & \underline{0.934} & \underline{0.997} & \underline{0.961} & 0.531 & 0.557 & 0.508 & 0.531\\
        Rodríguez-Ruiz et al. \cite{rodriguez2020one} & 0.824 & 0.786 & 0.991 & 0.877 & 0.764 & 0.795 & 0.929 & 0.857 & 0.660 & 0.616 & \underline{0.988} & 0.631\\
        GraphHist \cite{magelinski2020graph} & 0.774 & 0.731 & \textbf{1.000} & 0.845 & - & - & - & - & 0.513 & 0.513 & \textbf{0.991} & 0.676\\
        EvolveBot \cite{yang2013empirical} & 0.922 & 0.850 & 0.958 & 0.901 & - & - & - & - & 0.658 & 0.669 & 0.728 & 0.698\\
        Dehghan et al. \cite{dehghan2023detecting} & 0.621 & 0.962 & 0.839 & 0.883 & - & - & - & - & \underline{0.867} & \underline{0.947} & 0.822 & 0.762\\
        GCN \cite{kipf2016semi} & 0.964 & 0.956 & 0.988 & 0.972 & - & - & - & - & 0.775 & 0.752 & 0.876 & 0.809\\
        GAT \cite{velivckovic2017graph} & 0.969 & 0.961 & 0.991 & 0.976 & - & - & - & - & 0.833 & 0.814 & 0.895 & 0.853\\
        HGT \cite{hu2020heterogeneous} & 0.960 & 0.948 & 0.991 & 0.969 & - & - & - & - & \textbf{0.869} & 0.856 & 0.910 & \underline{0.882}\\
        SimpleHGN \cite{lv2021we} & 0.967 & 0.957 & 0.993 & 0.973 & - & - & - & - & \underline{0.867} & 0.848 & 0.921 & \textbf{0.883}\\
        BotRGCN \cite{feng2021botrgcn} & 0.965 & 0.955 & 0.992 & 0.973 & - & - & - & - & 0.858 & 0.845 & 0.902 & 0.873\\
        RGT \cite{feng2022heterogeneity} & \underline{0.972} & 0.964 & 0.992 & \underline{0.978} & - & - & - & - & 0.866 & 0.852 & 0.911 & 0.880\\ \hline
        OURS & \textbf{0.996} & \textbf{0.993} & 0.995 & \textbf{0.994} & \textbf{0.994} & \textbf{0.987} & 0.990 & \textbf{0.988} & 0.854 & 0.832 & 0.936 & 0.879 \\ 
    \end{tabular}}
    \caption{Comparison between different graph-based baselines from literature \cite{feng2022twibot}}
    \label{literature-results-comparation_graph}
\end{table}

\section{Conclusions}
\label{sec:Conclusions}

In this paper, we emphasize the necessity of leveraging a blend of account and content-based features for effective bot detection. Through the creation of a large set of features made up of literature features and proposed new features, a feature selection and a combination process, we managed, through a traditional random forest algorithm, to surpass the state of the art results across three distinct benchmark datasets.

Our findings highlighted the significance of defining bots through a fusion of account-based and content-based features. We observed that the importance of feature types varied based on the selection method employed, with account-based features proving more conducive to accurate classification. Furthermore, in distinguishing between different types of bots, we demonstrated the varying degrees of importance associated with different characteristics.

It is worth noting that in the era of generative AI, it has become easier to generate fake content or enhance fake account features, such as profile pictures. Therefore, systems that incorporate a variety of features, as proposed in our research, prove valuable in mitigating the proliferation of bot accounts.

As part of our future work, we aimed to enhance our system by incorporating graph-based methods to capture the entirety of account interactions within their environments. The inclusion of semantic information from content, along with exploring the impact of word embeddings, would enable the creation of bot prototypes. Additionally, we planned to incorporate user biographies and descriptions into the characteristic sets, further enriching the features for more nuanced bot detection.

\backmatter

\bmhead{Acknowledgements}
\label{Acknowledgments}
The research reported in this paper was supported by the DesinfoScan project: Grant TED2021-129402B-C21 funded by MCIU/AEI/10.13039/501100011033 and by the European Union NextGenerationEU/PRTR, and FederaMed project: Grant PID2021-123960OB-I00 funded by MCIU/AEI/10.13039/501100011033 and by ERDF/EU. Finally, the research reported in this paper is also funded by the European Union (BAG-INTEL project, grant agreement no. 101121309).

\section*{Declarations}

\begin{itemize}
\item \textbf{Funding} This research was funded by the European Union and the Spanish Ministry of Science, Innovation, and Universities.
\item \textbf{Conflict of Interest} The authors declare that they have no conflicts of interest relevant to the content of this article.
\item \textbf{Consent for Publication} All authors have reviewed and approved the manuscript and consent to its publication.
\item \textbf{Data Availability} Datasets used are available through public repositories.
\end{itemize}

\noindent

\begin{appendices}

\section{Account-based features}\label{secA1}

\begin{table}[H]
\centering
\setlength\tabcolsep{10pt}
\scriptsize
\begin{tabular}{|l|
>{\columncolor[HTML]{CCFFCC}}c |
>{\columncolor[HTML]{CCFFCC}}c |
>{\columncolor[HTML]{CCFFCC}}c |}
\hline
\cellcolor[HTML]{B4C7DC}Account-based features dataset & \multicolumn{1}{l|}{\cellcolor[HTML]{B4C7DC}C-15} & \multicolumn{1}{l|}{\cellcolor[HTML]{B4C7DC}C-17} & \multicolumn{1}{l|}{\cellcolor[HTML]{B4C7DC}T-20} \\ \hline
id                                             & {\color[HTML]{006600} V}                          & {\color[HTML]{006600} V}                          & {\color[HTML]{006600} V}                            \\ \hline
created\_at                                    & {\color[HTML]{006600} V}                          & {\color[HTML]{006600} V}                          & {\color[HTML]{006600} V}                            \\ \hline
description                                    & {\color[HTML]{006600} V}                          & {\color[HTML]{006600} V}                          & {\color[HTML]{006600} V}                            \\ \hline
entities                                       & \cellcolor[HTML]{FFCCCC}{\color[HTML]{CC0000} X}  & \cellcolor[HTML]{FFCCCC}{\color[HTML]{CC0000} X}  & {\color[HTML]{006600} V}                            \\ \hline
location                                       & {\color[HTML]{006600} V}                          & {\color[HTML]{006600} V}                          & {\color[HTML]{006600} V}                            \\ \hline
name                                           & {\color[HTML]{006600} V}                          & {\color[HTML]{006600} V}                          & {\color[HTML]{006600} V}                            \\ \hline
pinned\_tweet\_id                              & \cellcolor[HTML]{FFCCCC}{\color[HTML]{CC0000} X}  & \cellcolor[HTML]{FFCCCC}{\color[HTML]{CC0000} X}  & {\color[HTML]{006600} V}                            \\
\hline
\end{tabular}
\caption{Raw account-based features from each dataset. If the feature appears in the dataset, it is represented in table as a V, if not, it is represented as an X.}
\label{Raw-user-based-features-from-each-dataset}
\end{table}

\newpage

\centering
\setlength\tabcolsep{10pt}
\scriptsize
\begin{longtable}{|l|
>{\columncolor[HTML]{CCFFCC}}c |
>{\columncolor[HTML]{CCFFCC}}c |
>{\columncolor[HTML]{CCFFCC}}c |
>{\columncolor[HTML]{FFF2CC}}l |}

\hline
\cellcolor[HTML]{FFE994}Account-based features                       & \multicolumn{1}{l|}{\cellcolor[HTML]{FFE994}C-15} & \multicolumn{1}{l|}{\cellcolor[HTML]{FFE994}C-17} & \multicolumn{1}{l|}{\cellcolor[HTML]{FFE994}T-20} & \cellcolor[HTML]{FFE994}Source \\ \hline

name\_length                                                 & {\color[HTML]{006600} V}                          & {\color[HTML]{006600} V}                          & {\color[HTML]{006600} V}                            & INF                       \\ \hline
name\_digits\_count                                          & {\color[HTML]{006600} V}                          & {\color[HTML]{006600} V}                          & {\color[HTML]{006600} V}                            & INF                       \\ \hline
name\_contains\_bot                                          & {\color[HTML]{006600} V}                          & {\color[HTML]{006600} V}                          & {\color[HTML]{006600} V}                            & INF                       \\ \hline
name\_emoji\_count                                           & {\color[HTML]{006600} V}                          & {\color[HTML]{006600} V}                          & {\color[HTML]{006600} V}                            & INF                       \\ \hline
name\_mean\_bigram\_frequency                                & {\color[HTML]{006600} V}                          & {\color[HTML]{006600} V}                          & {\color[HTML]{006600} V}                            & INF                       \\ \hline
name\_entropy                                                & {\color[HTML]{006600} V}                          & {\color[HTML]{006600} V}                          & {\color[HTML]{006600} V}                            & INF                       \\
\end{longtable} \newpage

\section{ Content-based features}

\begin{table}[H]
\centering
\setlength\tabcolsep{10pt}
\scriptsize
\begin{tabular}{|l|
>{\columncolor[HTML]{CCFFCC}}c |
>{\columncolor[HTML]{CCFFCC}}c |
>{\columncolor[HTML]{FFCCCC}}c |}
\hline
\cellcolor[HTML]{BDD7EE} Content-based features dataset & \multicolumn{1}{l|}{\cellcolor[HTML]{BDD7EE}C-15} & \multicolumn{1}{l|}{\cellcolor[HTML]{BDD7EE}C-17} & \multicolumn{1}{l|}{\cellcolor[HTML]{BDD7EE}T-20} \\ \hline
author\_id/user\_id                              & {\color[HTML]{006600} V}                          & {\color[HTML]{006600} V}                          & {\color[HTML]{CC0000} X}                            \\ \hline
created\_at                                      & {\color[HTML]{006600} V}                          & {\color[HTML]{006600} V}                          & {\color[HTML]{CC0000} X}                            \\ \hline
num\_hashtags                                    & {\color[HTML]{006600} V}                          & {\color[HTML]{006600} V}                          & {\color[HTML]{CC0000} X}                            \\ \hline
num\_mentions                                    & {\color[HTML]{006600} V}                          & {\color[HTML]{006600} V}                          & {\color[HTML]{CC0000} X}                            \\ \hline
num\_urls                                        & {\color[HTML]{006600} V}                          & \cellcolor[HTML]{FFCCCC}{\color[HTML]{CC0000} X}  & {\color[HTML]{CC0000} X}                            \\ \hline
geo                                              & {\color[HTML]{006600} V}                          & {\color[HTML]{006600} V}                          & {\color[HTML]{CC0000} X}                            \\ \hline
id                                               & {\color[HTML]{006600} V}                          & {\color[HTML]{006600} V}                          & {\color[HTML]{CC0000} X}                            \\
\hline
\end{tabular}
\caption{Raw content-based features from each dataset. If the feature appears in the dataset, it is represented in table as a V, if not, it is represented as an X.}
\label{Raw-content-based-features-from-each-dataset}
\end{table}

\newpage
\centering
\setlength\tabcolsep{10pt}
\scriptsize
\begin{longtable}{|l|
>{\columncolor[HTML]{CCFFCC}}c |
>{\columncolor[HTML]{CCFFCC}}c |
>{\columncolor[HTML]{CCFFCC}}c |
>{\columncolor[HTML]{FFF2CC}}l |}
\hline
\cellcolor[HTML]{FFE699}Content-based features & \multicolumn{1}{l|}{\cellcolor[HTML]{FFE699}C-15} & \multicolumn{1}{l|}{\cellcolor[HTML]{FFE699}C-17} & \multicolumn{1}{l|}{\cellcolor[HTML]{FFE699}T-20} & \cellcolor[HTML]{FFE699}Source \\ \hline
average\_tweet\_length                   & {\color[HTML]{006600} V}                          & {\color[HTML]{006600} V}                          & {\color[HTML]{006600} V}                            & INF                       \\ \hline
average\_emoji\_count                    & {\color[HTML]{006600} V}                          & {\color[HTML]{006600} V}                          & {\color[HTML]{006600} V}                            & INF                       \\ \hline
num\_hashtags\_mean                      & {\color[HTML]{006600} V}                          & {\color[HTML]{006600} V}                          & {\color[HTML]{006600} V}                            & INF                       \\ \hline
num\_unique\_hashtags\_mean              & {\color[HTML]{006600} V}                          & {\color[HTML]{006600} V}                          & {\color[HTML]{006600} V}                            & INF                       \\ \hline
num\_urls\_mean                          & {\color[HTML]{006600} V}                          & {\color[HTML]{006600} V}                          & {\color[HTML]{006600} V}                            & INF                       \\ \hline
num\_unique\_urls\_mean                  & {\color[HTML]{006600} V}                          & {\color[HTML]{006600} V}                          & {\color[HTML]{006600} V}                            & INF                       \\
\end{longtable} 
\end{appendices}


\bibliography{sn-bibliography}


\begin{thebibliography}{89}
\ifx \bisbn   \undefined \def \bisbn  #1{ISBN #1}\fi
\ifx \binits  \undefined \def \binits#1{#1}\fi
\ifx \bauthor  \undefined \def \bauthor#1{#1}\fi
\ifx \batitle  \undefined \def \batitle#1{#1}\fi
\ifx \bjtitle  \undefined \def \bjtitle#1{#1}\fi
\ifx \bvolume  \undefined \def \bvolume#1{\textbf{#1}}\fi
\ifx \byear  \undefined \def \byear#1{#1}\fi
\ifx \bissue  \undefined \def \bissue#1{#1}\fi
\ifx \bfpage  \undefined \def \bfpage#1{#1}\fi
\ifx \blpage  \undefined \def \blpage #1{#1}\fi
\ifx \burl  \undefined \def \burl#1{\textsf{#1}}\fi
\ifx \doiurl  \undefined \def \doiurl#1{\url{https://doi.org/#1}}\fi
\ifx \betal  \undefined \def \betal{\textit{et al.}}\fi
\ifx \binstitute  \undefined \def \binstitute#1{#1}\fi
\ifx \binstitutionaled  \undefined \def \binstitutionaled#1{#1}\fi
\ifx \bctitle  \undefined \def \bctitle#1{#1}\fi
\ifx \beditor  \undefined \def \beditor#1{#1}\fi
\ifx \bpublisher  \undefined \def \bpublisher#1{#1}\fi
\ifx \bbtitle  \undefined \def \bbtitle#1{#1}\fi
\ifx \bedition  \undefined \def \bedition#1{#1}\fi
\ifx \bseriesno  \undefined \def \bseriesno#1{#1}\fi
\ifx \blocation  \undefined \def \blocation#1{#1}\fi
\ifx \bsertitle  \undefined \def \bsertitle#1{#1}\fi
\ifx \bsnm \undefined \def \bsnm#1{#1}\fi
\ifx \bsuffix \undefined \def \bsuffix#1{#1}\fi
\ifx \bparticle \undefined \def \bparticle#1{#1}\fi
\ifx \barticle \undefined \def \barticle#1{#1}\fi
\bibcommenthead
\ifx \bconfdate \undefined \def \bconfdate #1{#1}\fi
\ifx \botherref \undefined \def \botherref #1{#1}\fi
\ifx \url \undefined \def \url#1{\textsf{#1}}\fi
\ifx \bchapter \undefined \def \bchapter#1{#1}\fi
\ifx \bbook \undefined \def \bbook#1{#1}\fi
\ifx \bcomment \undefined \def \bcomment#1{#1}\fi
\ifx \oauthor \undefined \def \oauthor#1{#1}\fi
\ifx \citeauthoryear \undefined \def \citeauthoryear#1{#1}\fi
\ifx \endbibitem  \undefined \def \endbibitem {}\fi
\ifx \bconflocation  \undefined \def \bconflocation#1{#1}\fi
\ifx \arxivurl  \undefined \def \arxivurl#1{\textsf{#1}}\fi
\csname PreBibitemsHook\endcsname

\bibitem[\protect\citeauthoryear{Almars et~al.}{2022}]{almars2022users}
\begin{barticle}
\bauthor{\bsnm{Almars}, \binits{A.M.}},
\bauthor{\bsnm{Atlam}, \binits{E.-S.}},
\bauthor{\bsnm{Noor}, \binits{T.H.}},
\bauthor{\bsnm{ELmarhomy}, \binits{G.}},
\bauthor{\bsnm{Alagamy}, \binits{R.}},
\bauthor{\bsnm{Gad}, \binits{I.}}:
\batitle{Users opinion and emotion understanding in social media regarding covid-19 vaccine}.
\bjtitle{Computing}
\bvolume{104}(\bissue{6}),
\bfpage{1481}--\blpage{1496}
(\byear{2022})
\end{barticle}
\endbibitem

\bibitem[\protect\citeauthoryear{Linvill and Warren}{2020}]{linvill2020troll}
\begin{barticle}
\bauthor{\bsnm{Linvill}, \binits{D.L.}},
\bauthor{\bsnm{Warren}, \binits{P.L.}}:
\batitle{Troll factories: Manufacturing specialized disinformation on twitter}.
\bjtitle{Political Communication}
\bvolume{37}(\bissue{4}),
\bfpage{447}--\blpage{467}
(\byear{2020})
\end{barticle}
\endbibitem

\bibitem[\protect\citeauthoryear{Nisbet et~al.}{2021}]{nisbet2021presumed}
\begin{botherref}
\oauthor{\bsnm{Nisbet}, \binits{E.C.}},
\oauthor{\bsnm{Mortenson}, \binits{C.}},
\oauthor{\bsnm{Li}, \binits{Q.}}:
The presumed influence of election misinformation on others reduces our own satisfaction with democracy.
The Harvard Kennedy School Misinformation Review
(2021)
\end{botherref}
\endbibitem

\bibitem[\protect\citeauthoryear{Kennedy et~al.}{2022}]{kennedy2022repeat}
\begin{botherref}
\oauthor{\bsnm{Kennedy}, \binits{I.}},
\oauthor{\bsnm{Wack}, \binits{M.}},
\oauthor{\bsnm{Beers}, \binits{A.}},
\oauthor{\bsnm{Schafer}, \binits{J.S.}},
\oauthor{\bsnm{Garcia-Camargo}, \binits{I.}},
\oauthor{\bsnm{Spiro}, \binits{E.S.}},
\oauthor{\bsnm{Starbird}, \binits{K.}}:
Repeat spreaders and election delegitimization: A comprehensive dataset of misinformation tweets from the 2020 us election.
Journal of Quantitative Description: Digital Media
\textbf{2}
(2022)
\end{botherref}
\endbibitem

\bibitem[\protect\citeauthoryear{Freelon et~al.}{2022}]{freelon2022black}
\begin{barticle}
\bauthor{\bsnm{Freelon}, \binits{D.}},
\bauthor{\bsnm{Bossetta}, \binits{M.}},
\bauthor{\bsnm{Wells}, \binits{C.}},
\bauthor{\bsnm{Lukito}, \binits{J.}},
\bauthor{\bsnm{Xia}, \binits{Y.}},
\bauthor{\bsnm{Adams}, \binits{K.}}:
\batitle{Black trolls matter: Racial and ideological asymmetries in social media disinformation}.
\bjtitle{Social Science Computer Review}
\bvolume{40}(\bissue{3}),
\bfpage{560}--\blpage{578}
(\byear{2022})
\end{barticle}
\endbibitem

\bibitem[\protect\citeauthoryear{Shao et~al.}{2017}]{shao2017spread}
\begin{barticle}
\bauthor{\bsnm{Shao}, \binits{C.}},
\bauthor{\bsnm{Ciampaglia}, \binits{G.L.}},
\bauthor{\bsnm{Varol}, \binits{O.}},
\bauthor{\bsnm{Flammini}, \binits{A.}},
\bauthor{\bsnm{Menczer}, \binits{F.}}:
\batitle{The spread of fake news by social bots}.
\bjtitle{arXiv preprint arXiv:1707.07592}
\bvolume{96},
\bfpage{104}
(\byear{2017})
\end{barticle}
\endbibitem

\bibitem[\protect\citeauthoryear{Deseriis}{2017}]{deseriis2017hacktivism}
\begin{barticle}
\bauthor{\bsnm{Deseriis}, \binits{M.}}:
\batitle{Hacktivism: On the use of botnets in cyberattacks}.
\bjtitle{Theory, Culture \& Society}
\bvolume{34}(\bissue{4}),
\bfpage{131}--\blpage{152}
(\byear{2017})
\end{barticle}
\endbibitem

\bibitem[\protect\citeauthoryear{Hammi et~al.}{2019}]{hammi2019empirical}
\begin{barticle}
\bauthor{\bsnm{Hammi}, \binits{B.}},
\bauthor{\bsnm{Zeadally}, \binits{S.}},
\bauthor{\bsnm{Khatoun}, \binits{R.}}:
\batitle{An empirical investigation of botnet as a service for cyberattacks}.
\bjtitle{Transactions on emerging telecommunications technologies}
\bvolume{30}(\bissue{3}),
\bfpage{3537}
(\byear{2019})
\end{barticle}
\endbibitem

\bibitem[\protect\citeauthoryear{Hajli et~al.}{2022}]{hajli2022social}
\begin{barticle}
\bauthor{\bsnm{Hajli}, \binits{N.}},
\bauthor{\bsnm{Saeed}, \binits{U.}},
\bauthor{\bsnm{Tajvidi}, \binits{M.}},
\bauthor{\bsnm{Shirazi}, \binits{F.}}:
\batitle{Social bots and the spread of disinformation in social media: the challenges of artificial intelligence}.
\bjtitle{British Journal of Management}
\bvolume{33}(\bissue{3}),
\bfpage{1238}--\blpage{1253}
(\byear{2022})
\end{barticle}
\endbibitem

\bibitem[\protect\citeauthoryear{Miller and Busby-Earle}{2016}]{miller2016role}
\begin{bchapter}
\bauthor{\bsnm{Miller}, \binits{S.}},
\bauthor{\bsnm{Busby-Earle}, \binits{C.}}:
\bctitle{The role of machine learning in botnet detection}.
In: \bbtitle{2016 11th International Conference for Internet Technology and Secured Transactions (icitst)},
pp. \bfpage{359}--\blpage{364}
(\byear{2016}).
\bcomment{IEEE}
\end{bchapter}
\endbibitem

\bibitem[\protect\citeauthoryear{Nguyen et~al.}{2023}]{nguyen2023supervised}
\begin{botherref}
\oauthor{\bsnm{Nguyen}, \binits{H.-D.}},
\oauthor{\bsnm{Nguyen}, \binits{D.Q.}},
\oauthor{\bsnm{Nguyen}, \binits{C.-D.}},
\oauthor{\bsnm{To}, \binits{P.T.}},
\oauthor{\bsnm{Nguyen}, \binits{D.H.}},
\oauthor{\bsnm{Nguyen-Gia}, \binits{H.}},
\oauthor{\bsnm{Tran}, \binits{L.H.}},
\oauthor{\bsnm{Tran}, \binits{A.Q.}},
\oauthor{\bsnm{Dang-Hieu}, \binits{A.}},
\oauthor{\bsnm{Nguyen-Duc}, \binits{A.}}, et al.:
Supervised learning models for social bot detection: Literature review and benchmark.
Expert Systems with Applications,
122217
(2023)
\end{botherref}
\endbibitem

\bibitem[\protect\citeauthoryear{Gorwa and Guilbeault}{2020}]{gorwa2020unpacking}
\begin{barticle}
\bauthor{\bsnm{Gorwa}, \binits{R.}},
\bauthor{\bsnm{Guilbeault}, \binits{D.}}:
\batitle{Unpacking the social media bot: A typology to guide research and policy}.
\bjtitle{Policy \& Internet}
\bvolume{12}(\bissue{2}),
\bfpage{225}--\blpage{248}
(\byear{2020})
\end{barticle}
\endbibitem

\bibitem[\protect\citeauthoryear{Fan et~al.}{2020}]{fan2020social}
\begin{barticle}
\bauthor{\bsnm{Fan}, \binits{R.}},
\bauthor{\bsnm{Talavera}, \binits{O.}},
\bauthor{\bsnm{Tran}, \binits{V.}}:
\batitle{Social media bots and stock markets}.
\bjtitle{European Financial Management}
\bvolume{26}(\bissue{3}),
\bfpage{753}--\blpage{777}
(\byear{2020})
\end{barticle}
\endbibitem

\bibitem[\protect\citeauthoryear{Weng and Lin}{2022}]{weng2022public}
\begin{barticle}
\bauthor{\bsnm{Weng}, \binits{Z.}},
\bauthor{\bsnm{Lin}, \binits{A.}}:
\batitle{Public opinion manipulation on social media: Social network analysis of twitter bots during the covid-19 pandemic}.
\bjtitle{International journal of environmental research and public health}
\bvolume{19}(\bissue{24}),
\bfpage{16376}
(\byear{2022})
\end{barticle}
\endbibitem

\bibitem[\protect\citeauthoryear{Pastor-Galindo et~al.}{2022}]{pastor2022profiling}
\begin{barticle}
\bauthor{\bsnm{Pastor-Galindo}, \binits{J.}},
\bauthor{\bsnm{Marmol}, \binits{F.G.}},
\bauthor{\bsnm{P{\'e}rez}, \binits{G.M.}}:
\batitle{Profiling users and bots in twitter through social media analysis}.
\bjtitle{Information Sciences}
\bvolume{613},
\bfpage{161}--\blpage{183}
(\byear{2022})
\end{barticle}
\endbibitem

\bibitem[\protect\citeauthoryear{Abokhodair et~al.}{2015}]{abokhodair2015dissecting}
\begin{bchapter}
\bauthor{\bsnm{Abokhodair}, \binits{N.}},
\bauthor{\bsnm{Yoo}, \binits{D.}},
\bauthor{\bsnm{McDonald}, \binits{D.W.}}:
\bctitle{Dissecting a social botnet: Growth, content and influence in twitter}.
In: \bbtitle{Proceedings of the 18th ACM Conference on Computer Supported Cooperative Work \& Social Computing},
pp. \bfpage{839}--\blpage{851}
(\byear{2015})
\end{bchapter}
\endbibitem

\bibitem[\protect\citeauthoryear{Cresci}{2020}]{cresci2020decade}
\begin{barticle}
\bauthor{\bsnm{Cresci}, \binits{S.}}:
\batitle{A decade of social bot detection}.
\bjtitle{Communications of the ACM}
\bvolume{63}(\bissue{10}),
\bfpage{72}--\blpage{83}
(\byear{2020})
\end{barticle}
\endbibitem

\bibitem[\protect\citeauthoryear{Morstatter et~al.}{2016}]{morstatter2016new}
\begin{bchapter}
\bauthor{\bsnm{Morstatter}, \binits{F.}},
\bauthor{\bsnm{Wu}, \binits{L.}},
\bauthor{\bsnm{Nazer}, \binits{T.H.}},
\bauthor{\bsnm{Carley}, \binits{K.M.}},
\bauthor{\bsnm{Liu}, \binits{H.}}:
\bctitle{A new approach to bot detection: striking the balance between precision and recall}.
In: \bbtitle{2016 IEEE/ACM International Conference on Advances in Social Networks Analysis and Mining (ASONAM)},
pp. \bfpage{533}--\blpage{540}
(\byear{2016}).
\bcomment{IEEE}
\end{bchapter}
\endbibitem

\bibitem[\protect\citeauthoryear{Yang et~al.}{2020}]{yang2020scalable}
\begin{bchapter}
\bauthor{\bsnm{Yang}, \binits{K.-C.}},
\bauthor{\bsnm{Varol}, \binits{O.}},
\bauthor{\bsnm{Hui}, \binits{P.-M.}},
\bauthor{\bsnm{Menczer}, \binits{F.}}:
\bctitle{Scalable and generalizable social bot detection through data selection}.
In: \bbtitle{Proceedings of the AAAI Conference on Artificial Intelligence},
vol. \bseriesno{34},
pp. \bfpage{1096}--\blpage{1103}
(\byear{2020})
\end{bchapter}
\endbibitem

\bibitem[\protect\citeauthoryear{Assenmacher et~al.}{2020}]{assenmacher2020demystifying}
\begin{barticle}
\bauthor{\bsnm{Assenmacher}, \binits{D.}},
\bauthor{\bsnm{Clever}, \binits{L.}},
\bauthor{\bsnm{Frischlich}, \binits{L.}},
\bauthor{\bsnm{Quandt}, \binits{T.}},
\bauthor{\bsnm{Trautmann}, \binits{H.}},
\bauthor{\bsnm{Grimme}, \binits{C.}}:
\batitle{Demystifying social bots: On the intelligence of automated social media actors}.
\bjtitle{Social Media+ Society}
\bvolume{6}(\bissue{3}),
\bfpage{2056305120939264}
(\byear{2020})
\end{barticle}
\endbibitem

\bibitem[\protect\citeauthoryear{Ferrara et~al.}{2016}]{ferrara2016rise}
\begin{barticle}
\bauthor{\bsnm{Ferrara}, \binits{E.}},
\bauthor{\bsnm{Varol}, \binits{O.}},
\bauthor{\bsnm{Davis}, \binits{C.}},
\bauthor{\bsnm{Menczer}, \binits{F.}},
\bauthor{\bsnm{Flammini}, \binits{A.}}:
\batitle{The rise of social bots}.
\bjtitle{Communications of the ACM}
\bvolume{59}(\bissue{7}),
\bfpage{96}--\blpage{104}
(\byear{2016})
\end{barticle}
\endbibitem

\bibitem[\protect\citeauthoryear{Lopez-Joya et~al.}{2023}]{lopez2023bot}
\begin{bchapter}
\bauthor{\bsnm{Lopez-Joya}, \binits{S.}},
\bauthor{\bsnm{Diaz-Garcia}, \binits{J.A.}},
\bauthor{\bsnm{Ruiz}, \binits{M.D.}},
\bauthor{\bsnm{Martin-Bautista}, \binits{M.J.}}:
\bctitle{Bot detection in twitter: An overview}.
In: \bbtitle{International Conference on Flexible Query Answering Systems},
pp. \bfpage{131}--\blpage{144}
(\byear{2023}).
\bcomment{Springer}
\end{bchapter}
\endbibitem

\bibitem[\protect\citeauthoryear{Hayawi et~al.}{2022}]{hayawi2022deeprobot}
\begin{barticle}
\bauthor{\bsnm{Hayawi}, \binits{K.}},
\bauthor{\bsnm{Mathew}, \binits{S.}},
\bauthor{\bsnm{Venugopal}, \binits{N.}},
\bauthor{\bsnm{Masud}, \binits{M.M.}},
\bauthor{\bsnm{Ho}, \binits{P.-H.}}:
\batitle{Deeprobot: a hybrid deep neural network model for social bot detection based on user profile data}.
\bjtitle{Social Network Analysis and Mining}
\bvolume{12}(\bissue{1}),
\bfpage{43}
(\byear{2022})
\end{barticle}
\endbibitem

\bibitem[\protect\citeauthoryear{Mazza et~al.}{2019}]{mazza2019rtbust}
\begin{bchapter}
\bauthor{\bsnm{Mazza}, \binits{M.}},
\bauthor{\bsnm{Cresci}, \binits{S.}},
\bauthor{\bsnm{Avvenuti}, \binits{M.}},
\bauthor{\bsnm{Quattrociocchi}, \binits{W.}},
\bauthor{\bsnm{Tesconi}, \binits{M.}}:
\bctitle{Rtbust: Exploiting temporal patterns for botnet detection on twitter}.
In: \bbtitle{Proceedings of the 10th ACM Conference on Web Science},
pp. \bfpage{183}--\blpage{192}
(\byear{2019})
\end{bchapter}
\endbibitem

\bibitem[\protect\citeauthoryear{Heidari et~al.}{2020}]{heidari2020deep}
\begin{bchapter}
\bauthor{\bsnm{Heidari}, \binits{M.}},
\bauthor{\bsnm{Jones}, \binits{J.H.}},
\bauthor{\bsnm{Uzuner}, \binits{O.}}:
\bctitle{Deep contextualized word embedding for text-based online user profiling to detect social bots on twitter}.
In: \bbtitle{2020 International Conference on Data Mining Workshops (ICDMW)},
pp. \bfpage{480}--\blpage{487}
(\byear{2020}).
\bcomment{IEEE}
\end{bchapter}
\endbibitem

\bibitem[\protect\citeauthoryear{Pennington et~al.}{2014}]{pennington2014glove}
\begin{bchapter}
\bauthor{\bsnm{Pennington}, \binits{J.}},
\bauthor{\bsnm{Socher}, \binits{R.}},
\bauthor{\bsnm{Manning}, \binits{C.D.}}:
\bctitle{Glove: Global vectors for word representation}.
In: \bbtitle{Proceedings of the 2014 Conference on Empirical Methods in Natural Language Processing (EMNLP)},
pp. \bfpage{1532}--\blpage{1543}
(\byear{2014})
\end{bchapter}
\endbibitem

\bibitem[\protect\citeauthoryear{Sarzynska-Wawer et~al.}{2021}]{sarzynska2021detecting}
\begin{barticle}
\bauthor{\bsnm{Sarzynska-Wawer}, \binits{J.}},
\bauthor{\bsnm{Wawer}, \binits{A.}},
\bauthor{\bsnm{Pawlak}, \binits{A.}},
\bauthor{\bsnm{Szymanowska}, \binits{J.}},
\bauthor{\bsnm{Stefaniak}, \binits{I.}},
\bauthor{\bsnm{Jarkiewicz}, \binits{M.}},
\bauthor{\bsnm{Okruszek}, \binits{L.}}:
\batitle{Detecting formal thought disorder by deep contextualized word representations}.
\bjtitle{Psychiatry Research}
\bvolume{304},
\bfpage{114135}
(\byear{2021})
\end{barticle}
\endbibitem

\bibitem[\protect\citeauthoryear{Yin et~al.}{2024}]{yin2024person}
\begin{barticle}
\bauthor{\bsnm{Yin}, \binits{K.}},
\bauthor{\bsnm{Ding}, \binits{Z.}},
\bauthor{\bsnm{Dong}, \binits{Z.}},
\bauthor{\bsnm{Ji}, \binits{X.}},
\bauthor{\bsnm{Wang}, \binits{Z.}},
\bauthor{\bsnm{Chen}, \binits{D.}},
\bauthor{\bsnm{Li}, \binits{Y.}},
\bauthor{\bsnm{Yin}, \binits{G.}},
\bauthor{\bsnm{Wang}, \binits{Z.}}:
\batitle{Person re-identification method based on fine-grained feature fusion and self-attention mechanism}.
\bjtitle{Computing}
\bvolume{106}(\bissue{5}),
\bfpage{1681}--\blpage{1705}
(\byear{2024})
\end{barticle}
\endbibitem

\bibitem[\protect\citeauthoryear{Wang et~al.}{2024}]{wang2024feature}
\begin{barticle}
\bauthor{\bsnm{Wang}, \binits{Z.-F.}},
\bauthor{\bsnm{Yuan}, \binits{P.-Y.}},
\bauthor{\bsnm{Cao}, \binits{Z.-Y.}},
\bauthor{\bsnm{Zhang}, \binits{L.-Y.}}:
\batitle{Feature reduction of unbalanced data classification based on density clustering}.
\bjtitle{Computing}
\bvolume{106}(\bissue{1}),
\bfpage{29}--\blpage{55}
(\byear{2024})
\end{barticle}
\endbibitem

\bibitem[\protect\citeauthoryear{Peters and Van~Voorhis}{1940}]{peters1940chi}
\begin{botherref}
\oauthor{\bsnm{Peters}, \binits{C.C.}},
\oauthor{\bsnm{Van~Voorhis}, \binits{W.R.}}:
Chi square.
(1940)
\end{botherref}
\endbibitem

\bibitem[\protect\citeauthoryear{Shannon}{1948}]{shannon1948mathematical}
\begin{barticle}
\bauthor{\bsnm{Shannon}, \binits{C.E.}}:
\batitle{A mathematical theory of communication}.
\bjtitle{The Bell system technical journal}
\bvolume{27}(\bissue{3}),
\bfpage{379}--\blpage{423}
(\byear{1948})
\end{barticle}
\endbibitem

\bibitem[\protect\citeauthoryear{Fisher}{1936}]{fisher1936use}
\begin{barticle}
\bauthor{\bsnm{Fisher}, \binits{R.A.}}:
\batitle{The use of multiple measurements in taxonomic problems}.
\bjtitle{Annals of eugenics}
\bvolume{7}(\bissue{2}),
\bfpage{179}--\blpage{188}
(\byear{1936})
\end{barticle}
\endbibitem

\bibitem[\protect\citeauthoryear{Guyon et~al.}{2002}]{guyon2002gene}
\begin{barticle}
\bauthor{\bsnm{Guyon}, \binits{I.}},
\bauthor{\bsnm{Weston}, \binits{J.}},
\bauthor{\bsnm{Barnhill}, \binits{S.}},
\bauthor{\bsnm{Vapnik}, \binits{V.}}:
\batitle{Gene selection for cancer classification using support vector machines}.
\bjtitle{Machine learning}
\bvolume{46},
\bfpage{389}--\blpage{422}
(\byear{2002})
\end{barticle}
\endbibitem

\bibitem[\protect\citeauthoryear{Ferri et~al.}{1994}]{ferri1994comparative}
\begin{bchapter}
\bauthor{\bsnm{Ferri}, \binits{F.J.}},
\bauthor{\bsnm{Pudil}, \binits{P.}},
\bauthor{\bsnm{Hatef}, \binits{M.}},
\bauthor{\bsnm{Kittler}, \binits{J.}}:
\bctitle{Comparative study of techniques for large-scale feature selection}.
In: \bbtitle{Machine Intelligence and Pattern Recognition}
vol. \bseriesno{16},
pp. \bfpage{403}--\blpage{413}.
\bpublisher{Elsevier}, \blocation{???}
(\byear{1994})
\end{bchapter}
\endbibitem

\bibitem[\protect\citeauthoryear{Tibshirani}{1996}]{tibshirani1996regression}
\begin{barticle}
\bauthor{\bsnm{Tibshirani}, \binits{R.}}:
\batitle{Regression shrinkage and selection via the lasso}.
\bjtitle{Journal of the Royal Statistical Society Series B: Statistical Methodology}
\bvolume{58}(\bissue{1}),
\bfpage{267}--\blpage{288}
(\byear{1996})
\end{barticle}
\endbibitem

\bibitem[\protect\citeauthoryear{Breiman}{2001}]{breiman2001random}
\begin{barticle}
\bauthor{\bsnm{Breiman}, \binits{L.}}:
\batitle{Random forests}.
\bjtitle{Machine learning}
\bvolume{45},
\bfpage{5}--\blpage{32}
(\byear{2001})
\end{barticle}
\endbibitem

\bibitem[\protect\citeauthoryear{Mbona and Eloff}{2022}]{mbona2022feature}
\begin{barticle}
\bauthor{\bsnm{Mbona}, \binits{I.}},
\bauthor{\bsnm{Eloff}, \binits{J.H.}}:
\batitle{Feature selection using benford’s law to support detection of malicious social media bots}.
\bjtitle{Information Sciences}
\bvolume{582},
\bfpage{369}--\blpage{381}
(\byear{2022})
\end{barticle}
\endbibitem

\bibitem[\protect\citeauthoryear{Ilias and Roussaki}{2021}]{ilias2021detecting}
\begin{barticle}
\bauthor{\bsnm{Ilias}, \binits{L.}},
\bauthor{\bsnm{Roussaki}, \binits{I.}}:
\batitle{Detecting malicious activity in twitter using deep learning techniques}.
\bjtitle{Applied Soft Computing}
\bvolume{107},
\bfpage{107360}
(\byear{2021})
\end{barticle}
\endbibitem

\bibitem[\protect\citeauthoryear{Cardaioli et~al.}{2021}]{cardaioli2021sa}
\begin{bchapter}
\bauthor{\bsnm{Cardaioli}, \binits{M.}},
\bauthor{\bsnm{Conti}, \binits{M.}},
\bauthor{\bsnm{Di~Sorbo}, \binits{A.}},
\bauthor{\bsnm{Fabrizio}, \binits{E.}},
\bauthor{\bsnm{Laudanna}, \binits{S.}},
\bauthor{\bsnm{Visaggio}, \binits{C.A.}}:
\bctitle{It’sa matter of style: Detecting social bots through writing style consistency}.
In: \bbtitle{2021 International Conference on Computer Communications and Networks (ICCCN)},
pp. \bfpage{1}--\blpage{9}
(\byear{2021}).
\bcomment{IEEE}
\end{bchapter}
\endbibitem

\bibitem[\protect\citeauthoryear{Wu et~al.}{2021}]{wu2021novel}
\begin{barticle}
\bauthor{\bsnm{Wu}, \binits{Y.}},
\bauthor{\bsnm{Fang}, \binits{Y.}},
\bauthor{\bsnm{Shang}, \binits{S.}},
\bauthor{\bsnm{Jin}, \binits{J.}},
\bauthor{\bsnm{Wei}, \binits{L.}},
\bauthor{\bsnm{Wang}, \binits{H.}}:
\batitle{A novel framework for detecting social bots with deep neural networks and active learning}.
\bjtitle{Knowledge-Based Systems}
\bvolume{211},
\bfpage{106525}
(\byear{2021})
\end{barticle}
\endbibitem

\bibitem[\protect\citeauthoryear{Lee et~al.}{2011}]{lee2011seven}
\begin{bchapter}
\bauthor{\bsnm{Lee}, \binits{K.}},
\bauthor{\bsnm{Eoff}, \binits{B.}},
\bauthor{\bsnm{Caverlee}, \binits{J.}}:
\bctitle{Seven months with the devils: A long-term study of content polluters on twitter}.
In: \bbtitle{Proceedings of the International AAAI Conference on Web and Social Media},
vol. \bseriesno{5},
pp. \bfpage{185}--\blpage{192}
(\byear{2011})
\end{bchapter}
\endbibitem

\bibitem[\protect\citeauthoryear{Cresci et~al.}{2015}]{cresci2015fame}
\begin{barticle}
\bauthor{\bsnm{Cresci}, \binits{S.}},
\bauthor{\bsnm{Di~Pietro}, \binits{R.}},
\bauthor{\bsnm{Petrocchi}, \binits{M.}},
\bauthor{\bsnm{Spognardi}, \binits{A.}},
\bauthor{\bsnm{Tesconi}, \binits{M.}}:
\batitle{Fame for sale: Efficient detection of fake twitter followers}.
\bjtitle{Decision Support Systems}
\bvolume{80},
\bfpage{56}--\blpage{71}
(\byear{2015})
\end{barticle}
\endbibitem

\bibitem[\protect\citeauthoryear{Cresci et~al.}{2017}]{cresci2017paradigm}
\begin{bchapter}
\bauthor{\bsnm{Cresci}, \binits{S.}},
\bauthor{\bsnm{Di~Pietro}, \binits{R.}},
\bauthor{\bsnm{Petrocchi}, \binits{M.}},
\bauthor{\bsnm{Spognardi}, \binits{A.}},
\bauthor{\bsnm{Tesconi}, \binits{M.}}:
\bctitle{The paradigm-shift of social spambots: Evidence, theories, and tools for the arms race}.
In: \bbtitle{Proceedings of the 26th International Conference on World Wide Web Companion},
pp. \bfpage{963}--\blpage{972}
(\byear{2017})
\end{bchapter}
\endbibitem

\bibitem[\protect\citeauthoryear{Feng et~al.}{2021}]{feng2021twibot}
\begin{bchapter}
\bauthor{\bsnm{Feng}, \binits{S.}},
\bauthor{\bsnm{Wan}, \binits{H.}},
\bauthor{\bsnm{Wang}, \binits{N.}},
\bauthor{\bsnm{Li}, \binits{J.}},
\bauthor{\bsnm{Luo}, \binits{M.}}:
\bctitle{Twibot-20: A comprehensive twitter bot detection benchmark}.
In: \bbtitle{Proceedings of the 30th ACM International Conference on Information \& Knowledge Management},
pp. \bfpage{4485}--\blpage{4494}
(\byear{2021})
\end{bchapter}
\endbibitem

\bibitem[\protect\citeauthoryear{Daouadi et~al.}{2019}]{daouadi2019bot}
\begin{bchapter}
\bauthor{\bsnm{Daouadi}, \binits{K.E.}},
\bauthor{\bsnm{Reba{\"\i}}, \binits{R.Z.}},
\bauthor{\bsnm{Amous}, \binits{I.}}:
\bctitle{Bot detection on online social networks using deep forest}.
In: \bbtitle{Artificial Intelligence Methods in Intelligent Algorithms: Proceedings of 8th Computer Science On-line Conference 2019, Vol. 2 8},
pp. \bfpage{307}--\blpage{315}
(\byear{2019}).
\bcomment{Springer}
\end{bchapter}
\endbibitem

\bibitem[\protect\citeauthoryear{Beskow and Carley}{2018}]{beskow2018bot}
\begin{bchapter}
\bauthor{\bsnm{Beskow}, \binits{D.M.}},
\bauthor{\bsnm{Carley}, \binits{K.M.}}:
\bctitle{Bot conversations are different: leveraging network metrics for bot detection in twitter}.
In: \bbtitle{2018 IEEE/ACM International Conference on Advances in Social Networks Analysis and Mining (ASONAM)},
pp. \bfpage{825}--\blpage{832}
(\byear{2018}).
\bcomment{IEEE}
\end{bchapter}
\endbibitem

\bibitem[\protect\citeauthoryear{Przyby{\l}a and Soto}{2021}]{przybyla2021classification}
\begin{barticle}
\bauthor{\bsnm{Przyby{\l}a}, \binits{P.}},
\bauthor{\bsnm{Soto}, \binits{A.J.}}:
\batitle{When classification accuracy is not enough: Explaining news credibility assessment}.
\bjtitle{Information Processing \& Management}
\bvolume{58}(\bissue{5}),
\bfpage{102653}
(\byear{2021})
\end{barticle}
\endbibitem

\bibitem[\protect\citeauthoryear{Cresci et~al.}{2017}]{cresci2017social}
\begin{barticle}
\bauthor{\bsnm{Cresci}, \binits{S.}},
\bauthor{\bsnm{Di~Pietro}, \binits{R.}},
\bauthor{\bsnm{Petrocchi}, \binits{M.}},
\bauthor{\bsnm{Spognardi}, \binits{A.}},
\bauthor{\bsnm{Tesconi}, \binits{M.}}:
\batitle{Social fingerprinting: detection of spambot groups through dna-inspired behavioral modeling}.
\bjtitle{IEEE Transactions on Dependable and Secure Computing}
\bvolume{15}(\bissue{4}),
\bfpage{561}--\blpage{576}
(\byear{2017})
\end{barticle}
\endbibitem

\bibitem[\protect\citeauthoryear{Wu et~al.}{2020}]{wu2020using}
\begin{barticle}
\bauthor{\bsnm{Wu}, \binits{B.}},
\bauthor{\bsnm{Liu}, \binits{L.}},
\bauthor{\bsnm{Yang}, \binits{Y.}},
\bauthor{\bsnm{Zheng}, \binits{K.}},
\bauthor{\bsnm{Wang}, \binits{X.}}:
\batitle{Using improved conditional generative adversarial networks to detect social bots on twitter}.
\bjtitle{IEEE Access}
\bvolume{8},
\bfpage{36664}--\blpage{36680}
(\byear{2020})
\end{barticle}
\endbibitem

\bibitem[\protect\citeauthoryear{Diaz-Garcia et~al.}{2022}]{diaz2022noface}
\begin{barticle}
\bauthor{\bsnm{Diaz-Garcia}, \binits{J.A.}},
\bauthor{\bsnm{Ruiz}, \binits{M.D.}},
\bauthor{\bsnm{Martin-Bautista}, \binits{M.J.}}:
\batitle{Noface: A new framework for irrelevant content filtering in social media according to credibility and expertise}.
\bjtitle{Expert Systems with Applications}
\bvolume{208},
\bfpage{118063}
(\byear{2022})
\end{barticle}
\endbibitem

\bibitem[\protect\citeauthoryear{Diaz-Garcia et~al.}{2021}]{diaz2021comparative}
\begin{bchapter}
\bauthor{\bsnm{Diaz-Garcia}, \binits{J.A.}},
\bauthor{\bsnm{Ruiz}, \binits{M.D.}},
\bauthor{\bsnm{Martin-Bautista}, \binits{M.J.}}:
\bctitle{A comparative study of word embeddings for the construction of a social media expert filter}.
In: \bbtitle{International Conference on Flexible Query Answering Systems},
pp. \bfpage{196}--\blpage{208}
(\byear{2021}).
\bcomment{Springer}
\end{bchapter}
\endbibitem

\bibitem[\protect\citeauthoryear{Yang et~al.}{2013}]{yang2013empirical}
\begin{barticle}
\bauthor{\bsnm{Yang}, \binits{C.}},
\bauthor{\bsnm{Harkreader}, \binits{R.}},
\bauthor{\bsnm{Gu}, \binits{G.}}:
\batitle{Empirical evaluation and new design for fighting evolving twitter spammers}.
\bjtitle{IEEE Transactions on Information Forensics and Security}
\bvolume{8}(\bissue{8}),
\bfpage{1280}--\blpage{1293}
(\byear{2013})
\end{barticle}
\endbibitem

\bibitem[\protect\citeauthoryear{Joulin et~al.}{2016a}]{joulin2016bag}
\begin{botherref}
\oauthor{\bsnm{Joulin}, \binits{A.}},
\oauthor{\bsnm{Grave}, \binits{E.}},
\oauthor{\bsnm{Bojanowski}, \binits{P.}},
\oauthor{\bsnm{Mikolov}, \binits{T.}}:
Bag of tricks for efficient text classification.
arXiv preprint arXiv:1607.01759
(2016)
\end{botherref}
\endbibitem

\bibitem[\protect\citeauthoryear{Joulin et~al.}{2016b}]{joulin2016fasttext}
\begin{botherref}
\oauthor{\bsnm{Joulin}, \binits{A.}},
\oauthor{\bsnm{Grave}, \binits{E.}},
\oauthor{\bsnm{Bojanowski}, \binits{P.}},
\oauthor{\bsnm{Douze}, \binits{M.}},
\oauthor{\bsnm{J{\'e}gou}, \binits{H.}},
\oauthor{\bsnm{Mikolov}, \binits{T.}}:
Fasttext.zip: Compressing text classification models.
arXiv preprint arXiv:1612.03651
(2016)
\end{botherref}
\endbibitem

\bibitem[\protect\citeauthoryear{Venkatesh and Anuradha}{2019}]{venkatesh2019review}
\begin{barticle}
\bauthor{\bsnm{Venkatesh}, \binits{B.}},
\bauthor{\bsnm{Anuradha}, \binits{J.}}:
\batitle{A review of feature selection and its methods}.
\bjtitle{Cybernetics and information technologies}
\bvolume{19}(\bissue{1}),
\bfpage{3}--\blpage{26}
(\byear{2019})
\end{barticle}
\endbibitem

\bibitem[\protect\citeauthoryear{Kudugunta and Ferrara}{2018}]{kudugunta2018deep}
\begin{barticle}
\bauthor{\bsnm{Kudugunta}, \binits{S.}},
\bauthor{\bsnm{Ferrara}, \binits{E.}}:
\batitle{Deep neural networks for bot detection}.
\bjtitle{Information Sciences}
\bvolume{467},
\bfpage{312}--\blpage{322}
(\byear{2018})
\end{barticle}
\endbibitem

\bibitem[\protect\citeauthoryear{Beskow and Carley}{2019}]{beskow2019its}
\begin{barticle}
\bauthor{\bsnm{Beskow}, \binits{D.M.}},
\bauthor{\bsnm{Carley}, \binits{K.M.}}:
\batitle{Its all in a name: detecting and labeling bots by their name}.
\bjtitle{Computational and mathematical organization theory}
\bvolume{25},
\bfpage{24}--\blpage{35}
(\byear{2019})
\end{barticle}
\endbibitem

\bibitem[\protect\citeauthoryear{Abreu et~al.}{2020}]{abreu2020twitter}
\begin{bchapter}
\bauthor{\bsnm{Abreu}, \binits{J.V.F.}},
\bauthor{\bsnm{Ralha}, \binits{C.G.}},
\bauthor{\bsnm{Gondim}, \binits{J.J.C.}}:
\bctitle{Twitter bot detection with reduced feature set}.
In: \bbtitle{2020 IEEE International Conference on Intelligence and Security Informatics (ISI)},
pp. \bfpage{1}--\blpage{6}
(\byear{2020}).
\bcomment{IEEE}
\end{bchapter}
\endbibitem

\bibitem[\protect\citeauthoryear{Cresci et~al.}{2016}]{cresci2016dna}
\begin{barticle}
\bauthor{\bsnm{Cresci}, \binits{S.}},
\bauthor{\bsnm{Di~Pietro}, \binits{R.}},
\bauthor{\bsnm{Petrocchi}, \binits{M.}},
\bauthor{\bsnm{Spognardi}, \binits{A.}},
\bauthor{\bsnm{Tesconi}, \binits{M.}}:
\batitle{Dna-inspired online behavioral modeling and its application to spambot detection}.
\bjtitle{IEEE Intelligent Systems}
\bvolume{31}(\bissue{5}),
\bfpage{58}--\blpage{64}
(\byear{2016})
\end{barticle}
\endbibitem

\bibitem[\protect\citeauthoryear{Wei and Nguyen}{2019}]{wei2019twitter}
\begin{bchapter}
\bauthor{\bsnm{Wei}, \binits{F.}},
\bauthor{\bsnm{Nguyen}, \binits{U.T.}}:
\bctitle{Twitter bot detection using bidirectional long short-term memory neural networks and word embeddings}.
In: \bbtitle{2019 First IEEE International Conference on Trust, Privacy and Security in Intelligent Systems and Applications (TPS-ISA)},
pp. \bfpage{101}--\blpage{109}
(\byear{2019}).
\bcomment{IEEE}
\end{bchapter}
\endbibitem

\bibitem[\protect\citeauthoryear{Guo et~al.}{2021}]{guo2021social}
\begin{barticle}
\bauthor{\bsnm{Guo}, \binits{Q.}},
\bauthor{\bsnm{Xie}, \binits{H.}},
\bauthor{\bsnm{Li}, \binits{Y.}},
\bauthor{\bsnm{Ma}, \binits{W.}},
\bauthor{\bsnm{Zhang}, \binits{C.}}:
\batitle{Social bots detection via fusing bert and graph convolutional networks}.
\bjtitle{Symmetry}
\bvolume{14}(\bissue{1}),
\bfpage{30}
(\byear{2021})
\end{barticle}
\endbibitem

\bibitem[\protect\citeauthoryear{Liu et~al.}{2019}]{liu2019roberta}
\begin{botherref}
\oauthor{\bsnm{Liu}, \binits{Y.}},
\oauthor{\bsnm{Ott}, \binits{M.}},
\oauthor{\bsnm{Goyal}, \binits{N.}},
\oauthor{\bsnm{Du}, \binits{J.}},
\oauthor{\bsnm{Joshi}, \binits{M.}},
\oauthor{\bsnm{Chen}, \binits{D.}},
\oauthor{\bsnm{Levy}, \binits{O.}},
\oauthor{\bsnm{Lewis}, \binits{M.}},
\oauthor{\bsnm{Zettlemoyer}, \binits{L.}},
\oauthor{\bsnm{Stoyanov}, \binits{V.}}:
Roberta: A robustly optimized bert pretraining approach.
arXiv preprint arXiv:1907.11692
(2019)
\end{botherref}
\endbibitem

\bibitem[\protect\citeauthoryear{Raffel et~al.}{2020}]{raffel2020exploring}
\begin{barticle}
\bauthor{\bsnm{Raffel}, \binits{C.}},
\bauthor{\bsnm{Shazeer}, \binits{N.}},
\bauthor{\bsnm{Roberts}, \binits{A.}},
\bauthor{\bsnm{Lee}, \binits{K.}},
\bauthor{\bsnm{Narang}, \binits{S.}},
\bauthor{\bsnm{Matena}, \binits{M.}},
\bauthor{\bsnm{Zhou}, \binits{Y.}},
\bauthor{\bsnm{Li}, \binits{W.}},
\bauthor{\bsnm{Liu}, \binits{P.J.}}:
\batitle{Exploring the limits of transfer learning with a unified text-to-text transformer}.
\bjtitle{The Journal of Machine Learning Research}
\bvolume{21}(\bissue{1}),
\bfpage{5485}--\blpage{5551}
(\byear{2020})
\end{barticle}
\endbibitem

\bibitem[\protect\citeauthoryear{Efthimion et~al.}{2018}]{efthimion2018supervised}
\begin{barticle}
\bauthor{\bsnm{Efthimion}, \binits{P.G.}},
\bauthor{\bsnm{Payne}, \binits{S.}},
\bauthor{\bsnm{Proferes}, \binits{N.}}:
\batitle{Supervised machine learning bot detection techniques to identify social twitter bots}.
\bjtitle{SMU Data Science Review}
\bvolume{1}(\bissue{2}),
\bfpage{5}
(\byear{2018})
\end{barticle}
\endbibitem

\bibitem[\protect\citeauthoryear{Kantepe and Ganiz}{2017}]{kantepe2017preprocessing}
\begin{bchapter}
\bauthor{\bsnm{Kantepe}, \binits{M.}},
\bauthor{\bsnm{Ganiz}, \binits{M.C.}}:
\bctitle{Preprocessing framework for twitter bot detection}.
In: \bbtitle{2017 International Conference on Computer Science and Engineering (ubmk)},
pp. \bfpage{630}--\blpage{634}
(\byear{2017}).
\bcomment{IEEE}
\end{bchapter}
\endbibitem

\bibitem[\protect\citeauthoryear{Miller et~al.}{2014}]{miller2014twitter}
\begin{barticle}
\bauthor{\bsnm{Miller}, \binits{Z.}},
\bauthor{\bsnm{Dickinson}, \binits{B.}},
\bauthor{\bsnm{Deitrick}, \binits{W.}},
\bauthor{\bsnm{Hu}, \binits{W.}},
\bauthor{\bsnm{Wang}, \binits{A.H.}}:
\batitle{Twitter spammer detection using data stream clustering}.
\bjtitle{Information Sciences}
\bvolume{260},
\bfpage{64}--\blpage{73}
(\byear{2014})
\end{barticle}
\endbibitem

\bibitem[\protect\citeauthoryear{Varol et~al.}{2017}]{varol2017online}
\begin{bchapter}
\bauthor{\bsnm{Varol}, \binits{O.}},
\bauthor{\bsnm{Ferrara}, \binits{E.}},
\bauthor{\bsnm{Davis}, \binits{C.}},
\bauthor{\bsnm{Menczer}, \binits{F.}},
\bauthor{\bsnm{Flammini}, \binits{A.}}:
\bctitle{Online human-bot interactions: Detection, estimation, and characterization}.
In: \bbtitle{Proceedings of the International AAAI Conference on Web and Social Media},
vol. \bseriesno{11},
pp. \bfpage{280}--\blpage{289}
(\byear{2017})
\end{bchapter}
\endbibitem

\bibitem[\protect\citeauthoryear{Kouvela et~al.}{2020}]{kouvela2020bot}
\begin{bchapter}
\bauthor{\bsnm{Kouvela}, \binits{M.}},
\bauthor{\bsnm{Dimitriadis}, \binits{I.}},
\bauthor{\bsnm{Vakali}, \binits{A.}}:
\bctitle{Bot-detective: An explainable twitter bot detection service with crowdsourcing functionalities}.
In: \bbtitle{Proceedings of the 12th International Conference on Management of Digital EcoSystems},
pp. \bfpage{55}--\blpage{63}
(\byear{2020})
\end{bchapter}
\endbibitem

\bibitem[\protect\citeauthoryear{Ferreira Dos~Santos et~al.}{2019}]{ferreira2019uncovering}
\begin{bchapter}
\bauthor{\bsnm{Ferreira Dos~Santos}, \binits{E.}},
\bauthor{\bsnm{Carvalho}, \binits{D.}},
\bauthor{\bsnm{Ruback}, \binits{L.}},
\bauthor{\bsnm{Oliveira}, \binits{J.}}:
\bctitle{Uncovering social media bots: a transparency-focused approach}.
In: \bbtitle{Companion Proceedings of The 2019 World Wide Web Conference},
pp. \bfpage{545}--\blpage{552}
(\byear{2019})
\end{bchapter}
\endbibitem

\bibitem[\protect\citeauthoryear{Echeverría et~al.}{2018}]{echeverri2018lobo}
\begin{bchapter}
\bauthor{\bsnm{Echeverría}, \binits{J.}},
\bauthor{\bsnm{De~Cristofaro}, \binits{E.}},
\bauthor{\bsnm{Kourtellis}, \binits{N.}},
\bauthor{\bsnm{Leontiadis}, \binits{I.}},
\bauthor{\bsnm{Stringhini}, \binits{G.}},
\bauthor{\bsnm{Zhou}, \binits{S.}}:
\bctitle{Lobo: Evaluation of generalization deficiencies in twitter bot classifiers}.
In: \bbtitle{Proceedings of the 34th Annual Computer Security Applications Conference},
pp. \bfpage{137}--\blpage{146}
(\byear{2018})
\end{bchapter}
\endbibitem

\bibitem[\protect\citeauthoryear{Fazil et~al.}{2021}]{fazil2021deepsbd}
\begin{barticle}
\bauthor{\bsnm{Fazil}, \binits{M.}},
\bauthor{\bsnm{Sah}, \binits{A.K.}},
\bauthor{\bsnm{Abulaish}, \binits{M.}}:
\batitle{Deepsbd: a deep neural network model with attention mechanism for socialbot detection}.
\bjtitle{IEEE Transactions on Information Forensics and Security}
\bvolume{16},
\bfpage{4211}--\blpage{4223}
(\byear{2021})
\end{barticle}
\endbibitem

\bibitem[\protect\citeauthoryear{Wu et~al.}{2023}]{wu2023bottrinet}
\begin{bchapter}
\bauthor{\bsnm{Wu}, \binits{J.}},
\bauthor{\bsnm{Ye}, \binits{X.}},
\bauthor{\bsnm{Man}, \binits{Y.}}:
\bctitle{Bottrinet: A unified and efficient embedding for social bots detection via metric learning}.
In: \bbtitle{2023 11th International Symposium on Digital Forensics and Security (ISDFS)},
pp. \bfpage{1}--\blpage{6}
(\byear{2023}).
\bcomment{IEEE}
\end{bchapter}
\endbibitem

\bibitem[\protect\citeauthoryear{Heidari et~al.}{2022}]{heidari2022online}
\begin{botherref}
\oauthor{\bsnm{Heidari}, \binits{M.}},
\oauthor{\bsnm{Jones~Jr}, \binits{J.H.}},
\oauthor{\bsnm{Uzuner}, \binits{O.}}:
Online user profiling to detect social bots on twitter.
arXiv preprint arXiv:2203.05966
(2022)
\end{botherref}
\endbibitem

\bibitem[\protect\citeauthoryear{Feng et~al.}{2022}]{feng2022twibot}
\begin{barticle}
\bauthor{\bsnm{Feng}, \binits{S.}},
\bauthor{\bsnm{Tan}, \binits{Z.}},
\bauthor{\bsnm{Wan}, \binits{H.}},
\bauthor{\bsnm{Wang}, \binits{N.}},
\bauthor{\bsnm{Chen}, \binits{Z.}},
\bauthor{\bsnm{Zhang}, \binits{B.}},
\bauthor{\bsnm{Zheng}, \binits{Q.}},
\bauthor{\bsnm{Zhang}, \binits{W.}},
\bauthor{\bsnm{Lei}, \binits{Z.}},
\bauthor{\bsnm{Yang}, \binits{S.}}, \betal:
\batitle{Twibot-22: Towards graph-based twitter bot detection}.
\bjtitle{Advances in Neural Information Processing Systems}
\bvolume{35},
\bfpage{35254}--\blpage{35269}
(\byear{2022})
\end{barticle}
\endbibitem

\bibitem[\protect\citeauthoryear{Moghaddam and Abbaspour}{2022}]{moghaddam2022friendship}
\begin{barticle}
\bauthor{\bsnm{Moghaddam}, \binits{S.H.}},
\bauthor{\bsnm{Abbaspour}, \binits{M.}}:
\batitle{Friendship preference: Scalable and robust category of features for social bot detection}.
\bjtitle{IEEE Transactions on Dependable and Secure Computing}
\bvolume{20}(\bissue{2}),
\bfpage{1516}--\blpage{1528}
(\byear{2022})
\end{barticle}
\endbibitem

\bibitem[\protect\citeauthoryear{Ali~Alhosseini et~al.}{2019}]{ali2019detect}
\begin{bchapter}
\bauthor{\bsnm{Ali~Alhosseini}, \binits{S.}},
\bauthor{\bsnm{Bin~Tareaf}, \binits{R.}},
\bauthor{\bsnm{Najafi}, \binits{P.}},
\bauthor{\bsnm{Meinel}, \binits{C.}}:
\bctitle{Detect me if you can: Spam bot detection using inductive representation learning}.
In: \bbtitle{Companion Proceedings of the 2019 World Wide Web Conference},
pp. \bfpage{148}--\blpage{153}
(\byear{2019})
\end{bchapter}
\endbibitem

\bibitem[\protect\citeauthoryear{Knauth}{2019}]{knauth2019language}
\begin{bchapter}
\bauthor{\bsnm{Knauth}, \binits{J.}}:
\bctitle{Language-agnostic twitter-bot detection}.
In: \bbtitle{Proceedings of the International Conference on Recent Advances in Natural Language Processing (RANLP 2019)},
pp. \bfpage{550}--\blpage{558}
(\byear{2019})
\end{bchapter}
\endbibitem

\bibitem[\protect\citeauthoryear{Beskow and Carley}{2020}]{beskow2020you}
\begin{botherref}
\oauthor{\bsnm{Beskow}, \binits{D.M.}},
\oauthor{\bsnm{Carley}, \binits{K.M.}}:
You are known by your friends: Leveraging network metrics for bot detection in twitter.
Open Source Intelligence and Cyber Crime: Social Media Analytics,
53--88
(2020)
\end{botherref}
\endbibitem

\bibitem[\protect\citeauthoryear{Feng et~al.}{2021}]{feng2021satar}
\begin{bchapter}
\bauthor{\bsnm{Feng}, \binits{S.}},
\bauthor{\bsnm{Wan}, \binits{H.}},
\bauthor{\bsnm{Wang}, \binits{N.}},
\bauthor{\bsnm{Li}, \binits{J.}},
\bauthor{\bsnm{Luo}, \binits{M.}}:
\bctitle{Satar: A self-supervised approach to twitter account representation learning and its application in bot detection}.
In: \bbtitle{Proceedings of the 30th ACM International Conference on Information \& Knowledge Management},
pp. \bfpage{3808}--\blpage{3817}
(\byear{2021})
\end{bchapter}
\endbibitem

\bibitem[\protect\citeauthoryear{Yang et~al.}{2022}]{yang2022botometer}
\begin{barticle}
\bauthor{\bsnm{Yang}, \binits{K.-C.}},
\bauthor{\bsnm{Ferrara}, \binits{E.}},
\bauthor{\bsnm{Menczer}, \binits{F.}}:
\batitle{Botometer 101: Social bot practicum for computational social scientists}.
\bjtitle{Journal of Computational Social Science}
\bvolume{5}(\bissue{2}),
\bfpage{1511}--\blpage{1528}
(\byear{2022})
\end{barticle}
\endbibitem

\bibitem[\protect\citeauthoryear{Rodr{\'\i}guez-Ruiz et~al.}{2020}]{rodriguez2020one}
\begin{barticle}
\bauthor{\bsnm{Rodr{\'\i}guez-Ruiz}, \binits{J.}},
\bauthor{\bsnm{Mata-S{\'a}nchez}, \binits{J.I.}},
\bauthor{\bsnm{Monroy}, \binits{R.}},
\bauthor{\bsnm{Loyola-Gonzalez}, \binits{O.}},
\bauthor{\bsnm{L{\'o}pez-Cuevas}, \binits{A.}}:
\batitle{A one-class classification approach for bot detection on twitter}.
\bjtitle{Computers \& Security}
\bvolume{91},
\bfpage{101715}
(\byear{2020})
\end{barticle}
\endbibitem

\bibitem[\protect\citeauthoryear{Magelinski et~al.}{2020}]{magelinski2020graph}
\begin{bchapter}
\bauthor{\bsnm{Magelinski}, \binits{T.}},
\bauthor{\bsnm{Beskow}, \binits{D.}},
\bauthor{\bsnm{Carley}, \binits{K.M.}}:
\bctitle{Graph-hist: Graph classification from latent feature histograms with application to bot detection}.
In: \bbtitle{Proceedings of the AAAI Conference on Artificial Intelligence},
vol. \bseriesno{34},
pp. \bfpage{5134}--\blpage{5141}
(\byear{2020})
\end{bchapter}
\endbibitem

\bibitem[\protect\citeauthoryear{Dehghan et~al.}{2023}]{dehghan2023detecting}
\begin{barticle}
\bauthor{\bsnm{Dehghan}, \binits{A.}},
\bauthor{\bsnm{Siuta}, \binits{K.}},
\bauthor{\bsnm{Skorupka}, \binits{A.}},
\bauthor{\bsnm{Dubey}, \binits{A.}},
\bauthor{\bsnm{Betlen}, \binits{A.}},
\bauthor{\bsnm{Miller}, \binits{D.}},
\bauthor{\bsnm{Xu}, \binits{W.}},
\bauthor{\bsnm{Kami{\'n}ski}, \binits{B.}},
\bauthor{\bsnm{Pra{\l}at}, \binits{P.}}:
\batitle{Detecting bots in social-networks using node and structural embeddings}.
\bjtitle{Journal of Big Data}
\bvolume{10}(\bissue{1}),
\bfpage{119}
(\byear{2023})
\end{barticle}
\endbibitem

\bibitem[\protect\citeauthoryear{Kipf and Welling}{2016}]{kipf2016semi}
\begin{botherref}
\oauthor{\bsnm{Kipf}, \binits{T.N.}},
\oauthor{\bsnm{Welling}, \binits{M.}}:
Semi-supervised classification with graph convolutional networks.
arXiv preprint arXiv:1609.02907
(2016)
\end{botherref}
\endbibitem

\bibitem[\protect\citeauthoryear{Veli{\v{c}}kovi{\'c} et~al.}{2017}]{velivckovic2017graph}
\begin{botherref}
\oauthor{\bsnm{Veli{\v{c}}kovi{\'c}}, \binits{P.}},
\oauthor{\bsnm{Cucurull}, \binits{G.}},
\oauthor{\bsnm{Casanova}, \binits{A.}},
\oauthor{\bsnm{Romero}, \binits{A.}},
\oauthor{\bsnm{Lio}, \binits{P.}},
\oauthor{\bsnm{Bengio}, \binits{Y.}}:
Graph attention networks.
arXiv preprint arXiv:1710.10903
(2017)
\end{botherref}
\endbibitem

\bibitem[\protect\citeauthoryear{Hu et~al.}{2020}]{hu2020heterogeneous}
\begin{bchapter}
\bauthor{\bsnm{Hu}, \binits{Z.}},
\bauthor{\bsnm{Dong}, \binits{Y.}},
\bauthor{\bsnm{Wang}, \binits{K.}},
\bauthor{\bsnm{Sun}, \binits{Y.}}:
\bctitle{Heterogeneous graph transformer}.
In: \bbtitle{Proceedings of the Web Conference 2020},
pp. \bfpage{2704}--\blpage{2710}
(\byear{2020})
\end{bchapter}
\endbibitem

\bibitem[\protect\citeauthoryear{Lv et~al.}{2021}]{lv2021we}
\begin{bchapter}
\bauthor{\bsnm{Lv}, \binits{Q.}},
\bauthor{\bsnm{Ding}, \binits{M.}},
\bauthor{\bsnm{Liu}, \binits{Q.}},
\bauthor{\bsnm{Chen}, \binits{Y.}},
\bauthor{\bsnm{Feng}, \binits{W.}},
\bauthor{\bsnm{He}, \binits{S.}},
\bauthor{\bsnm{Zhou}, \binits{C.}},
\bauthor{\bsnm{Jiang}, \binits{J.}},
\bauthor{\bsnm{Dong}, \binits{Y.}},
\bauthor{\bsnm{Tang}, \binits{J.}}:
\bctitle{Are we really making much progress? revisiting, benchmarking and refining heterogeneous graph neural networks}.
In: \bbtitle{Proceedings of the 27th ACM SIGKDD Conference on Knowledge Discovery \& Data Mining},
pp. \bfpage{1150}--\blpage{1160}
(\byear{2021})
\end{bchapter}
\endbibitem

\bibitem[\protect\citeauthoryear{Feng et~al.}{2021}]{feng2021botrgcn}
\begin{bchapter}
\bauthor{\bsnm{Feng}, \binits{S.}},
\bauthor{\bsnm{Wan}, \binits{H.}},
\bauthor{\bsnm{Wang}, \binits{N.}},
\bauthor{\bsnm{Luo}, \binits{M.}}:
\bctitle{Botrgcn: Twitter bot detection with relational graph convolutional networks}.
In: \bbtitle{Proceedings of the 2021 IEEE/ACM International Conference on Advances in Social Networks Analysis and Mining},
pp. \bfpage{236}--\blpage{239}
(\byear{2021})
\end{bchapter}
\endbibitem

\bibitem[\protect\citeauthoryear{Feng et~al.}{2022}]{feng2022heterogeneity}
\begin{bchapter}
\bauthor{\bsnm{Feng}, \binits{S.}},
\bauthor{\bsnm{Tan}, \binits{Z.}},
\bauthor{\bsnm{Li}, \binits{R.}},
\bauthor{\bsnm{Luo}, \binits{M.}}:
\bctitle{Heterogeneity-aware twitter bot detection with relational graph transformers}.
In: \bbtitle{Proceedings of the AAAI Conference on Artificial Intelligence},
vol. \bseriesno{36},
pp. \bfpage{3977}--\blpage{3985}
(\byear{2022})
\end{bchapter}
\endbibitem

\end{thebibliography}

\end{document}